\RequirePackage
		[
			abort,
			l2tabu,
			orthodox,
		]
		{nag}
\RequirePackage{import}

\makeatletter
\def\ece#1#2{\expandafter#1\csname#2\endcsname}%
\def\setproperty#1#2#3{\ece\protected@edef{#1@p#2}{\unexpanded{#3}}}%
\def\getproperty#1#2{%
  \expandafter\ifx\csname#1@p#2\endcsname\relax
  \else \csname#1@p#2\endcsname
  \fi
}%
\def\ifthenelsepropertydefined#1#2#3#4{%
  \expandafter\ifx\csname#1@p#2\endcsname\relax
  #4
  \else#3
  \fi
}%
\def\ifpropertydefined#1#2#3{%
    \ifthenelsepropertydefined{#1}{#2}{#3}{}
}
\def\ifpropertyundefined#1#2#3{%
    \ifthenelsepropertydefined{#1}{#2}{}{#3}
}
\def\raiseifpropertyundefined#1#2#3{%
    \ifpropertyundefined{#2}{#3}{\PackageError{#1}{Property #2 #3 needs to be defined. Put \@backslashchar setproperty{#2}{#3} to your settings file}{Grep for your property :)}}
}
\def\setpropertyifundefined#1#2#3{%
    \ifpropertyundefined{#1}{#2}{%
        \setproperty{#1}{#2}{#3}
    }{}
}
\def\setpropertyifundefinedwithusageinfo#1#2#3{%
    \setpropertyifundefined{#1}{#2}{TODO>> Put \backslash{}setproperty\{#1\}\{#2\}\{<your value here, e.g. ``#3``>\} into content/settings.tex <<}
}
\def\ifthenelseproperty#1#2#3#4{\providetoggle{#1@p#2}\settoggle{#1@p#2}{\getproperty{#1}{#2}}\iftoggle{#1@p#2}{#3}{#4}}
\def\ifthenelsepropertyequal#1#2#3#4#5{\ifthenelse{\equal{\getproperty{#1}{#2}}{#3}}{#4}{#5}}

\makeatother

\makeatletter
\newcommand\RequirePackageWithOption[2][]{%
    \@ifpackageloaded{#2}{%
        \PassOptionsToPackage{#1}{#2}
    }{%
        \RequirePackage[#1]{#2}
    }
}

\newcommand\ImportDefault[1]{
    \IfFileExists{content/#1.tex}{
        \typeout{----- LOAD OWN #1 -----}
        \import{content/}{#1}
    }{
        \typeout{----- LOAD DEFAULT #1 -----}
        \ifthenelsepropertydefined{default}{#1}{%
            \import{templates/\getproperty{default}{#1}/}{#1}
        }{%
            \import{templates/Default/}{#1}
        }
    }
}
\newcommand\ImportPackage[1]{
    \IfFileExists{content/packages/#1.tex}{
        \input{content/packages/#1}
    }{%
        \input{preamble/packages/#1}
    }
}
\makeatother

\documentclass
			[
				a4paper,
				DIV=calc,
				BCOR=8mm,
				headinclude,
				bibliography=totoc,
				listof=totoc,
				index=totoc,
				british,
				twoside,
				10pt,
				openright,
				captions=tableheading,
				chapterprefix,
                twocolumn,
			]
			{scrartcl}

\newif\ifKOMA
\KOMAtrue

\setproperty{default}{titlepage}{Paper}
\setproperty{default}{appendix}{Paper}
\setproperty{default}{content}{Paper}

\setproperty{compilation}{appendix}{true}
\setproperty{compilation}{lof}{false}
\setproperty{compilation}{lot}{false}
\setproperty{compilation}{toc}{false}

\setproperty{compilation}{complementglyphs}{false}

\setproperty{compilation}{final}{true}
\setproperty{compilation}{comment}{footnote}

\usepackage{authblk}
\setproperty{coauthors}{W7Xteaminclude}{True}
\usepackage[]{algorithm2e}

\usepackage{multirow}

\usepackage[columnwise,switch]{lineno}

\usepackage
		[
			ngerman,
			main=british
		]
		{babel}

\RequirePackageWithOption
		[
			log-declarations=false 
		]
		{xparse}

\usepackage{xpatch}
\usepackage{shellesc}

\setpropertyifundefined{color}{main}{black}
\setpropertyifundefined{color}{secondary}{black}

\setpropertyifundefinedwithusageinfo{author}{firstname}{J.R.R.}
\setpropertyifundefinedwithusageinfo{author}{familyname}{Tolkien}
\setpropertyifundefinedwithusageinfo{author}{acadtitle}{Prof.~Dr.~h.\,c.}
\setpropertyifundefinedwithusageinfo{author}{addressstreet}{Way of no return 77}
\setpropertyifundefinedwithusageinfo{author}{addresscity}{12345 Hobbingen}
\AtBeginDocument{%
    \setpropertyifundefined{author}{address}{\getproperty{author}{addressstreet}\\\getproperty{author}{addresscity}}
}
\setpropertyifundefinedwithusageinfo{author}{dateofbirth}{01.01.1970}
\setpropertyifundefinedwithusageinfo{author}{placeofbirth}{Helms Klamm, Rohan}
\setpropertyifundefinedwithusageinfo{author}{nationality}{german}
\setpropertyifundefinedwithusageinfo{author}{mobile}{+47 173 123456}
\setpropertyifundefinedwithusageinfo{author}{phone}{+53 001 123456}
\setpropertyifundefinedwithusageinfo{author}{faxnr}{+53 001 123456}
\setpropertyifundefinedwithusageinfo{author}{email}{minna@barnhelm.com}
\setpropertyifundefinedwithusageinfo{author}{orcid}{1234567}
\setpropertyifundefinedwithusageinfo{author}{signature}{\backslash{}igraph\{path to your signature here>\}}
\setpropertyifundefined{author}{affiliationindices}{1}

\setpropertyifundefinedwithusageinfo{affiliations}{1}{Max-Planck-Institute for Plasma Physics, 17491 Greifswald, Germany}

\setpropertyifundefined{document}{date}{\today}
\setpropertyifundefinedwithusageinfo{document}{title}{The Lord of the Rings}
\setpropertyifundefinedwithusageinfo{document}{titlehead}{There and back again}
\setpropertyifundefinedwithusageinfo{document}{subject}{Novel}
\setpropertyifundefinedwithusageinfo{document}{keywords}{ring,mordor}
\setpropertyifundefinedwithusageinfo{document}{location}{Rivendale}
\setpropertyifundefinedwithusageinfo{document}{publishers}{Elrond Verlag}
\setpropertyifundefinedwithusageinfo{document}{dedication}{For my precious}
\setpropertyifundefinedwithusageinfo{document}{logos}{\backslash{}igraph\{<path to your logo here>\}}
\setpropertyifundefinedwithusageinfo{document}{type}{book}
\setpropertyifundefinedwithusageinfo{document}{doi}{145125.1234150}

\setpropertyifundefined{compilation}{externalfolder}{tikz/}
\setpropertyifundefined{compilation}{externalize}{false}
\setpropertyifundefined{compilation}{pdfa}{false}

\setpropertyifundefined{compilation}{final}{false}

\AtEndPreamble{%
	\ifthenelseproperty{compilation}{pdfa}{\setpropertyifundefined{compilation}{externalize}{false}}{}
}

\setpropertyifundefined{compilation}{minionfonts}{false}
\setpropertyifundefined{compilation}{libertinusfonts}{false}
\setpropertyifundefined{compilation}{complementglyphs}{true}
\setpropertyifundefined{compilation}{fancychapterheadings}{false}
\setpropertyifundefined{compilation}{microtype}{false}
\setpropertyifundefined{compilation}{fontspec}{false}

\setpropertyifundefinedwithusageinfo{letter}{recipientname}{Sauron Mustermann}
\setpropertyifundefinedwithusageinfo{letter}{recipientaddress}{Ring Institute\\Tower 1\\12345 Mordor}
\setpropertyifundefinedwithusageinfo{letter}{subject}{Application for free Ork position}
\setpropertyifundefinedwithusageinfo{letter}{opening}{Dear Madam or Sir,}
\setpropertyifundefined{letter}{closing}{Sincerely yours,}

\setpropertyifundefined{thesis}{faculty}{FACULTY}
\setpropertyifundefined{thesis}{degree}{DEGREE}
\setpropertyifundefined{thesis}{defensedate}{DEFENSEDATE}
\setpropertyifundefined{thesis}{submissionyear}{YEAR OF SUBMISSION}

\makeatletter
\@ifundefined{ifKOMA}{%
    \newif\ifKOMA%
	\KOMAfalse%
}{}

\@ifclassloaded{book}{%
    \setpropertyifundefined{compilation}{titlepage}{true}
    \setpropertyifundefined{compilation}{abstract}{true}
    \setpropertyifundefined{compilation}{toc}{true}
    \setpropertyifundefined{compilation}{bibliography}{true}
    \setpropertyifundefined{compilation}{lof}{true}
    \setpropertyifundefined{compilation}{lot}{true}
    \setpropertyifundefined{compilation}{appendix}{true}
    \setpropertyifundefined{compilation}{affidavit}{true}
    \setpropertyifundefined{compilation}{clsdefineschapter}{true}
}{}
\@ifclassloaded{report}{%
    \setpropertyifundefined{compilation}{titlepage}{true}
    \setpropertyifundefined{compilation}{abstract}{true}
    \setpropertyifundefined{compilation}{toc}{true}
    \setpropertyifundefined{compilation}{bibliography}{true}
    \setpropertyifundefined{compilation}{lof}{true}
    \setpropertyifundefined{compilation}{lot}{true}
    \setpropertyifundefined{compilation}{appendix}{true}
    \setpropertyifundefined{compilation}{affidavit}{true}
    \setpropertyifundefined{compilation}{clsdefineschapter}{true}
}{}
\@ifclassloaded{article}{%
    \setpropertyifundefined{compilation}{titlepage}{true}
    \setpropertyifundefined{compilation}{abstract}{true}
    \setpropertyifundefined{compilation}{toc}{true}
    \setpropertyifundefined{compilation}{bibliography}{true}
}{}
\@ifclassloaded{scrbook}{%
    \setpropertyifundefined{compilation}{frontmatter}{true}
    \setpropertyifundefined{compilation}{titlepage}{true}
    \setpropertyifundefined{compilation}{abstract}{true}
    \setpropertyifundefined{compilation}{toc}{true}
    \setpropertyifundefined{compilation}{bibliography}{true}
    \setpropertyifundefined{compilation}{lof}{true}
    \setpropertyifundefined{compilation}{lot}{true}
    \setpropertyifundefined{compilation}{appendix}{true}
    \setpropertyifundefined{compilation}{affidavit}{true}
    \setpropertyifundefined{compilation}{clsdefineschapter}{true}
}{}
\@ifclassloaded{scrreprt}{%
    \setpropertyifundefined{compilation}{titlepage}{true}
    \setpropertyifundefined{compilation}{abstract}{true}
    \setpropertyifundefined{compilation}{toc}{true}
    \setpropertyifundefined{compilation}{bibliography}{true}
    \setpropertyifundefined{compilation}{lof}{true}
    \setpropertyifundefined{compilation}{lot}{true}
    \setpropertyifundefined{compilation}{appendix}{true}
    \setpropertyifundefined{compilation}{affidavit}{true}
    \setpropertyifundefined{compilation}{clsdefineschapter}{true}
}{}
\@ifclassloaded{scrartcl}{%
    \setpropertyifundefined{compilation}{titlepage}{true}
    \setpropertyifundefined{compilation}{abstract}{true}
    \setpropertyifundefined{compilation}{toc}{true}
    \setpropertyifundefined{compilation}{bibliography}{true}
}{}
\@ifclassloaded{moderncv}{%
    \setpropertyifundefined{compilation}{titlepage}{false}
    \setpropertyifundefined{compilation}{abstract}{false}
    \setpropertyifundefined{compilation}{toc}{false}
    \setpropertyifundefined{compilation}{bibliography}{false}
    \setpropertyifundefined{compilation}{glossaries}{false}
}{}
\setpropertyifundefined{compilation}{frontmatter}{false}
\setpropertyifundefined{compilation}{titlepage}{false}
\setpropertyifundefined{compilation}{abstract}{false}
\setpropertyifundefined{compilation}{toc}{false}
\setpropertyifundefined{compilation}{los}{false}
\setpropertyifundefined{compilation}{bibliography}{false}
\setpropertyifundefined{compilation}{lof}{false}
\setpropertyifundefined{compilation}{lot}{false}
\setpropertyifundefined{compilation}{lol}{false}
\setpropertyifundefined{compilation}{appendix}{false}
\setpropertyifundefined{compilation}{listofpublications}{false}
\setpropertyifundefined{compilation}{acknowledgement}{false}
\setpropertyifundefined{compilation}{affidavit}{false}
\setpropertyifundefined{compilation}{clsdefineschapter}{false}
\setpropertyifundefined{compilation}{acronyms}{false}
\setpropertyifundefined{compilation}{glossaries}{true}

\makeatother

\usepackage{ifluatex}

\ifluatex

	\newcommand{\FirstWord}[1]{\luaexec{tex.print(#1)}}
	
\else
\fi

\RequirePackageWithOption[
			fleqn 
		]
		{amsmath}

\usepackage{amsthm}
\usepackage{calc}
\ifluatex
\else
    \usepackage
		[
		]
		{amsfonts}

    \usepackage
		[
		]
		{amssymb}

\fi

\ifluatex
	\usepackage
		[
			quiet
		]
		{fontspec}

\setproperty{compilation}{fontspec}{true}

	\usepackage
		[
			math-style=ISO,
			bold-style=ISO,
			nabla=upright,
		]
		{unicode-math}
	\AtEndPreamble{
        \ifthenelseproperty{compilation}{minionfonts}{
			\setmathfont{Cambria Math}
			\ifthenelseproperty{compilation}{complementglyphs}{%
				\setmathfont[range=\mathup/{num,latin,Latin,greek,Greek}]{Minion Pro}
				\setmathfont[range=\mathbfup/{num,latin,Latin,greek,Greek}]{MinionPro-Bold}
				\setmathfont[range=\mathit/{num,latin,Latin,greek,Greek}]{MinionPro-It}
				\setmathfont[range=\mathbfit/{num,latin,Latin,greek,Greek}]{MinionPro-BoldIt}
				\setmathfont[range={"0025,"002B,"002C,"002D,"002E,"2212,"2248,"007E,`±,`×,`·,`÷,`¬,`∂,`∆,`∕,`∞,`⌐,`<,`>,`=,`≥,`≤,`−}]{Minion Pro}
				\setmathfont[range={"2211,"221A}]{Latin Modern Math}
			}{}
			\renewcommand{\mathcal}[1]{{\textit{\addfontfeatures{Contextuals=Swash}#1}}}
			\newfontface{\MinionProSwash}{MinionPro-It}
					[
						Contextuals=Swash,
						Ligatures={TeX,Common,Rare,Historic,Contextual,Required}
					]
			\typeout{----- USING MINION FONTS -----}
		}{
		    \ifthenelseproperty{compilation}{libertinusfonts}{
		    	\setmainfont{Libertinus Serif}
		[
			BoldFont          = {* Bold},
  			ItalicFont        = {* Italic},
  			BoldItalicFont    = {* Bold Italic},
			SmallCapsFeatures = {Renderer=Basic},
			Ligatures         = {TeX,Common},
			Scale             = MatchLowercase,
			RawFeature=-lnum;+pnum;+onum;
		]

		    	\setsansfont{Libertinus Sans}
		[
			BoldFont          = {* Bold},
  			ItalicFont        = {* Italic},
  			BoldItalicFont    = {* -Bold Italic},
			SmallCapsFeatures = {Renderer=Basic},
			Ligatures         = {TeX,Common},
			Scale             = MatchLowercase,
			Numbers           = {Lining,Monospaced},
		]
		    	\setmonofont{Libertinus Mono}
		[
			BoldFont          = {* Bold},
  			ItalicFont        = {* Italic},
  			BoldItalicFont    = {* Bold Italic},
			SmallCapsFeatures = {Renderer=Basic},
			Ligatures         = {Common},
			Scale             = MatchLowercase,
			Numbers           = {Lining,Monospaced},
		]
		    	\setmathfont{Cambria Math}
				\ifthenelseproperty{compilation}{complementglyphs}{%
					\setmathfont[range=\mathup/{num,latin,Latin,greek,Greek}]{LibertinusSerif}
					\setmathfont[range=\mathbfup/{num,latin,Latin,greek,Greek}]{LibertinusSerif-Bold}
					\setmathfont[range=\mathit/{num,latin,Latin,greek,Greek}]{LibertinusSerif-Italic}
					\setmathfont[range=\mathbfit/{num,latin,Latin,greek,Greek}]{LibertinusSerif-BoldItalic}
					\setmathfont[range={"0025,"002B,"002C,"002D,"002E,"2212,"2248,"007E,"2208,"221D,"2A85,`±,`×,`·,`÷,`¬,`∂,`∆,`∕,`∞,`⌐,`<,`>,`=,`≥,`≤,`−,`;,}]{Libertinus Math}
					\setmathfont[range={"2211,"221A}]{Latin Modern Math}
					\setmathfont[range=\mathcal/{num,latin,Latin,greek,Greek},Contextuals=Swash]{Cambria Math}
				}{}
		    	\newfontface{\LibertinusSwash}{LibertinusSerif-Italic}
		    			[
		    				Contextuals=Swash,
		    				Ligatures={TeX,Common,Rare,Historic,Contextual,Required}
		    			]
		    	\newfontface{\LibertinusInitials}{LibertinusSerif-Initials}
		    	\typeout{----- USING LIBERTINUS FONTS -----}
		    }{
                \setmathfont{latinmodern-math.otf}
				\ifthenelseproperty{compilation}{complementglyphs}{%
                    \setmathfont[range=\setminus]{Asana Math}
				}{}
            }
        }
	}
\else
	\usepackage
		[
			T1
		]
		{fontenc}
	\usepackage
		[
			utf8
		]
		{inputenc}
	\usepackage{lmodern}
\fi

\usepackage{fontawesome5} 

\ifluatex
    \usepackage[a-3b,pdf17]{pdfx}
\usepackage{xmpincl}

\ifluatex
    \pdfvariable omitcidset=1
\else
\fi

\ifluatex
	\newcommand{\replaceNBSP}[1]{\luadirect{s,_ = string.gsub("\luatexluaescapestring{#1}", "\luatexluaescapestring{~}", " "); tex.print(-2,s)}}
\else
	\newcommand{\replaceNBSP}[1]{} 
\fi

\AtBeginDocument{
	\newwrite\metadatafile
	\immediate\openout\metadatafile=\jobname.xmpdata
	\immediate\write\metadatafile{\unexpanded{\Title}{\expanded{\replaceNBSP{\getproperty{document}{title}}}}}
	\immediate\write\metadatafile{\unexpanded{\Author}{\expanded{\getproperty{author}{firstname} \getproperty{author}{familyname}}}}
	\immediate\write\metadatafile{\unexpanded{\Subject}{\expanded{\getproperty{document}{subject}}}}
	\immediate\write\metadatafile{\unexpanded{\Keywords}{\getproperty{document}{keywords}}}
	\immediate\write\metadatafile{\unexpanded{\PublicationType}{\expanded{\getproperty{document}{type}}}}
	\immediate\write\metadatafile{\unexpanded{\Doi}{\expanded{\getproperty{document}{doi}}}}
	\immediate\closeout\metadatafile
}

\else
\fi

\usepackage
		[
			locale=UK,
			separate-uncertainty,
			retain-zero-exponent=true,
			binary-units, 
		]
		{siunitx}

\DeclareSIUnit{\arbitraryunit}{a.\,u.}

\DeclareSIUnit{\permille}{‰}

\DeclareSIUnit{\sample}{S}

\DeclareSIPrefix{\Femto}{f\kern0.1ex}{-15}

\DeclareSIUnit{\pb}{\pico\barn}
\DeclareSIUnit{\fb}{\femto\barn}
\DeclareSIUnit{\nb}{\nano\barn}

\usepackage{chemfig}

\usepackage{chemmacros}

\usechemmodule{all}

\renewtagform{reaction}[]{(}{)}

\AtBeginEnvironment{reactions}{}
\AtEndEnvironment{reactions}{}

\usepackage{chemformula}

\setchemformula
		{
			arrow-style={
							>=stealth',
							line cap=round,
							thick
						},
		}

\ifthenelseproperty{compilation}{minionfonts}{
	\setchemformula
			{
				arrow-style={
								>=stealth',
								line cap=round,
								thick
							},
				font-spec={[Numbers={Proportional,Lining}]MinionPro}
			}
}{}

\ifthenelseproperty{compilation}{libertinusfonts}{
	\setchemformula
			{
				arrow-style={
								>=stealth',
								line cap=round,
								thick
							},
				font-spec={[Numbers={Proportional,Lining}]LibertinusSerif}
			}
}{}

\RenewChemArrow{->}{\draw[chemarrow,->] (cf_arrow_start) -- (cf_arrow_end);}

\AtEndPreamble{
\usepackage{scrhack}
}

\AtEndPreamble{
	\ifthenelseproperty{compilation}{pdfa}{
	}{
		\usepackage{intopdf}
		
	}
}

\ifthenelseproperty{compilation}{final}{%
    \PassOptionsToPackage{disable}{todonotes}
}{}

\usepackage%
		[
			colorinlistoftodos, 
		]
		{todonotes}

\newcommand*{\TODO}[2][inline]{%
    \todo[#1]{#2}
}

\ifthenelseproperty{compilation}{final}{%
    \PassOptionsToPackage{final}{changes}
}{%
    \typeout{---------------------------------------------------}
    \typeout{---------------------------------------------------}
    \typeout{---------------------------------------------------}
    \typeout{---------------------------------------------------}
    \typeout{---------------------------------------------------}
    \typeout{---------------------------------------------------}
    \typeout{\getproperty{compilation}{comment}}
    \ifpropertydefined{compilation}{comment}{%
        \PassOptionsToPackage{commentmarkup=\getproperty{compilation}{comment}}{changes}
    }
}

\usepackage%
		[
		]
		{changes}

\usepackage{seqsplit}

\usepackage{booktabs}

\ifthenelseproperty{compilation}{fontspec}{%
	\AtBeginEnvironment{tabular}{\addfontfeatures{Numbers={Monospaced}}}
    }{}

\usepackage{xcolor} 

\PassOptionsToPackage{dvipsnames,svgnames,x11names,rgb}{xcolor}
\definecolor{darkgreen}{RGB}{0,80,0} 
\definecolor{darkred}{RGB}{80,0,0} 

\usepackage{graphicx}

\usepackage{pgf}  

\SendSettingsToPgf

\usepackage
		[
			format=plain, 
			indention=1em, 
			textfont={color={black},small}, 
			width=0.925\linewidth, 
			hypcap=true, 
		]
		{caption}

\AtBeginDocument
			{
				\captionsetup
						{
                            labelfont={color=\getproperty{color}{main},small,sf,bf}, 
						}
			}

\usepackage{wrapfig}

\usepackage
		[
			hypcap=true, 
			textfont={color={black},small}, 
			width=0.925\textwidth, 
			indention=1em, 
		]
		{subcaption}

\usepackage{tikz}

\usetikzlibrary
			{%
				arrows,
				arrows.meta,
				bending,
				backgrounds,
				calc,
				fit,
				fadings,
				graphs,
				intersections,
				shadings,
				shapes,
				shapes.misc,
				shapes.geometric,
 				patterns,
 				plotmarks,
				chains,  
 				matrix,
 				through,
 				positioning,  
 				scopes,
				spy,
				decorations.shapes,
				decorations.text,
				decorations.pathmorphing,
				decorations.pathreplacing,
				decorations.markings,
				intersections,
			}
\ifluatex
    \usetikzlibrary
			{%
				graphdrawing,
			}
\fi

\makeatletter
\pgfkeys{/pgf/.cd,
  rectangle corner radius/.initial=3pt
}
\newif\ifpgf@rectanglewrc@donecorner@
\def\pgf@rectanglewithroundedcorners@docorner#1#2#3#4{%
  \edef\pgf@marshal{%
    \noexpand\pgfintersectionofpaths
      {%
        \noexpand\pgfpathmoveto{\noexpand\pgfpoint{\the\pgf@xa}{\the\pgf@ya}}%
        \noexpand\pgfpathlineto{\noexpand\pgfpoint{\the\pgf@x}{\the\pgf@y}}%
      }%
      {%
        \noexpand\pgfpathmoveto{\noexpand\pgfpointadd
          {\noexpand\pgfpoint{\the\pgf@xc}{\the\pgf@yc}}%
          {\noexpand\pgfpoint{#1}{#2}}}%
        \noexpand\pgfpatharc{#3}{#4}{\cornerradius}%
      }%
    }%
  \pgf@process{\pgf@marshal\pgfpointintersectionsolution{1}}%
  \pgf@process{\pgftransforminvert\pgfpointtransformed{}}%
  \pgf@rectanglewrc@donecorner@true
}
\pgfdeclareshape{rectangle with rounded corners}
{
  \inheritsavedanchors[from=rectangle] 
  \inheritanchor[from=rectangle]{north}
  \inheritanchor[from=rectangle]{north west}
  \inheritanchor[from=rectangle]{north east}
  \inheritanchor[from=rectangle]{center}
  \inheritanchor[from=rectangle]{west}
  \inheritanchor[from=rectangle]{east}
  \inheritanchor[from=rectangle]{mid}
  \inheritanchor[from=rectangle]{mid west}
  \inheritanchor[from=rectangle]{mid east}
  \inheritanchor[from=rectangle]{base}
  \inheritanchor[from=rectangle]{base west}
  \inheritanchor[from=rectangle]{base east}
  \inheritanchor[from=rectangle]{south}
  \inheritanchor[from=rectangle]{south west}
  \inheritanchor[from=rectangle]{south east}

  \savedmacro\cornerradius{%
    \edef\cornerradius{\pgfkeysvalueof{/pgf/rectangle corner radius}}%
  }

  \backgroundpath{%
    \northeast\advance\pgf@y-\cornerradius\relax
    \pgfpathmoveto{}%
    \pgfpatharc{0}{90}{\cornerradius}%
    \northeast\pgf@ya=\pgf@y\southwest\advance\pgf@x\cornerradius\relax\pgf@y=\pgf@ya
    \pgfpathlineto{}%
    \pgfpatharc{90}{180}{\cornerradius}%
    \southwest\advance\pgf@y\cornerradius\relax
    \pgfpathlineto{}%
    \pgfpatharc{180}{270}{\cornerradius}%
    \northeast\pgf@xa=\pgf@x\advance\pgf@xa-\cornerradius\southwest\pgf@x=\pgf@xa
    \pgfpathlineto{}%
    \pgfpatharc{270}{360}{\cornerradius}%
    \northeast\advance\pgf@y-\cornerradius\relax
    \pgfpathlineto{}%
  }

  \anchor{south west}{%
    \southwest
    \pgfmathparse{\cornerradius*(1-1/sqrt(2))}
    \advance\pgf@x\pgfmathresult pt
    \advance\pgf@y\pgfmathresult pt}

  \anchor{north east}{%
    \northeast
    \pgfmathparse{\cornerradius*(1-1/sqrt(2))}
    \advance\pgf@x-\pgfmathresult pt
    \advance\pgf@y-\pgfmathresult pt}

  \anchor{north west}{%
    \southwest\pgf@xa=\pgf@x
    \northeast\pgf@x=\pgf@xa
    \pgfmathparse{\cornerradius*(1-1/sqrt(2))}
    \advance\pgf@x\pgfmathresult pt
    \advance\pgf@y-\pgfmathresult pt}

  \anchor{south east}{%
    \southwest\pgf@ya=\pgf@y
    \northeast\pgf@y=\pgf@ya
    \pgfmathparse{\cornerradius*(1-1/sqrt(2))}
    \advance\pgf@x-\pgfmathresult pt
    \advance\pgf@y\pgfmathresult pt}

  \anchor{before north east}{\northeast\advance\pgf@y-\cornerradius}
  \anchor{after north east}{\northeast\advance\pgf@x-\cornerradius}
  \anchor{before north west}{\southwest\pgf@xa=\pgf@x\advance\pgf@xa\cornerradius
    \northeast\pgf@x=\pgf@xa}
  \anchor{after north west}{\northeast\pgf@ya=\pgf@y\advance\pgf@ya-\cornerradius
    \southwest\pgf@y=\pgf@ya}
  \anchor{before south west}{\southwest\advance\pgf@y\cornerradius}
  \anchor{after south west}{\southwest\advance\pgf@x\cornerradius}
  \anchor{before south east}{\northeast\pgf@xa=\pgf@x\advance\pgf@xa-\cornerradius
    \southwest\pgf@x=\pgf@xa}
  \anchor{after south east}{\southwest\pgf@ya=\pgf@y\advance\pgf@ya\cornerradius
    \northeast\pgf@y=\pgf@ya}

  \anchorborder{%
    \pgf@xb=\pgf@x
    \pgf@yb=\pgf@y%
    \southwest%
    \pgf@xa=\pgf@x
    \pgf@ya=\pgf@y%
    \northeast%
    \advance\pgf@x by-\pgf@xa%
    \advance\pgf@y by-\pgf@ya%
    \pgf@xc=.5\pgf@x
    \pgf@yc=.5\pgf@y%
    \advance\pgf@xa by\pgf@xc
    \advance\pgf@ya by\pgf@yc%
    \edef\pgf@marshal{%
      \noexpand\pgfpointborderrectangle
      {\noexpand\pgfqpoint{\the\pgf@xb}{\the\pgf@yb}}
      {\noexpand\pgfqpoint{\the\pgf@xc}{\the\pgf@yc}}%
    }%
    \pgf@process{\pgf@marshal}%
    \advance\pgf@x by\pgf@xa%
    \advance\pgf@y by\pgf@ya%
    \pgfextract@process\borderpoint{}%
    \pgf@rectanglewrc@donecorner@false
    \southwest\pgf@xc=\pgf@x\pgf@yc=\pgf@y
    \advance\pgf@xc\cornerradius\relax\advance\pgf@yc\cornerradius\relax 
    \borderpoint
    \ifdim\pgf@x<\pgf@xc\relax\ifdim\pgf@y<\pgf@yc\relax
      \pgf@rectanglewithroundedcorners@docorner{-\cornerradius}{0pt}{180}{270}%
    \fi\fi
    \ifpgf@rectanglewrc@donecorner@\else
      \southwest\pgf@yc=\pgf@y\relax\northeast\pgf@xc=\pgf@x\relax
      \advance\pgf@xc-\cornerradius\relax\advance\pgf@yc\cornerradius\relax
      \borderpoint
      \ifdim\pgf@x>\pgf@xc\relax\ifdim\pgf@y<\pgf@yc\relax
       \pgf@rectanglewithroundedcorners@docorner{0pt}{-\cornerradius}{270}{360}%
      \fi\fi
    \fi
    \ifpgf@rectanglewrc@donecorner@\else
      \northeast\pgf@xc=\pgf@x\relax\pgf@yc=\pgf@y\relax
      \advance\pgf@xc-\cornerradius\relax\advance\pgf@yc-\cornerradius\relax
      \borderpoint
      \ifdim\pgf@x>\pgf@xc\relax\ifdim\pgf@y>\pgf@yc\relax
       \pgf@rectanglewithroundedcorners@docorner{\cornerradius}{0pt}{0}{90}%
      \fi\fi
    \fi
    \ifpgf@rectanglewrc@donecorner@\else
      \northeast\pgf@yc=\pgf@y\relax\southwest\pgf@xc=\pgf@x\relax
      \advance\pgf@xc\cornerradius\relax\advance\pgf@yc-\cornerradius\relax
      \borderpoint
      \ifdim\pgf@x<\pgf@xc\relax\ifdim\pgf@y>\pgf@yc\relax
       \pgf@rectanglewithroundedcorners@docorner{0pt}{\cornerradius}{90}{180}%
      \fi\fi
    \fi
  }
}
\makeatother

\usepackage{pgfplots}
\usepackage{pgfplotstable}

\pgfplotsset{compat=newest}

\usetikzlibrary{pgfplots.groupplots}

\usepgflibrary{decorations.pathmorphing}
\usepgfplotslibrary{fillbetween,colormaps,external,colorbrewer}

\pgfplotsset{aspect ratio/.code args={#1:#2}{%
  \def\axisdefaultwidth{#1 cm}%
  \def\axisdefaultheight{#2 cm}},
}

\pgfplotsset{cycle list/Dark2-8}
\pgfplotsset{colormap/viridis}
\pgfkeys{/pgf/number format/.cd,1000 sep={\,}}

\AtEndPreamble{\ifthenelseproperty{compilation}{externalize}{
		\tikzexternalize[prefix=\getproperty{compilation}{externalfolder}]
		\typeout{----- USING TIKZ EXTERNALIZE -----}
	}{
		\typeout{----- NOT USING TIKZ EXTERNALIZE -----}
	}
}

\usepackage{listings}

\definecolor{javagray}{rgb}{0.55, 0.52, 0.54} 
\definecolor{javared}{rgb}{0.6,0,0} 
\definecolor{javagreen}{rgb}{0.25,0.5,0.35} 
\definecolor{javapurple}{rgb}{0.5,0,0.35} 
\definecolor{javadocblue}{rgb}{0.25,0.35,0.75} 
\definecolor{javaLila}{RGB}{127,0,85}

\let\origthelstnumber\thelstnumber

\makeatletter
\newcommand*\Suppressnumber{%
  \lst@AddToHook{OnNewLine}{%
    \let\thelstnumber\relax%
  }%
}
\newcommand\Reactivatenumber[1]{%
  \global\c@lstnumber#1%
  \global\advance\c@lstnumber\m@ne\relax%
  \lst@AddToHook{OnNewLine}{%
  \let\thelstnumber\origthelstnumber%
  }%
}
\makeatother

\makeatletter
\lst@InputCatcodes
\def\lst@DefEC{%
 \lst@CCECUse \lst@ProcessLetter
^^80^^81^^82^^83^^84^^85^^86^^87^^88^^89^^8a^^8b^^8c^^8d^^8e^^8f%
  ^^90^^91^^92^^93^^94^^95^^96^^97^^98^^99^^9a^^9b^^9c^^9d^^9e^^9f%
  ^^a0^^a1^^a2^^a3^^a4^^a5^^a6^^a7^^a8^^a9^^aa^^ab^^ac^^ad^^ae^^af%
  ^^b0^^b1^^b2^^b3^^b4^^b5^^b6^^b7^^b8^^b9^^ba^^bb^^bc^^bd^^be^^bf%
  ^^c0^^c1^^c2^^c3^^c4^^c5^^c6^^c7^^c8^^c9^^ca^^cb^^cc^^cd^^ce^^cf%
  ^^d0^^d1^^d2^^d3^^d4^^d5^^d6^^d7^^d8^^d9^^da^^db^^dc^^dd^^de^^df%
  ^^e0^^e1^^e2^^e3^^e4^^e5^^e6^^e7^^e8^^e9^^ea^^eb^^ec^^ed^^ee^^ef%
  ^^f0^^f1^^f2^^f3^^f4^^f5^^f6^^f7^^f8^^f9^^fa^^fb^^fc^^fd^^fe^^ff%
	^^^^03b8^^^^03c8^^^^03b7^^^^03bc^^^^03c3^^^^03b1^^^^03a9^^^^03b6%
	^^^^03c9^^^^03b4^^^^03c0%
^^00}
\lst@RestoreCatcodes
\makeatother

\RequirePackageWithOption{hyperref}

\hypersetup
		{
			pdfencoding=unicode,
			pdfpagemode=UseOutlines,
			bookmarksopenlevel=0,
			pdfpagelayout=TwoPageRight,
			hidelinks,
			pdfduplex=DuplexFlipLongEdge,
		}

\usepackage{epigraph}

\usepackage
		[
			math,
			toc,
		]
		{blindtext}

\usepackage
		[
			noabbrev
		]
		{cleveref}

\crefname{line}{line}{lines}
\creflabelformat{line}{#2#1#3}

\crefname{reaction}{reaction}{reactions}
\creflabelformat{reaction}{#2(#1)#3}

\crefname{listing}{code}{codes}

\usepackage
		[
			strict=true, 
			autostyle=true, 
			german=guillemets, 
		]
		{csquotes}

\setquotestyle{german} 

\usepackage{lettrine} 

\usepackage
		[
			hyperref=true,
			backend=biber, 
			url=true, 
			abbreviate=false,
			isbn=false, 
			doi=true, 
			backref=false, 
			urldate=long, 
			style=numeric-comp,
			maxbibnames=3, 
			giveninits=true,
			block=none,
            sorting=none,
			defernumbers=true,
		]
		{biblatex}

\makeatletter
\@ifpackageloaded{seqsplit}{
	\DeclareFieldFormat{eid}{Art.\addnbspace\seqsplit{#1}}
}{}
\makeatother

\DeclareSourcemap{
  \maps[datatype=bibtex]{
    \map{
      \step[fieldsource=doi,final]
      \step[fieldset=url,null]
    }
  }
}

\DeclareFieldFormat{journaltitle}{\mkbibemph{#1},} 
\DeclareFieldFormat[inbook,thesis]{title}{\mkbibemph{#1}\addperiod} 
\DeclareFieldFormat[article,misc]{title}{\enquote{#1}} 
\DeclareFieldFormat[article,inproceedings,incollection]{volume}{Vol.~#1} 
\DeclareFieldFormat{edition}{\ifinteger{#1}{\mkbibordedition{#1}\addnbthinspace{}ed.}{#1\isdot}}

\DeclareFieldFormat{eprint:urn}{\textsc{URN}\addcolon\space\ifhyperref{\href{http://www.nbn-resolving.org/#1}{\nolinkurl{#1}}}{\nolinkurl{#1}}}

\DeclareSortingTemplate{ndymdt}{
 	\sort{
		\field{presort}
	}
	\sort[final]{
		\field{sortkey}
	}
	\sort[direction=descending]{
		\field{sortyear}
		\field{year}
		\literal{9999}
	}
	\sort[direction=descending]{
		\field[padside=left,padwidth=2,padchar=0]{month}
		\literal{99}
	}
	\sort[direction=descending]{
		\field[padside=left,padwidth=2,padchar=0]{day}
		\literal{99}
	}
}

\DeclareBibliographyCategory{dontbib}
\let\printbibliographyold\printbibliography%
\renewcommand{\printbibliography}[1][]{\typeout{----- printbibliography ------}\printbibliographyold[notcategory=dontbib,#1]}%
\makeatletter
\newcommand{\tempmaxup}[1]{\def\blx@maxcitenames{99}#1}
\DeclareCiteCommand{\fullcitecontribution}[\tempmaxup]{
    \usebibmacro{prenote}
    \addtocategory{dontbib}{\thefield{entrykey}}
}{
    \textbf{\usebibmacro{maintitle+title}}
    \newline\nopunct\newblock
    \usebibmacro{author}
    \newline\nopunct\newblock
    \usebibmacro{journal+issuetitle}, \usebibmacro{doi+eprint+url} \usebibmacro{addendum+pubstate}
}{
    \multicitedelim
}{
    \usebibmacro{postnote}
}
\makeatother

\usepackage
		[
			automark,
			headsepline, 
		]
		{scrlayer-scrpage}

\AtBeginDocument{%
    \ifKOMA
        \addtokomafont{pagehead}{\color{\getproperty{color}{main}}}
        \ifthenelseproperty{compilation}{clsdefineschapter}{%
            \addtokomafont{chapter}{\color{\getproperty{color}{main}}}
        }{}
        \addtokomafont{section}{\color{\getproperty{color}{main}}}
        \addtokomafont{subsection}{\color{\getproperty{color}{main}}}
        \addtokomafont{subsubsection}{\color{\getproperty{color}{main}}}
        \addtokomafont{paragraph}{\color{\getproperty{color}{main}}}

        \addtokomafont{footnoterule}{\color{\getproperty{color}{secondary}}}
        \addtokomafont{headsepline}{\color{\getproperty{color}{secondary}}}

    \fi
}

\usepackage{bookmark}

\bookmarksetup
			{
				open,
				addtohook={
					\ifnum\bookmarkget{level}<1
						\bookmarksetup{bold}
					\fi
				},
			}

\PassOptionsToPackage{automake=immediate}{glossaries-extra}

\usepackage%
		[
			nonumberlist, 
			acronym, 
			toc, 
		]
		{glossaries-extra}

\usepackage{glossary-mcols}

\setabbreviationstyle[acronym]{long-short}

\newglossary[slg]{symbols}{syi}{syg}{List of Symbols} 

\makeglossaries

\newglossarystyle{customGlossaryStyleForListOfSymbols}{%
    \renewenvironment{theglossary}{%
        \begin{longtable}{p{0.12\textwidth}p{\glsdescwidth}p{\glspagelistwidth}}
    }{%
        \end{longtable}
    }
    
    \renewcommand*{\glsgroupheading}[1]{}

}

\newglossarystyle{customListOfSymbols}{
    \setglossarystyle{customGlossaryStyleForListOfSymbols}
    
    \renewenvironment{theglossary}{%
        \begin{longtable}{p{0.12\textwidth}p{\glsdescwidth}p{\glspagelistwidth}}
    }{%
        \end{longtable}
    }
}

\newglossarystyle{customListOfAbbreviations}{%
    \renewenvironment{theglossary}{%
        \begin{longtable}{lp{\glsdescwidth}}
    }{%
        \end{longtable}
    }
    
    \renewcommand*{\glsgroupheading}[1]{}

}

\makeatletter
\newglossarystyle{customIndex}{
    \renewenvironment{theglossary}{%
        \setlength{\parindent}{0pt}
        \setlength{\parskip}{0pt plus 0.3pt}
        \let\item\@idxitem
    }{%
    }
    
    \renewcommand*{\glsgroupheading}[1]{}
    \renewcommand*{\glossaryentryfield}[5]{%
    \item\glstarget{##1}{##2}
        \ifx\relax##4\relax
        \else
            \space(##4)
        \fi
        \dotfill ##3\glspostdescription \space ##5
    }
    \renewcommand*{\glossarysubentryfield}[6]{%
        \ifcase##1\relax
        \item
        \or
            \subitem
        \else
            \subsubitem
        \fi
        \glstarget{##2}{##3}
        \ifx\relax##5\relax
        \else
            \space(##5)
        \fi
        \dotfill ##4\glspostdescription\space ##6
    }
    
    \renewcommand*{\glsgroupheading}[1]{%
        \item\textbf{\glsgetgrouptitle{##1}}\indexspace
    }
}
\makeatother

\newglossarystyle{documentationLabelStyle}{
    \renewenvironment{theglossary}{%
        \begin{longtable}{p{.2\textwidth}p{.3\textwidth}p{.5\textwidth}}
    }{%
        \end{longtable}
    }%

}

\newglossarystyle{documentationSymbolsStyle}{
    \renewenvironment{theglossary}{%
        \begin{longtable}{p{0.12\textwidth}p{.2\textwidth}p{.48\textwidth}p{.2\textwidth}}
    }{%
        \end{longtable}
    }

}

\newglossarystyle{publicationLabelStyle}{
	\renewenvironment{theglossary}{%
		\begin{longtable}{p{.2\textwidth}p{.8\textwidth}}
		}{%
		\end{longtable}
	}%

}

\newglossarystyle{publicationSymbolsStyle}{
}

\usepackage{scrtime}

\usepackage{ragged2e}

\usepackage{marginnote}

\DeclareDocumentCommand{\myMarginnote} 
					{
				 		O{0cm} O{c} m 
				 	}{
						\marginnote
								{
									\ifthispageodd
											{
												\RaggedRight 
											}{
												\RaggedLeft 
											}
									\raisebox{#1}[#1]
											{
												\begin{minipage}[#2]{\marginparwidth}
													\RaggedRight 
													\color{\getMainColor}
													\lineskiplimit=-\maxdimen
													\normalfont\sffamily
													#3\end{minipage}}
											}
					}

\ifluatex
	\usepackage[luatex]{pdflscape}
\else
\fi

\usepackage
		[
			final, 
		]
		{pdfpages}

\pretocmd{\includepdf}{%
    \ifthenelseproperty{compilation}{externalize}{%
        \tikzset{external/optimize=false}%
    }{}
}{}{}

\makeatletter
\@ifpackageloaded{cleveref}{%
    \newcounter{articlenumber}
    \crefname{articlenumber}{article}{articles}
    \Crefname{articlenumber}{Article}{Articles}

}
\makeatother

\usepackage{enumitem}

\setlist[description]{leftmargin=*,style=sameline}

\ifluatex

	\usepackage{luacode}

\usepackage
		[
			draft, 
		]
		{impnattypo}

\usepackage{selnolig}

    \usepackage{pdftexcmds}
    \makeatletter
    \let\pdfstrcmp\pdf@strcmp
    \makeatother
\else
\fi

\def\clap#1{\hbox to 0pt{\hss#1\hss}}

\newlength{\heightOfX}
\AtBeginDocument{\settoheight{\heightOfX}{X}} 

\newlength{\fsize}

\newcommand{\nobreakbefore}
	{%
		\relax
		\ifvmode
		\else
			\ifhmode
				\ifdim\lastskip > 0pt\relax
					\unskip\nobreakspace
				\fi
    		\fi
  		\fi
	}
\let\oldcite\cite
\renewcommand{\cite}{\nobreakbefore\oldcite}
\let\oldref\ref
\renewcommand{\ref}{\nobreakbefore\oldref}

\newcommand{\disabledprotrusion}[1]
		{
			\ifKOMA
                \ifthenelseproperty{compilation}{fontspec}{%
					\begingroup
						\addfontfeatures{Numbers={Lining,Monospaced}}
                }{}
			\fi
            \ifthenelseproperty{compilation}{microtype}{%
				\microtypesetup{protrusion=false}
            }{}
			#1
            \ifthenelseproperty{compilation}{microtype}{%
				\microtypesetup{protrusion=true}
            }{}
			\ifKOMA
                \ifthenelseproperty{compilation}{fontspec}{%
					\endgroup
                }{}
			\fi
		}

\makeatletter
\newcommand{\removeifnextchar}[3]
		{
			\begingroup
			\ltx@LocToksA{\endgroup#2}
			\ltx@LocToksB{\endgroup#3}
			\ltx@ifnextchar{#1}
				{
					\def\next{\the\ltx@LocToksA}
					\afterassignment\next
					\let\scratch= %
				}{
					\the\ltx@LocToksB
				}
		}
\makeatother

\newcommand{\Signature}[2]
		{%
			\vspace{2cm}%
			\noindent%
            \begin{tabular*}{\textwidth}{@{\extracolsep{0pt}} l @{\extracolsep{\fill}} r @{\extracolsep{0pt}}}%
#1, \getproperty{document}{date}	& \rule{0.33\textwidth}{1pt} 	\\
& \raggedleft{\textsc{#2}}
			\end{tabular*}
			\vspace{1cm}
		}

\AtBeginDocument{%
	\newif\ifKOMAandFancyChapterHeadings%
	\KOMAandFancyChapterHeadingsfalse%
    \ifKOMA%
        \ifthenelseproperty{compilation}{fancychapterheadings}{%
            \RedeclareSectionCommand[
                    beforeskip=\dimexpr3.3\baselineskip+1\parskip\relax,
                    innerskip=0pt,
                    afterskip=1.725\baselineskip plus .115\baselineskip minus .192\baselineskip,
                    afterindent=false
                ]%
                {chapter}%
            \KOMAandFancyChapterHeadingstrue%
        }{}%
    \fi%
    \ifKOMAandFancyChapterHeadings%
    \else%
    \fi%
}

\ifKOMA
    \ifthenelseproperty{compilation}{fontspec}{%
		\DeclareTOCStyleEntry[%
							entryformat={%
                                \addfontfeatures{Numbers={Lining,Monospaced}}\bfseries\sffamily\color{\getproperty{color}{main}}
										},
							pagenumberformat={%
											\addfontfeatures{Numbers={Lining,Monospaced}}\bfseries\sffamily
										},
						]
						{default}
						{chapter}
		\DeclareTOCStyleEntry[%
							entryformat={%
											\addfontfeatures{Numbers={Lining,Monospaced}}
										},
							pagenumberformat={%
											\addfontfeatures{Numbers={Lining,Monospaced}}
										},
						]
						{default}
						{section}
		\DeclareTOCStyleEntry[%
							entryformat={%
											\addfontfeatures{Numbers={Lining,Monospaced}}
										},
							pagenumberformat={%
											\addfontfeatures{Numbers={Lining,Monospaced}}
										},
						]
						{default}
						{subsection}
		\DeclareTOCStyleEntry[%
							entryformat={%
											\addfontfeatures{Numbers={Lining,Monospaced}}
										},
							pagenumberformat={%
											\addfontfeatures{Numbers={Lining,Monospaced}}
										},
						]
						{default}
						{subsubsection}
		\DeclareTOCStyleEntry[%
							entryformat={%
											\addfontfeatures{Numbers={Lining,Monospaced}}
										},
							pagenumberformat={%
											\addfontfeatures{Numbers={Lining,Monospaced}}
										},
						]
						{tocline}
						{figure}
		\DeclareTOCStyleEntry[%
							entryformat={%
											\addfontfeatures{Numbers={Lining,Monospaced}}
										},
							pagenumberformat={%
											\addfontfeatures{Numbers={Lining,Monospaced}}
										},
						]
						{tocline}
						{table}
    }{}
\fi

\makeatletter
\newcommand{\headlesschapter}[1]{%
  \begingroup
  \let\@makechapterhead\@gobble 
  \chapter{#1}
  \endgroup
}
\makeatother

\makeatletter
\newcommand\appendgraphicspath[1]{%
  \g@addto@macro\Ginput@path{#1}%
}
\makeatother

\RequirePackage{xkeyval}
\RequirePackage{xstring}
\RequirePackage{xfp}
\RequirePackage{ifthen}
\RequirePackage{tikzscale} 
\makeatletter
\define@key{igraph}{width}{\def\igraph@width{#1}}
\define@key{igraph}{svgwidth}{\def\igraph@svgwidth{#1}}
\define@key{igraph}{angle}{\def\igraph@angle{#1}}
\define@key{igraph}{draft}{\def\igraph@draft{#1}}
\define@key{igraph}{scale}{\def\igraph@scale{#1}}
\define@key{igraph}{height}{\def\igraph@height{#1}}
\define@boolkey{igraph}{scaletext}{\def\igraph@scaletext{#1}}
\newcommand{\igraph}[2][]{%
    \filename@parse{#2}
    \setkeys{igraph}{width=\linewidth} 
    \setkeys{igraph}{svgwidth=\linewidth} 
    \setkeys{igraph}{height=\empty} 
    \setkeys{igraph}{scaletext=false} 
    \setkeys{igraph}{#1}
    \ifthenelse{%
        \equal{\filename@ext}{pgf}%
    }{%
        \typeout{----- INCLUDE pgf @ #2 ------}%
        \let\pgfimageWithoutPath\pgfimage%
        \renewcommand{\pgfimage}[2][]{\typeout{----- INCLUDE pgfimgage @ ##2 ------}\pgfimageWithoutPath[##1]{\filename@area/##2}}%
        \resizebox{\igraph@width}{!}{\input{#2}}%
    }{%
        \ifnum\pdfstrcmp{\filename@ext}{pdf_tex}=0
            \typeout{----- INCLUDE pdf_tex @ #2 ------}%
            \IfSubStr{#1}{height}{%
                \PackageError{igraph}{pdf_tex does not allow the height attribute}{}%
                }{}%
            \ifthenelse{%
                \boolean{\igraph@scaletext}%
            }{%
                \def\svgwidth{\igraph@svgwidth}%
                \resizebox{\igraph@width}{!}{%
                    \import{\filename@area}{\filename@base.\filename@ext}%
                }%
            }{%
                \def\svgwidth{\igraph@width}%
                \import{\filename@area}{\filename@base.\filename@ext}%
            }%
        \else
            \ifthenelse{%
                \equal{\filename@ext}{svg}%
            }{%
                \typeout{----- INCLUDE svg @ #2 ------}%
                \PackageError{igraph}{svg not implemented!}{}%
            }{%
                \ifthenelse{%
                    \equal{\filename@ext}{tikz}%
                }{%
                    \typeout{----- INCLUDE tikz @ #2 ------}%
	                \tikzsetnextfilename{\filename@base}%
                    \begin{minipage}{\igraph@width}%
                        \input{#2}%
                    \end{minipage}%
                }{%
                    \typeout{----- INCLUDE with includegraphics @ #2 ------}%
                    \IfSubStr{#1}{height}{%
                        \includegraphics[#1]{#2}%
                    }{%
                        \includegraphics[width=\igraph@width, #1]{#2}%
                    }%
                }%
            }%
        \fi
    }%
}%
\makeatother

\newlength{\imageh}
\newlength{\imaged}
\newlength{\imagew}
\newcommand{\setimageh}[1]{
 \settoheight{\imageh}{\usebox{#1}}
}
\newcommand{\setimagew}[1]{
 \settowidth{\imagew}{\usebox{#1}}
}
\newcommand{\setimaged}[1]{
 \settodepth{\imaged}{\usebox{#1}}
}
\newsavebox{\imagesavebox}
\newcommand{\setimagedimensions}[1]{
    \savebox{\imagesavebox}{\igraph{#1}}
    \setimageh{\imagesavebox}
    \setimagew{\imagesavebox}
    \setimaged{\imagesavebox}
}

\makeatletter
\newcommand*{\storedata}[2]{%
  \count@=0 %
  \@tfor\@tmp:=#2\do{%
    \advance\count@\@ne
    \expandafter\let\csname data:\the\count@:#1\endcsname\@tmp
  }%
  \expandafter\edef\csname data:0:#1\endcsname{\the\count@}%
}
\newcommand*{\getdata}[2]{%
  \@ifundefined{data:0:#2}{%
    \@latex@error{Undefined data `#2'}\@ehc
  }{%
    \expandafter\@getdata\expandafter{%
      \the\numexpr
        \ifnum\numexpr(#1)<\z@
          \@nameuse{data:0:#2}+1+%
        \fi
        (#1)%
      \relax
    }{#2}{#1}%
  }%
}
\newcommand*{\@getdata}[3]{%
  \ifnum#1<\z@
    \@getdata@error{\the\numexpr(#3)\relax}{#2}%
  \else
    \ifnum#1>\@nameuse{data:0:#2} %
      \@getdata@error{#1}{#2}%
    \else
      \@nameuse{data:#1:#2}%
    \fi
  \fi
}
\newcommand*{\@getdata@error}[2]{%
  \@latex@error{%
    Wrong data selector #1 for `#2',\MessageBreak
    which only contains \@nameuse{data:0:#2} item(s)%
  }\@ehc
}
\makeatother

\storedata{mydata}{{one}{two}{three}}
\storedata{mylongdata}{
  {one} {two} {three} {four} {five} {six} {seven} {eight} {nine} {ten}
  {eleven} {twelve} {thirteen} {fourteen} {fifteen} {sixteen}
  {seventeen} {eighteen} {nineteen} {twenty} {twenty-one} {twenty-two}
  {twenty-three} {twenty-four}
}

\makeatletter
\RequirePackage{forarray}
\RequirePackage{calc}
\RequirePackage{listings}
\newlength\figratiosum
\newlength\figratio
\newlength\figheight
\newlength\figwidth
\newcounter{npaths}

\newcommand{\catcodeigraph}{\begingroup
  \catcode`_=11 \igraph}
\define@key{multigraph}{width}{\def\multigraph@width{#1}}
\define@key{multigraph}{labels}{\def\multigraph@labels{#1}}
\newcommand{\domultigraph}[3][]{
    \setkeys{multigraph}{width=\linewidth - 1em} 
    \setkeys{multigraph}{labels=\empty} 
    \setkeys{multigraph}{#1}

    \setcounter{npaths}{0}
    \ForEachX{;}{%
        \stepcounter{npaths}
    }{#2}

    \IfSubStr{#1}{labels}{%
        \def\subfigurelabels{}
        \ForEachX{;}{%
            \edef\subfigurelabels{\subfigurelabels"\thislevelitem"}
            \ifnum\thislevelcount=\value{npaths}
            \else
                \edef\subfigurelabels{\subfigurelabels,}
            \fi
        }{\multigraph@labels}
    }{}

    \def\paths{}
    \setlength\figratiosum{0pt}
    \def\figratios{}
    \ForEachX{;}{%
        \edef\paths{\paths"\thislevelitem"}
        \ifnum\thislevelcount=\value{npaths}
        \else
            \edef\paths{\paths,}
        \fi

        \setimagedimensions{\thislevelitem}
        \setlength\figratio{1pt*\ratio{\imagew}{\imageh}}
        \addtolength{\figratiosum}{\figratio}
        \edef\figratios{\figratios"\the\figratio"}
        \ifnum\thislevelcount=\value{npaths}
        \else
            \edef\figratios{\figratios,}
        \fi
    }{#2}

    \storedata{subcaptions}{#3}

    \setlength{\figheight}{1pt*\ratio{\multigraph@width}{\figratiosum}}

    \addtocounter{npaths}{-1}  
    \foreach \i in {0,...,\value{npaths}}{%
        \pgfmathparse{{\figratios}[\i]}
        \setlength\figwidth{\figheight}
        \pgfmathparse{\figwidth * \pgfmathresult}
        \setlength{\figwidth}{\pgfmathresult pt}
        \begin{subfigure}{\figwidth}
            \centering
            \pgfmathparse{{\paths}[\i]}
            \typeout{----- Width for multigraph figure \pgfmathresult : \the\figwidth ------}
            \catcodeigraph{\pgfmathresult}\endgroup
            \subcaption{\getdata{\i+1}{subcaptions}}
            \IfSubStr{#1}{labels}{%
                \pgfmathparse{{\subfigurelabels}[\i]}
                \label{\pgfmathresult}
            }{}
        \end{subfigure}
    }
    \endgroup
}
\makeatother

\newcommand{\makeup}[1]{%
    \ensuremath{
        \ifluatex
            \symup{#1}
        \else
            \mathrm{#1}
        \fi
    }
}
\newcommand{\makebf}[1]{%
    \ensuremath{
        \ifluatex
            \symbf{#1}
        \else
            \mathbf{#1}
        \fi
    }
}

\newcommand{\eg}{e.\,g.\xspace}

\newcommand{\ie}{i.\,e.\xspace}

\newcommand{\real}{\ensuremath{\mathbb{R}}}

\newcommand{\naturals}{\ensuremath{\mathbb{N}}}  

\newcommand{\vect}[1]{\ensuremath{\vec{#1}}} 
\newcommand{\matr}[1]{\ensuremath{\makebf{#1}}}

\makeatletter
\newcommand{\Grad}[2][\@nil]{%
    \def\tmp{#1}%
    \ifx\tmp\@nnil
    	\ensuremath{\vect{\nabla} #2}
    \else
    	\ensuremath{\vect{\nabla}_{\!\!#1} #2}
    \fi}
\makeatother

\newcommand{\dx}[3][\empty]
		{
			\if{#1}\equal{\empty}
				\frac{\mathrm{d}#2}{\mathrm{d}#3}
			\else
				\frac{\mathrm{d}^{#1}#2}{\mathrm{d}#3^{#1}}
		}
\newcommand{\pdx}[3][\empty]
		{
			\if{#1}\equal{\empty}
				\frac{\partial#3}{\partial#2}
			\else
				\frac{\partial^{#1}#3}{\partial#2^{#1}}
		}
\makeatletter
\providecommand*{\diff}{\@ifnextchar^{\DIfF}{\DIfF^{}}}
\def\DIfF^#1{\mathop{\mathrm{\mathstrut d}}\nolimits^{#1}\gobblespace}
\def\gobblespace{\futurelet\diffarg\opspace}
\def\opspace
		{
			\let\DiffSpace\!
			\ifx\diffarg(
				\let\DiffSpace\relax
			\else
				\ifx\diffarg[%
					\let\DiffSpace\relax
				\else
					\ifx\diffarg\{%
						\let\DiffSpace\relax
					\fi
				\fi
			\fi
			\DiffSpace
		}
\makeatother

\newcommand{\orderof}[1]{\ensuremath{\mathcal{O}(#1)}}

\newcommand{\tightoverset}[2]{\mathop{#2}\limits^{\vbox to -.5ex{\kern-0.75ex\hbox{$\! #1$}\vss}}} 

\DeclareSIUnit[number-unit-product = \,]{\permille}{\textperthousand}

\newacronym{statistics}{Statistics}{\nopostdesc}
\newacronym[parent=statistics,plural=MCs]{MC}{MC}{Monte Carlo}
\newacronym[parent=statistics,plural=MCMCs]{MCMC}{MCMC}{Markov chain Monte Carlo}
\newacronym[parent=statistics]{MAP}{MAP}{maximum a posteriori}
\newacronym[parent=statistics,plural=LAs,longplural={Laplace approximations}]{LA}{LA}{Laplace approximation}
\newacronym[parent=statistics,first=kernel density estimate (KDE) \cite{Parzen1962,Silverman1986},plural=KDEs,firstplural=kernel density estimates (KDEs) \cite{Parzen1962,Silverman1986},longplural={kernel density estimates}]{KDE}{KDE}{kernel density estimate}
\newacronym[parent=statistics,plural=PDFs,longplural={probability density functions}]{PDF}{PDF}{probability density function}
\newacronym[parent=statistics,plural=GPs,longplural={Gaussian processes}]{GP}{GP}{Gaussian process}
\newacronym[parent=statistics,plural=LGIs,longplural={linear Gaussian inversions}]{LGI}{LGI}{linear Gaussian inversion}
\newacronym[parent=statistics,plural=LSs,longplural={least-squares}]{LS}{LS}{least-square}
\newacronym[parent=statistics]{MaLi}{Max-$\mathcal{L}$}{maximum likelihood}
\newacronym[parent=statistics,plural=iid,longplural={independent and identically distributed}]{iid}{iid}{independent and identically distributed}
\newacronym[parent=statistics]{FWHM}{FWHM}{full-width half-maximum}
\newacronym[parent=statistics]{HDI}{HDI}{highest density interval}
\newacronym[parent=statistics]{MHA}{MHA}{Metropolis-Hastings algorithm}
\newacronym[parent=statistics]{KS}{KS}{Kolmogorov–Smirnov}
\newglossaryentry{PearsonCorrelationSample}{%
	parent=statistics,
	name=\ensuremath{r_{\text{xy}}},
	type=symbols,
	sort=correlation,
	description={Pearson sample correlation coefficient},
	symbol={},
}
\newacronym[parent=statistics,plural=HPs,longplural={hyper-parameters}]{HP}{HP}{hyper-parameter}

\newacronym{dynamicalSystems}{Dynamical systems}{\nopostdesc}
\newacronym[parent=dynamicalSystems]{KAM}{KAM}{Kolmogorov–Arnold–Moser}

\newacronym{methods}{Methods}{\nopostdesc}
\newacronym[parent=methods]{FP}{FP}{Function parametrization}
\newacronym[parent=methods]{MISO}{MISO}{multiple-input single-output}
\newacronym[parent=methods]{MIMO}{MIMO}{multiple-input multiple-output}

\newacronym{functionalAnalysis}{Functional analysis}{\nopostdesc}
\newacronym[parent=functionalAnalysis,plural=FCs,longplural={Fourier coefficients}]{FC}{FC}{Fourier coefficient}

\newacronym{GeneralPhysics}{General Physics}{\nopostdesc}
\newacronym[parent=GeneralPhysics]{IR}{IR}{Infrared}
\newacronym[parent=GeneralPhysics]{EM}{EM}{electro-magnetic}

\newacronym{Organisational}{Organisational}{\nopostdesc}
\newacronym[parent=Organisational]{OP}{OP}{operation phase}

\newacronym[parent=Organisational]{KKS}{KKS}{\emph{Kraftwerk-Kennzeichensystem} Reference Designation System for Power Plants}
\newacronym[parent=Organisational]{PLM}{PLM}{Project Lifecycle Management}

\newacronym{plasmaTheory}{Plasma Theory}{\nopostdesc}
\newacronym[parent=plasmaTheory,plural=tokamaks]{tokamak}{tokamak}{тороидальная камера в магнитных катушках}
\newacronym[parent=plasmaTheory]{MHD}{MHD}{magnetohydrodynamic}
\newacronym[parent=plasmaTheory]{LCFS}{LCFS}{last closed flux surface}
\newacronym[parent=plasmaTheory]{LCMS}{LCMS}{last closed magnetic surface}
\newacronym[parent=plasmaTheory]{ISS95}{ISS95}{international stellarator scaling 1995}
\newacronym[parent=plasmaTheory]{ISS04}{ISS04}{international stellarator scaling 2004}
\newacronym[parent=plasmaTheory]{NTM}{NTM}{neoclassical tearing modes}
\newacronym[parent=plasmaTheory]{HPP}{HPP}{heat pulse propagation}

\newacronym{plasmaPhysics}{Plasma Physics}{\nopostdesc}

\newacronym[parent=plasmaPhysics]{Omode}{O~mode}{ordinary mode}
\newacronym[parent=plasmaPhysics]{Xmode}{X~mode}{extraordinary mode}
\newacronym[parent=plasmaPhysics]{EBW}{EBW}{electron Bernstein wave}
\newacronym[parent=plasmaPhysics]{TM}{TM}{transverse magnetic}
\newacronym[parent=plasmaPhysics]{TEM}{TEM}{transverse electromagnetic}
\newacronym[parent=plasmaPhysics,plural=EEDFs,firstplural=electron energy distribution functions (EEDFs)]{EEDF}{EEDF}{electron energy distribution function}
\newacronym[parent=plasmaPhysics]{ECE}{ECE}{electron cyclotron emission}

\newacronym[parent=plasmaPhysics]{LFS}{LFS}{low field side}
\newacronym[parent=plasmaPhysics]{HFS}{HFS}{high field side}
\newacronym[parent=plasmaPhysics]{ELM}{ELM}{edge localized mode}
\newacronym[parent=plasmaPhysics]{MARFE}{MARFE}{multifaceted asymmetric radiation from the edge}
\newacronym[parent=plasmaPhysics]{Serpens}{Serpens}{self-regulated plasma edge beneath the last-closed-flux-surface}
\newacronym[parent=plasmaPhysics,plural=SOLs,firstplural=scrape-off layers]{SOL}{SOL}{scrape-off layer}
\newacronym[parent=plasmaPhysics]{ECRH}{ECRH}{electron cyclotron resonance heating}
\newacronym[parent=plasmaPhysics]{ECCD}{ECCD}{electron cyclotron current drive}
\newacronym[parent=plasmaPhysics]{NBCD}{NBCD}{neutral beam current drive}
\newacronym[parent=plasmaPhysics]{ICRH}{ICRH}{ion cyclotron radiation heating}
\newacronym[parent=plasmaPhysics]{FL}{FL}{field line}
\newacronym[parent=plasmaPhysics]{PFR}{PFR}{private flux region}

\newacronym[parent=plasmaPhysics]{see}{\textsc{SEE}}{secondary electron emission}
\newacronym[parent=plasmaPhysics]{SE}{\textsc{SE}}{sheath edge}
\newacronym[parent=plasmaPhysics]{SL}{SL}{strike line}
\newacronym[parent=plasmaPhysics]{PWI}{PWI}{plasma wall interaction}

\newacronym{fusionReactorComponent}{Nuclear Fusion Machine Components}{\nopostdesc}
\newacronym[parent=fusionReactorComponent,plural=PFCs,firstplural=plasma facing components (PFCs)]{PFC}{PFC}{plasma facing component}
\newacronym[parent=fusionReactorComponent,]{VV}{VV}{vaccum vessel}
\newacronym[parent=fusionReactorComponent,plural=IVCs,firstplural=in-vessel components (IVCs)]{IVC}{IVC}{in-vessel component}
\newacronym[parent=fusionReactorComponent,plural=TDUs,firstplural=test divertor units (TDUs)]{TDU}{TDU}{test divertor unit}
\newacronym[parent=fusionReactorComponent,plural=HHFDs, firstplural=high heatflux divertors (HHFDs)]{HHFD}{HHFD}{high heatflux divertor}
\newacronym[parent=fusionReactorComponent,]{TDUSE}{TDUSE}{test divertor unit scraper element}
\newacronym[parent=fusionReactorComponent,]{PG}{PG}{Pumping Gap}
\newacronym[parent=fusionReactorComponent,]{TE}{TE}{Target Element}

\newacronym[parent=fusionReactorComponent]{NBI}{NBI}{neutral beam injection}
\newacronym[parent=fusionReactorComponent,plural=PIs,firstplural=pellet injections (PIs)]{PI}{PI}{pellet injection}

\newacronym{plasmaTheoryCode}{Plasma Theory Codes}{\nopostdesc}
\newacronym[parent=plasmaTheoryCode]{FLD}{FLD}{field line diffusion}
\newacronym[parent=plasmaTheoryCode]{FLT}{FLT}{field line tracing}
\newacronym[parent=plasmaTheoryCode]{LOWENDEL}{LOWENDEL}{``loads on Wendelstein''}
\newacronym[parent=plasmaTheoryCode]{VMEC}{VMEC}{variational moments equilibrium code}
\newacronym[parent=plasmaTheoryCode]{EXTENDER}{EXTENDER}{virtual casing principle code}
\newacronym[parent=plasmaTheoryCode]{PIES}{PIES}{Princeton iterative equilibrium solver}
\newacronym[parent=plasmaTheoryCode]{SPEC}{SPEC}{stepped-pressure equilibrium code}
\newacronym[parent=plasmaTheoryCode]{EMC3E}{EMC3-Eirene}{Edge Monte Carlo 3D code coupled to the neutral transport code EIRENE}
\newacronym[parent=plasmaTheoryCode]{TRAVIS}{TRAVIS}{tracing visualized}
\newacronym[parent=plasmaTheoryCode]{SPECE}{SPECE}{solver for plasma electron cyclotron emission} 
\newacronym[parent=plasmaTheoryCode]{IDA}{IDA}{integrated data analysis}
\newacronym[parent=plasmaTheoryCode]{EFIT}{EFIT}{equilibrium fitting}
\newacronym[parent=plasmaTheoryCode]{Theo}{THEODOR}{Thermal Energy onto Divertor}
\newacronym[parent=plasmaTheoryCode]{STRAHL}{STRAHL}{Radial impurity transport and emission code}
\newacronym[parent=plasmaTheoryCode]{ROSE}{ROSE}{ROSE Optimises Stellarator Equilibria}
\newacronym[parent=plasmaTheoryCode]{V3FIT}{V3FIT}{three-dimensional equilibrium reconstruction}
\newacronym[parent=plasmaTheoryCode]{BSM}{BSM}{Bayesian scientific modeling}

\newacronym{database}{Databases}{\nopostdesc}
\newacronym[parent=database]{ADAS}{ADAS}{Atomic Data and Analysis Structure}

\newacronym{plasmaDiagnostic}{Plasma Diagnostics}{\nopostdesc}
\newacronym[parent=plasmaDiagnostic]{MPM}{MPM}{Multi-purpose manipulator}
\newacronym[parent=plasmaDiagnostic]{TS}{TS}{Thomson scattering}
\newacronym[parent=plasmaDiagnostic]{CTS}{CTS}{collective Thomson scattering}
\newacronym[parent=plasmaDiagnostic]{XICS}{XICS}{X-ray imaging crystal spectroscopy}
\newacronym[parent=plasmaDiagnostic]{CXRS}{CXRS}{charge exchange recombination spectroscopy}
\newacronym[parent=plasmaDiagnostic]{LiBES}{LiBES}{lithium beam emission spectroscopy}
\newacronym[parent=plasmaDiagnostic]{DCI}{DCI}{deuterium cyanide laser interferometry}
\newacronym[parent=plasmaDiagnostic,plural=RFMs,firstplural=reflectometers (RFMs)]{RFM}{RFM}{reflectometer}
\newacronym[parent=plasmaDiagnostic]{LP}{LP}{Langmuir probe}
\newacronym[parent=plasmaDiagnostic,sort=Ha-Camera]{HAC}{$H_\alpha$-C}{$H_\alpha$-Camera}
\newacronym[parent=plasmaDiagnostic,sort=Ha-Spectrometer]{HAS}{$H_\alpha$-S}{$H_\alpha$-Spectrometer}
\newacronym[parent=plasmaDiagnostic]{IRC}{IRC}{Infrared-Camera}
\newacronym[parent=plasmaDiagnostic]{HB}{HeBS}{Helium beam spectroscopy}
\newacronym[parent=plasmaDiagnostic]{Hexos}{HEXOS}{High efficiency extreme ultraviolet overview spectrometer}

\newacronym{gasDiagnostic}{Gas Diagnostics}{\nopostdesc}
\newacronym[parent=gasDiagnostic]{DRGA}{DRGA}{Diagnostic residual gas analyser}
\newacronym[parent=gasDiagnostic]{LIF}{LIF}{Laser induced fluorescence}

\newacronym{fusionExperiment}{Nuclear Fusion Experiments}{\nopostdesc}

\newacronym[parent=fusionExperiment,sort=W7]{W7}{\protect\mbox{W7}}{Wendelstein~\protect\mbox{7}}

\newacronym[parent=fusionExperiment,sort=W7-A]{W7A}{\protect\mbox{W7-A}}{Wendelstein~\protect\mbox{7-A}}

\newacronym[parent=fusionExperiment,sort=W7-AS]{W7AS}{\protect\mbox{W7-AS}}{Wendelstein~\protect\mbox{7-AS}}

\newacronym[parent=fusionExperiment,sort=W7-X]{W7X}{\protect\mbox{W7-X}}{Wendelstein~\protect\mbox{7-X}}

\newacronym[parent=fusionExperiment]{NCSX}{NCSX}{National Compact Stellarator Experiment}

\newacronym[parent=fusionExperiment]{CNT}{CNT}{Columbia Non-neutral Torus}

\newacronym[parent=fusionExperiment,sort=LHD]{LHD}{\protect\mbox{LHD}}{large helical device}

\newacronym[parent=fusionExperiment]{WEGA}{WEGA}{Wendelstein Experiment in Greifswald für die Ausbildung} 

\newacronym[parent=fusionExperiment]{HSX}{HSX}{Helically Symmetric Experiment}

\newacronym[parent=fusionExperiment]{JET}{JET}{Joint European Torus}

\newacronym[parent=fusionExperiment]{MAST}{MAST}{Mega-Ampere Spherical Tokamak}

\newacronym[parent=fusionExperiment]{AUG}{AUG}{ASDEX Upgrade}

\newacronym[parent=fusionExperiment]{ASDEX}{ASDEX}{axialsymmetrisches Divertor-Experiment}

\newacronym[parent=fusionExperiment]{EAST}{EAST}{experimental advanced superconducting tokamak}

\newacronym[parent=fusionExperiment]{ITER}{ITER}{lat.~\textit{the way} (formerly International Thermonuclear Experimental Reactor)}

\newacronym[parent=fusionExperiment]{AlcCM}{Alcator C-Mod}{Alto Campo Toro C-Mod} 

\newacronym[parent=fusionExperiment,sort=D]{D3D}{\mbox{DIII-D}}{Doublet III-D}

\newacronym[parent=fusionExperiment,sort=JT60]{JT60}{\mbox{JT-60}}{Japan Torus-60}
\newacronym[parent=fusionExperiment,sort=JT60SA]{JT60SA}{\mbox{JT-60SA}}{Japan Torus-60 Super Upgrade}

\newacronym[parent=fusionExperiment,sort=WEST]{WEST}{\mbox{WEST}}{Tungsten Environment in Steady-state Tokamak (formerly Tore Supra)}

\newacronym[parent=fusionExperiment,sort=TCV]{TCV}{\mbox{TCV}}{Tokamak à configuration variable}

\newacronym[parent=fusionExperiment,sort=TEXT]{TEXT}{\mbox{TEXT}}{Texas Tokamak}
\newacronym[parent=fusionExperiment,sort=TEXTU]{TEXTU}{\mbox{TEXT-U}}{Texas Tokamak - Upgrade}

\newacronym[parent=fusionExperiment,sort=Textor]{Textor}{\mbox{Textor}}{Tokamak Experiment for Technology Oriented Research}

\newacronym[parent=fusionExperiment,sort=Helias]{Helias}{\mbox{Helias}}{helical advanced stellarator}

\newacronym{plasmaLab}{Plasma Physics Laboratories}{\nopostdesc}

\newacronym[parent=plasmaLab]{PPPL}{PPPL}{Princeton Plasma Physics Laporatory}

\newacronym[parent=plasmaLab]{IPP}{IPP}{Max-Planck Institute for Plasma Physics}

\newacronym[sort=W7-X]{magneticConfig}{\mbox{W7-X} Magnetic Configurations}{\nopostdesc}
\newacronym[parent=magneticConfig]{EIM}{\textsc{EIM}}{'Standard' magnetic configuration in the centre of accessible space}
\newacronym[parent=magneticConfig]{DBM}{\textsc{DBM}}{Low iota magnetic configuration}

\newacronym{nuclearReaction}{Nuclear Reactions}{\nopostdesc}
\newacronym[parent=nuclearReaction]{DT}{DT}{Deuterium-Tritium}

\newglossaryentry{physics}{%
    name={General physics quantitites},
    type=symbols,
    description={\nopostdesc},
    symbol={},
}
\newglossaryentry{n}{%
    parent=physics,
    name=\ensuremath{n},
	type=symbols,
	sort=density,
	description={Particle density, $n = N / V$},
    symbol={\si{\per\cubic\meter}}
}
\newglossaryentry{T}{%
    parent=physics,
    name=\ensuremath{T},
	type=symbols,
	sort=temperature,
	description={Temperature},
    symbol={\si{\eV}}
}
\newglossaryentry{energy}{%
    parent=physics,
    name=\ensuremath{W},
	type=symbols,
	sort=energy,
	description={Plasma kinetic energy},
    symbol={\si{\kg\square\meter\per\square\second}}
}
\newglossaryentry{cs}{%
    parent=physics,
    name=\ensuremath{\sigma},
	type=symbols,
	sort=cross-section,
	description={Reaction cross-section},
    symbol={\si{\meter\square}}
}
\newglossaryentry{v}{%
    parent=physics,
    name=\ensuremath{\vect{v}},
	type=symbols,
	sort=velocity,
	description={Particle velocity},
    symbol={\si{\meter\per\second}}
}
\newglossaryentry{me}{%
    parent=physics,
	name=\ensuremath{m_{\text{e}}},
	type=symbols,
	sort= mass ,
	description={Electron mass},
    symbol={\si{\kilo\gram}},
}
\newglossaryentry{mp}{%
    parent=physics,
	name=\ensuremath{m_{\text{p}}},
	type=symbols,
	sort= mass ,
	description={Proton mass},
    symbol={\si{\kilo\gram}},
}

\newglossaryentry{B}{%
    parent=physics,
    name=\ensuremath{\vect{B}},
	type=symbols,
	sort=electromagnetism,
	description={Magnetic field},
    symbol={\si{\tesla} = \si{\kg\per\ampere\per\square\second}}
}
\newglossaryentry{E}{%
    parent=physics,
    name=\ensuremath{\vect{E}},
	type=symbols,
	sort=electromagnetism,
	description={Electric field},
    symbol={\si{\kg\meter\per\ampere\per\cubic\second}}
}
\newglossaryentry{chargedensity}{%
    parent=physics,
    name=\ensuremath{\rho},
	type=symbols,
	sort=electromagnetism,
	description={Electric charge density},
    symbol={\si{\ampere\second\per\cubic\meter}}
}
\newglossaryentry{currentdensity}{%
    parent=physics,
    name=\ensuremath{\vect{j}},
	type=symbols,
	sort=electromagnetism,
	description={Electric current density},
    symbol={\si{\ampere\per\square\meter}}
}
\newglossaryentry{resistance}{%
    parent=physics,
    name=\ensuremath{R},
	type=symbols,
	sort=impedance,
	description={Electric resistance},
    symbol={\si{\ohm}}
}
\newglossaryentry{inductance}{%
    parent=physics,
    name=\ensuremath{L},
	type=symbols,
	sort=inductance,
	description={Inductance},
    symbol={\si{\henry} = \si{\kg\square\meter\per\square\ampere\per\square\second}}
}
\newglossaryentry{larmor}{%
    parent=physics,
	name=\ensuremath{\rho},
	type=symbols,
	sort={Larmor radius},
	description={Larmor radius},
	symbol={\si{\meter}}
}

\newglossaryentry{plasmaphysics}{%
    name={Plasma physics quantitites},
    type=symbols,
    description={\nopostdesc},
    symbol={},
}

\newglossaryentry{vperp}{%
    parent=plasmaphysics,
	name=\ensuremath{v_{\perp}},
	type=symbols,
	sort=velocity,
	description={Field-perpendicular velocity},
	symbol={\si{\meter\per\second}}
}
\newglossaryentry{flowvelocity}{%
    parent=plasmaphysics,
    name=\ensuremath{\vect{v_{\text{f}}}},
	type=symbols,
	sort=velocity,
	description={Plasma flow velocity},
    symbol={\si{\meter\per\second}}
}
\newglossaryentry{p}{%
    parent=plasmaphysics,
    name=\ensuremath{p},
	type=symbols,
	sort=pressure,
	description={Plasma pressure},
    symbol={\si{\kg\per\meter\per\square\second}}
}
\newglossaryentry{V}{%
    parent=plasmaphysics,
    name=\ensuremath{V},
	type=symbols,
	sort=volume,
	description={Plasma volume},
    symbol={\si{\cubic\meter}}
}
\newglossaryentry{viscosity}{%
    parent=plasmaphysics,
    name=\ensuremath{\matr{\pi}},
	type=symbols,
	sort=viscosity,
	description={Plasma viscosity tensor},
    symbol={\si{\kg\per\meter\per\second}}
}
\newglossaryentry{Te}{%
    parent=plasmaphysics,
    name=\ensuremath{T_{\mathrm{e}}},
	type=symbols,
	sort=temperature,
	description={Electron temperature},
    symbol={\si{\eV}}
}
\newglossaryentry{Ti}{%
    parent=plasmaphysics,
	name=\ensuremath{T_{\mathrm{i}}},
	type=symbols,
	sort=temperature,
	description={Ion temperature},
    symbol={\si{\eV}}
}
\newglossaryentry{Tn}{%
    parent=plasmaphysics,
	name=\ensuremath{T_{\mathrm{n}}},
	type=symbols,
	sort=temperature,
	description={Neutral temperature},
	symbol={\si{\eV}}
}
\newglossaryentry{ni}{%
    parent=plasmaphysics,
    name=\ensuremath{n_{\mathrm{i}}},
	type=symbols,
	sort=density,
	description={Ion density},
    symbol={\si{\per\cubic\meter}}
}
\newglossaryentry{ne}{%
    parent=plasmaphysics,
    name=\ensuremath{n_{\mathrm{e}}},
	type=symbols,
	sort=density,
	description={Electron density},
    symbol={\si{\per\cubic\meter}}
}
\newglossaryentry{nn}{%
    parent=plasmaphysics,
	name=\ensuremath{n_{\mathrm{n}}},
	type=symbols,
	sort=density,
	description={Neutral density},
	symbol={\si{\per\cubic\meter}}
}

\newglossaryentry{nimp}{%
    parent=plasmaphysics,
	name=\ensuremath{n_{\mathrm{imp}}},
	type=symbols,
	sort=density,
	description={Impurity density},
	symbol={\si{\per\cubic\meter}}
}

\newglossaryentry{te}{%
    parent=plasmaphysics,
    name=\ensuremath{\tau_{\mathrm{E}}},
	type=symbols,
	sort=time,
	description={Energy confinement time},
    symbol={\si{\second}}
}
\newglossaryentry{radialShift}{%
    parent=plasmaphysics,
    name=\ensuremath{\makeup{\Delta} R},
	type=symbols,
	sort=confinement,
	description={Radial shift of the magnetic axis},
    symbol={\si{\meter}}
}
\newglossaryentry{plasmaBeta}{%
    parent=plasmaphysics,
    name=\ensuremath{\beta},
	type=symbols,
	sort=confinement,
	description={Plasma beta},
    symbol={}
}
\newglossaryentry{Itor}{%
    parent=plasmaphysics,
    name=\ensuremath{I_{\mathrm{tor}}},
	type=symbols,
	sort=confinement,
	description={Toroidal plasma current},
    symbol={\si{\ampere}}
}
\newglossaryentry{Ibs}{%
    parent=plasmaphysics,
    name=\ensuremath{I_{\mathrm{bs}}},
	type=symbols,
	sort=confinement,
	description={Bootstrap current},
    symbol={\si{\ampere}}
}
\newglossaryentry{Ips}{%
    parent=plasmaphysics,
    name=\ensuremath{I_{\mathrm{ps}}},
	type=symbols,
	sort=confinement,
	description={Pfirsch-Schlueter current},
    symbol={\si{\ampere}}
}
\newcommand{\quer}[1]{\mathrel{\hbox{-}\mkern-6.55mu #1}}
\newglossaryentry{ibar}{%
    parent=plasmaphysics,
    name=\ensuremath{\quer{\iota}},
	type=symbols,
	sort=confinement,
	description={Rotational transform},
    symbol={}
}

\newglossaryentry{shear}{%
	parent=plasmaphysics,
	name=\ensuremath{s},
	type=symbols,
	sort=confinement,
	description={Shear, radial derivative of \ensuremath{\quer{\iota}}},
	symbol={}
}

\newglossaryentry{ibarCF}{%
    parent=plasmaphysics,
    name=\ensuremath{\quer{\iota}_{\mathrm{CF}}},
	type=symbols,
	sort=confinement,
	description={Current free rotational transform},
    symbol={}
}

\newglossaryentry{Vp}{%
    parent=plasmaphysics,
    name=\ensuremath{V_\mathrm{p}},
	type=symbols,
	sort=potential,
	description={Plasma potential},
    symbol={\si{\volt}}
}
\newglossaryentry{Vf}{%
    parent=plasmaphysics,
    name=\ensuremath{V_\mathrm{f}},
	type=symbols,
	sort=potential,
	description={Floating potential},
    symbol={\si{\volt}}
}
\newglossaryentry{Isat}{%
    parent=plasmaphysics,
    name=\ensuremath{I_\text{sat}},
	type=symbols,
	sort=current,
	description={Ion saturation current},
    symbol={\si{\ampere}}
}	
\newglossaryentry{esat}{%
	parent=plasmaphysics,
	name=\ensuremath{I_{e,\text{sat}}},
	type=symbols,
	sort=current,
	description={Electron saturation current},
	symbol={\si{\ampere}}
}	
\newglossaryentry{jsat}{%
    parent=plasmaphysics,
    name=\ensuremath{j_\text{sat}},
	type=symbols,
	sort=current,
	description={Ion saturation current density},
    symbol={\si{\ampere\per\square\meter}}
}
\newglossaryentry{csound}{%
    parent=plasmaphysics,
    name=\ensuremath{c_\mathrm{s}},
	type=symbols,
	sort=velocity,
	description={Ion sound speed},
    symbol={\si{\meter\per\second}}
}
\newglossaryentry{Vbias}{%
    parent=plasmaphysics,
    name=\ensuremath{V_\mathrm{bias}},
	type=symbols,
	sort=potential,
	description={Bias voltage},
    symbol={\si{\volt}}
}
\newglossaryentry{r}{%
    parent=plasmaphysics,
    name=\ensuremath{r},
	type=symbols,
	sort=radius,
	description={Minor radius},
    symbol={\si{\meter}}
}
\newglossaryentry{ra}{%
    parent=plasmaphysics,
    name=\ensuremath{r_a},
	type=symbols,
	sort=radius,
	description={Minor radius of the last closed flux surface},
    symbol={\si{\meter}}
}
\newglossaryentry{reff}{%
    parent=plasmaphysics,
    name=\ensuremath{r_{\mathrm{eff}}},
	type=symbols,
	sort=radius,
	description={Effective minor radius},
    symbol={\si{\meter}}
}
\newglossaryentry{R}{%
    parent=plasmaphysics,
    name=\ensuremath{R},
	type=symbols,
	sort=radius,
	description={Major radius},
    symbol={\si{\meter}}
}
\newglossaryentry{normminrad}{%
	parent=plasmaphysics,
	name=\ensuremath{\varrho},
	type=symbols,
	sort=radius,
	description={Normalised minor radius},
	symbol={}
}
\newglossaryentry{aspectRatio}{%
    parent=plasmaphysics,
    name=\ensuremath{\epsilon},
	type=symbols,
	sort=radius,
	description={Aspect ratio},
    symbol={\si{}}
}
\newglossaryentry{Nfp}{%
    parent=plasmaphysics,
    name=\ensuremath{N_{\text{fp}}},
	type=symbols,
	sort=number,
	description={Number of field periods},
    symbol={\si{}},
}
\newglossaryentry{phiEdge}{%
    parent=plasmaphysics,
    name=\ensuremath{\phi_{\mathrm{edge}}},
	type=symbols,
	sort=magnetic flux,
	description={Total enclosed magnetic toroidal flux},
    symbol={\si{\kg\square\meter\per\square\second\per\ampere}}
}

\newglossaryentry{Pconv}{%
    parent=plasmaphysics,
    name=\ensuremath{P_{\mathrm{conv}}},
	type=symbols,
	sort=power,
	description={Convective power},
    symbol={\si{\watt}}
}
\newglossaryentry{Prad}{%
    parent=plasmaphysics,
    name=\ensuremath{P_{\mathrm{rad}}},
	type=symbols,
	sort=power,
	description={Radiated power},
    symbol={\si{\watt}}
}
\newglossaryentry{Pheat}{%
    parent=plasmaphysics,
    name=\ensuremath{P_{\mathrm{heat}}},
	type=symbols,
	sort=power,
	description={Total heating power},
    symbol={\si{\watt}}
}
\newglossaryentry{Lc}{%
    parent=plasmaphysics,
    name=\ensuremath{L_{\mathrm{c}}},
	type=symbols,
	sort=length,
	description={Connection length},
    symbol={\si{\meter}}
}
\newglossaryentry{heatflux}{%
    parent=plasmaphysics,
    name=\ensuremath{q},
	type=symbols,
	sort=power flux,
	description={Heat flux or heat load},
    symbol={\si{\MW\per\square\meter}}
}
\newglossaryentry{Zav}{%
    parent=plasmaphysics,
    name=\ensuremath{Z_{\mathrm{av}}},
	type=symbols,
	sort=ion charge,
	description={Average ion charge},
    symbol={\si{}},
}
\newglossaryentry{Zeff}{%
    parent=plasmaphysics,
    name=\ensuremath{Z_{\mathrm{eff}}},
	type=symbols,
	sort=ion charge,
	description={Effective ion charge},
    symbol={\si{}},
}
\newglossaryentry{iongyro}{%
    parent=plasmaphysics,
    name=\ensuremath{\rho_{\text{i}}},
	type=symbols,
	sort=radius,
	description={Ion gyro (Lamor) radius},
    symbol={\si{\meter}},
}
\newglossaryentry{mi}{%
    parent=plasmaphysics,
    name=\ensuremath{m_{\text{i}}},
	type=symbols,
	sort=mass,
	description={Ion mass},
    symbol={\si{\kilo\gram}},
}
\newglossaryentry{ndl}{%
	parent=plasmaphysics,
	name=\ensuremath{n\text{d}\ell},
	type=symbols,
	sort=density,
	description={Line integrated density},
	symbol={\si{\per\square\meter}},
}
\newglossaryentry{Wdia}{%
	parent=plasmaphysics,
	name=\ensuremath{W_\text{dia}},
	type=symbols,
	sort=energy,
	description={Diamagnetic energy},
	symbol={\si{\joule}},
}

\newglossaryentry{Halpha}{%
    parent=plasmaphysics,
	name=\ensuremath{\text{H}_{α}\xspace},
	type=symbols,
	sort= Wavelengths ,
	description={Hydrogen $\alpha$ transition line, Balmer series transition $n=3 \rightarrow n=2$, \SI{656.5}{\nano\meter}},
	symbol=\si{\nano\meter},
}
\newglossaryentry{sxb}{%
    parent=plasmaphysics,
	name=\ensuremath{\text{S/XB}},
	type=symbols,
	sort= Coefficients,
	description={Ratio of ionisation, excitation and branching ratio. Inverse photons per neutral.},
	symbol=,
}
\newglossaryentry{particleFlux}{%
    parent=plasmaphysics,
	name=\ensuremath{\Gamma},
	type=symbols,
	sort= Particle flux,
	description={Particle flux},
	symbol=\si{\per\second\per\square\meter},
}
\newglossaryentry{stf}{%
    parent=plasmaphysics,
	name=\ensuremath{\gamma_{\text{s}}},
	type=symbols,
	sort=Sheath transmission coefficient,
	description={Sheath transmission coefficient},
    symbol={\si{}},
}
\newglossaryentry{frad}{%
    parent=plasmaphysics,
	name=\ensuremath{f_{\text{rad}}},
	type=symbols,
	sort=Fraction,
	description={Radiated power fraction},
    symbol={\si{}},
}
\newglossaryentry{mfpi}{%
	parent=plasmaphysics,
	name=\ensuremath{\lambda_\text{mfp,i}},
	type=symbols,
	sort=Length,
	description={Mean free path length of ionisation},
	symbol={\si{\meter}},
}
\newglossaryentry{frec}{%
	parent=plasmaphysics,
	name=\ensuremath{f_\text{rec}},
	type=symbols,
	sort=Fraction,
	description={Fraction of ions recycled as neutrals at PFCs},
	symbol={\si{}},
}
\newglossaryentry{Pecrh}{%
	parent=plasmaphysics,
	name=\ensuremath{P_\text{ECRH}},
	type=symbols,
	sort=Power,
	description={ECRH Power},
	symbol={\si{\watt}},
}

\newacronym{engineering}{Engineering}{\nopostdesc}
\newacronym[parent=engineering]{MPCDF}{MPCDF}{Max Planck computing and data facility}

\newacronym[parent=engineering]{CAD}{CAD}{computer-aided design}
\newacronym[parent=engineering]{SOA}{SOA}{service-oriented architecture}
\newacronym[parent=engineering]{ESB}{ESB}{enterprise service bus}
\newacronym[parent=engineering]{HPC}{HPC}{high-performance computing}
\newacronym[parent=engineering,plural=GPUs,longplural={graphical processing units}]{GPU}{GPU}{graphical processing unit}

\newacronym[parent=engineering]{DAQ}{DAQ}{data acquisition}
\newacronym[parent=engineering]{ADC}{ADC}{analog-to-digital converter}
\newacronym[parent=engineering]{LO}{LO}{local oscillator}
\newacronym[parent=engineering]{PLL}{PLL}{phase locked loop}
\newacronym[parent=engineering]{CPU}{CPU}{central processing unit}
\newacronym[parent=engineering]{OAMP}{OAmp}{operational amplifier}
\newacronym[parent=engineering]{DAMP}{DAmp}{differential amplifier}
\newacronym[parent=engineering]{PCB}{PCB}{printed circuit board}
\newacronym[parent=engineering]{CMRR}{CMRR}{common mode rejection ratio}
\newacronym[parent=engineering]{IC}{IC}{integrated circuit}
\newacronym[parent=engineering]{CCD}{CCD}{charge-coupled device}
\newacronym[parent=engineering]{PMT}{PMT}{photo-multiplyer tube}
\newacronym[parent=engineering]{ROI}{ROI}{region of interest}
\newacronym[parent=engineering,plural=LOSs,longplural={lines of sight}]{LOS}{LOS}{line of sight}
\newacronym[parent=engineering]{EDICAM}{EDICAM}{Event Detection Intelligent Camera}

\newacronym[parent=engineering]{CFC}{CFC}{carbon fibre reinforced carbon}

\newacronym{machineLearning}{Machine Learning}{\nopostdesc}
\newacronym[parent=machineLearning]{ML}{ML}{machine learning}
\newacronym[parent=machineLearning]{RL}{RL}{reinforcement learning}
\newacronym[parent=machineLearning,plural=NNs,firstplural=artificial neural networks (NNs)]{NN}{NN}{artificial neural network}
\newacronym[parent=machineLearning]{EI}{EI}{expected improvement}

\newacronym[parent=machineLearning,plural=FF-FCs,firstplural=feed-forward fully-connected neural networks (FF-FC)]{FFFC}{FF-FC}{feed-forward fully-connected neural network}
\newacronym[parent=machineLearning,plural=CNNs,firstplural=convolutional neural networks (CNNs)]{CNN}{CNN}{convolutional neural network}
\newacronym[parent=machineLearning,plural=DCNNs,firstplural=deep convolutional neural networks (DCNNs)]{DCNN}{DCNN}{deep convolutional neural network}
\newacronym[parent=machineLearning,plural=IRNNs,firstplural=Inception-ResNets (IRNNs)]{IRNN}{IRNN}{Inception-ResNet}
\newacronym[parent=machineLearning,plural=DINNs,firstplural=deep inception neural networks (DINNs)]{DINN}{DINN}{deep inception neural network}
\newacronym[parent=machineLearning,plural=GANs,firstplural=generative adversarial neural networks (GANs)]{GAN}{GAN}{generative adversarial neural network}
\newacronym[parent=machineLearning,plural=DQNs,firstplural=deep Q-networks]{DQN}{DQN}{deep Q-network}

\newacronym[parent=machineLearning]{MNIST}{MNIST}{MNIST} 

\newacronym[parent=machineLearning]{ReLU}{ReLU\xspace}{rectified linear unit}
\newacronym[parent=machineLearning]{SeLU}{SeLU\xspace}{scaled exponential linear unit}

\newacronym[parent=machineLearning]{rmse}{rmse\xspace}{root-mean-square error}
\newacronym[parent=machineLearning]{mse}{mse\xspace}{mean-squared error}
\newacronym[parent=machineLearning]{msd}{msd\xspace}{mean-squared deviation}
\newacronym[parent=machineLearning]{nmse}{nmse\xspace}{normalized mean-squared error}
\newacronym[parent=machineLearning]{rmsd}{rmsd\xspace}{root-mean-squared deviation}
\newacronym[parent=machineLearning]{nrmsd}{nrmsd\xspace}{normalized root-mean-squared deviation}
\newacronym[parent=machineLearning]{nrmse}{nrmse\xspace}{normalized root-mean-squared error}
\newacronym[parent=machineLearning]{mae}{mae\xspace}{mean absolute error}

\newacronym[parent=machineLearning]{TPE}{TPE}{tree-structured Parzen estimator}
\newacronym[parent=machineLearning]{NAS}{NAS}{neural architecture search}
\newacronym[parent=machineLearning]{NES}{NES}{neural ensemble search}
\newacronym[parent=machineLearning]{SMBO}{SMBO}{sequential model-based optimization}

\newglossaryentry{ml}{%
    name={Machine learning quantitites},
    type=symbols,
    description={\nopostdesc},
    symbol={},
}
\newglossaryentry{lossfunction}{%
    name=\ensuremath{\mathcal{L}},
	type=symbols,
	parent=ml,
	sort=loss,
	description={Loss function},
    symbol={\si{}}
}

\setproperty{document}{title}{Proof of concept of a fast surrogate model of the VMEC code via neural networks in Wendelstein 7-X scenarios}
\setproperty{author}{firstname}{Andrea}
\setproperty{author}{familyname}{Merlo}
\setproperty{coauthor1}{firstname}{Daniel}
\setproperty{coauthor1}{familyname}{Böckenhoff}
\setproperty{coauthor1}{affiliationindices}{1}
\setproperty{coauthor2}{firstname}{Jonathan}
\setproperty{coauthor2}{familyname}{Schilling}
\setproperty{coauthor2}{affiliationindices}{1}
\setproperty{coauthor3}{firstname}{Udo}
\setproperty{coauthor3}{familyname}{Höfel}
\setproperty{coauthor3}{affiliationindices}{1}
\setproperty{coauthor4}{firstname}{Sehyun}
\setproperty{coauthor4}{familyname}{Kwak}
\setproperty{coauthor4}{affiliationindices}{1}
\setproperty{coauthor5}{firstname}{Jakob}
\setproperty{coauthor5}{familyname}{Svensson}
\setproperty{coauthor5}{affiliationindices}{1}
\setproperty{coauthor6}{firstname}{Andrea}
\setproperty{coauthor6}{familyname}{Pavone}
\setproperty{coauthor6}{affiliationindices}{1}
\setproperty{coauthor7}{firstname}{Samuel Aaron}
\setproperty{coauthor7}{familyname}{Lazerson}
\setproperty{coauthor7}{affiliationindices}{1}
\setproperty{coauthor8}{firstname}{Thomas Sunn}
\setproperty{coauthor8}{familyname}{Pedersen}
\setproperty{coauthor8}{affiliationindices}{1}
\setproperty{coauthors}{W7Xteaminclude}{True}
\setproperty{affiliations}{1}{Max-Planck-Institute for Plasma Physics, 17491 Greifswald, Germany}

\definechangesauthor[name={Andrea Merlo}, color=red]{AM}
\definechangesauthor[name={Daniel Böckenhoff}, color=blue]{DB}
\definechangesauthor[name={Jonathan Schilling}, color=orange]{JS}
\definechangesauthor[name={Thomas Sunn Pedersen}, color=magenta]{TP}

\newcommand{\D}{\ensuremath{\mathcal{D}}\xspace}
\newcommand{\Dnull}{\ensuremath{\mathbb{D}_{\text{config}}}\xspace}
\newcommand{\Dfinite}{\ensuremath{\mathbb{D}_{\beta}}\xspace}

\newcommand{\nullIota}{\textit{config-iota}\xspace}
\newcommand{\finiteIota}{\textit{$\beta$-iota}\xspace}
\newcommand{\nullSurfaces}{\textit{config-surfaces}\xspace}
\newcommand{\finiteSurfaces}{\textit{$\beta$-surfaces}\xspace}
\newcommand{\nullB}{\textit{config-B}\xspace}
\newcommand{\finiteB}{\textit{$\beta$-B}\xspace}

\newcommand{\Ns}{\ensuremath{N_{\text{s}}}\xspace}
\newcommand{\mpol}{\ensuremath{m_{\text{pol}}}\xspace}
\newcommand{\ntor}{\ensuremath{n_{\text{tor}}}\xspace}
\newcommand{\hatmpol}{\ensuremath{\hat{m}_{\text{pol}}}\xspace}
\newcommand{\hatntor}{\ensuremath{\hat{n}_{\text{tor}}}\xspace}
\newcommand{\hatNs}{\ensuremath{\hat{N}_{\text{s}}}\xspace}

\newcommand{\averagePlasmaBeta}{\ensuremath{\langle \gls{plasmaBeta} \rangle}\xspace}

\newcommand{\rmseStarIota}{\ensuremath{\text{\gls{rmse}}_{\gls{ibar}}^*}\xspace}
\newcommand{\rmseStarRZ}{\ensuremath{\text{\gls{rmse}}_{\text{R,Z}}^*}\xspace}
\newcommand{\rmseStarLambda}{\ensuremath{\text{\gls{rmse}}_{\lambda}^*}\xspace}
\newcommand{\rmseStarB}{\ensuremath{\text{\gls{rmse}}_{\text{B}}^*}\xspace}

\newcommand{\lCore}{\ensuremath{l_{\text{core}}}\xspace}
\newcommand{\lEdge}{\ensuremath{l_{\text{edge}}}\xspace}

\newcommand{\tTrain}{\ensuremath{t_{\text{train}}}\xspace}
\newcommand{\tPrediction}{\ensuremath{t_{\text{inference}}}\xspace}

\definecolor{lightNullColor}{HTML}{d1e5f0}
\definecolor{nullColor}{HTML}{67a9cf}
\definecolor{darkNullColor}{HTML}{2166ac}
\definecolor{lightFiniteColor}{HTML}{fddbc7}
\definecolor{finiteColor}{HTML}{ef8a62}
\definecolor{darkFiniteColor}{HTML}{b2182b}

\definecolor{iotaColor}{HTML}{0571b0}
\definecolor{bColor}{HTML}{008837}
\definecolor{rColor}{HTML}{66c2a5}
\definecolor{lColor}{HTML}{fc8d62}
\definecolor{zColor}{HTML}{8da0cb}

\newcommand{\subfigureWidth}{\linewidth - 1em}
\newcommand{\subfigureTwo}{0.5}
\newcommand{\subfigureThree}{0.33}

\newcommand{\tikzMarkSize}{3pt}

\newcommand{\lineWidth}{0.48pt}
\newcommand{\thickLineWidth}{1pt}
\newcommand{\legendLineLength}{15pt}

\DeclareRobustCommand\nullSquare{{\tikz{\node[rectangle,fill=darkNullColor](){};}}}
\DeclareRobustCommand\finiteSquare{{\tikz{\node[rectangle,fill=darkFiniteColor](){};}}}

\DeclareSIUnit{\nothing}{\relax}

\makeatletter
\let\blx@rerun@biber\relax
\makeatother

\begin{document}

    \typeout{----- BEGIN DOCUMENT -----}
    \typeout{}
    \typeout{--------------------------------}
    \typeout{----- Document properties: -----}
    \typeout{--------------------------------}
    \typeout{Font size: \the\fsize}
    \typeout{Text width: \the\textwidth}
    \typeout{--------------------------------}
    \typeout{--------------------------------}

    \ifthenelseproperty{compilation}{frontmatter}{%
	\frontmatter
}{}

\ifthenelseproperty{compilation}{titlepage}{%
    \ifKOMA
    \pagestyle{empty}%
    \title{\getproperty{document}{title}\thanks{This is the Accepted Manuscript version of an article accepted for publication in Nuclear Fusion. IOP Publishing Ltd is not responsible for any errors or omissions in this version of the manuscript or any version derived from it. This Accepted Manuscript is published under a CC BY licence. The Version of Record is available online at \url{https://doi.org/10.1088/1741-4326/ac1a0d}}}%
    \author[\getproperty{author}{affiliationindices}]{\getproperty{author}{firstname} \getproperty{author}{familyname}}
    \affil[1]{\getproperty{affiliations}{1}}
    
    \ifpropertydefined{coauthor1}{familyname}{
    \author[\getproperty{coauthor1}{affiliationindices}]{\getproperty{coauthor1}{firstname} \getproperty{coauthor1}{familyname}}}{}

    \ifpropertydefined{coauthor2}{familyname}{
    \author[\getproperty{coauthor2}{affiliationindices}]{\getproperty{coauthor2}{firstname} \getproperty{coauthor2}{familyname}}}{}

    \ifpropertydefined{coauthor3}{familyname}{
    \author[\getproperty{coauthor3}{affiliationindices}]{\getproperty{coauthor3}{firstname} \getproperty{coauthor3}{familyname}}}{}

    \ifpropertydefined{coauthor4}{familyname}{
    \author[\getproperty{coauthor4}{affiliationindices}]{\getproperty{coauthor4}{firstname} \getproperty{coauthor4}{familyname}}}{}

    \ifpropertydefined{coauthor5}{familyname}{
    \author[\getproperty{coauthor5}{affiliationindices}]{\getproperty{coauthor5}{firstname} \getproperty{coauthor5}{familyname}}}{}

    \ifpropertydefined{coauthor6}{familyname}{
    \author[\getproperty{coauthor6}{affiliationindices}]{\getproperty{coauthor6}{firstname} \getproperty{coauthor6}{familyname}}}{}

    \ifpropertydefined{coauthor7}{familyname}{
    \author[\getproperty{coauthor7}{affiliationindices}]{\getproperty{coauthor7}{firstname} \getproperty{coauthor7}{familyname}}}{}

    \ifpropertydefined{coauthor8}{familyname}{
    \author[\getproperty{coauthor8}{affiliationindices}]{\getproperty{coauthor8}{firstname} \getproperty{coauthor8}{familyname}}}{}

    \ifpropertydefined{coauthors}{W7Xteaminclude}{
    \author[ ]{the W7-X team}}{}
    
    \ifpropertydefined{affiliations}{2}{
        \affil[2]{\getproperty{affiliations}{2}}
    }{}

    \date{\today}%
    \maketitle%
    \pagestyle{scrheadings}
\else
	\pagestyle{empty}
	\title{\getproperty{document}{title}}
	\author{\getproperty{author}{firstname} \getproperty{author}{familyname}}
	\date{\getproperty{document}{date}}
	\maketitle
\fi

}{}

\ifthenelseproperty{compilation}{abstract}{%
	\section*{Abstract}
In magnetic confinement fusion research,
the achievement of high plasma pressure is key to reaching the goal of net energy production.
The \gls{MHD} model is used to self-consistently calculate the effects the plasma pressure induces on the magnetic field used to confine the plasma.
Such \gls{MHD} calculations --- usually done computationally --- serve as input for the assessment of a number of important physics questions.
The \gls{VMEC} is the most widely used to evaluate 3D ideal-\gls{MHD} equilibria,
as prominently present in stellarators.
However,
considering the computational cost,
it is rarely used in large-scale or online applications (\eg, \glsentrylong{BSM}, real-time plasma control).
Access to fast \gls{MHD} equilbria is a challenging problem in fusion research,
one which \glsentrylong{ML} could effectively address.
In this paper,
we present \gls{NN} models able to quickly compute the equilibrium magnetic field of \glsentrylong{W7X}.
Magnetic configurations that extensively cover the device operational space,
and plasma profiles with volume-averaged normalized plasma pressure \averagePlasmaBeta (\gls{plasmaBeta} = $\nicefrac{2 \mu_0 p}{B^2}$) up to \SI{5}{\percent} and non-zero net toroidal current are included in the data set.
By using convolutional layers,
the spectral representation of the magnetic flux surfaces can be efficiently computed with a single network.
To discover better models, a Bayesian hyper-parameter search is carried out,
and 3D \glsentrylongpl{CNN} are found to outperform \glsentrylongpl{FFFC}.
The achieved \glsentrylong{nrmse},
the ratio between the regression error and the spread of the data,
ranges from \SI{1}{\percent} to \SI{20}{\percent} across the different scenarios.
The model inference time for a single equilibrium is on the order of milliseconds.
Finally,
this work shows the feasibility of a fast \gls{NN} drop-in surrogate model for \gls{VMEC},
and it opens up new operational scenarios where target applications could make use of magnetic equilibria at unprecedented scales.
\glsresetall%

}{}

\ifthenelseproperty{compilation}{glossaries}{
	\glsresetall
}{}

\ifthenelseproperty{compilation}{toc}{%
    \disabledprotrusion{\tableofcontents}
}{}

\makeatletter
\@ifundefined{mainmatter}{}{\mainmatter}
\makeatother


\section{Introduction}\label{sec:introduction}

The computation of \gls{MHD} equilibria is central in magnetic confinement fusion,
where it represents the core component of most modeling and experimental applications.
In the stellarator community,
the 3D ideal-\gls{MHD} \gls{VMEC}~\cite{Hirshman1983} is the most widely used,
\eg,
to infer plasma parameters\deleted[id=AM]{ in classical data analysis and \gls{BSM} frameworks}~\cite{Langenberg2016,Bozhenkov2020},
to reconstruct magnetic equilibria~\cite{Hanson2009,Lazerson2015,Andreeva2019a,Howell2020,Lazerson2020},
and to design future devices~\cite{Mynick2002,Drevlak2019,Feng2020}.
\gls{VMEC} is also employed for equilibrium studies in perturbed\added[id=AM]{,} and hence non-2D\added[id=AM]{,} \replaced[id=AM]{axisymmetric}{tokamak} \replaced[id=AM]{configurations}{magnetic fields}~\cite{Terranova2013,Lazerson2013,Chapman2014,Lazerson2014,Schmitt2014,King2015,Lazerson2016a,Koliner2016,Wingen2017,Cianciosa2017,Cianciosa2018}.
However,
a single \gls{VMEC} equilibrium evaluation can take up to \orderof{\num{10}} minutes\footnote{Run time on the \gls{MPCDF} cluster ``DRACO'', using the \textit{small} partition and 16 cores.}
even on a \gls{HPC} facility,
especially for a reactor-relevant high-\gls{plasmaBeta} plasma configuration.
\Cref{tab:vmec-wall-time} reports the orders of magnitude of \gls{VMEC} total iterations and wall-clock time typically encountered in target applications.
The high computational cost limits an exhaustive exploration of the use case input space.
A parallel version of \gls{VMEC} has recently been developed~\cite{Seal2016},
however,
for example,
the wall-clock time of a single free boundary equilibrium reconstruction,
both in the case of a stellarator and a tokamak scenario,
is still on the order of hours~\cite{Seal2017,SchmittPC2021}.

\begin{table}
  \caption{
    Order of magnitude of \gls{VMEC} iterations and wall-clock time in target applications.
    \replaced[id=AM]{
      Fixed-boundary equilibria are considered in stellarator optimization,
      while \replaced[id=AM]{the inference of plasma parameters}{\gls{BSM}} and equilibrium reconstruction usually requires free-boundary equilibria.
      \gls{VMEC} computation time strongly depends on the run requirements
      (\eg, radial resolution, Fourier resolution, field periodicity, convergence tolerance),
      thus the \SI{e1}{\second} -- \SI{e4}{\second} range has been considered in this table.
    }{A single \gls{VMEC} run time of \mbox{\SI{e2}{\second}} and \mbox{\SI{e3}{\second}} is considered in case of fixed and free boundary mode, respectively.}
  }%
  \label{tab:vmec-wall-time}
  \small
  \centering
  \begin{tabular}{lcc}
    \toprule
    Application & Iterations & Time [\si{\second}]\\
    \midrule
    \replaced[id=AM]{Bayesian Inference}{\gls{BSM}}~\protect\cite{Hoefel2019a} & \num{e4} & \replaced[id=AM]{\num{e5} -- \num{e8}}{\num{e7}}\\
    Equilibrium reconstruction~\protect\cite{Hanson2009,SchmittPC2021} & \replaced[id=AM]{\num{e0}}{\num{e1}} & \replaced[id=AM]{\num{e1} -- \num{e4}}{\num{e4}}\\
    Stellarator optimization~\protect\cite{Paul2018} & \num{e3} & \replaced[id=AM]{\num{e4} -- \num{e7}}{\num{e5}}\\
    \bottomrule
  \end{tabular}
\end{table}

In this paper,
we use \glspl{NN} (see~\cref{sec:architectures}) as function approximators to build a fast surrogate model for \gls{VMEC}.
A reduction in run times of up to \num{6} orders of magnitude can be achieved.
The models are trained on \gls{VMEC} runs from two independent data sets:
\Dnull and \Dfinite (see~\ref{sec:beta-scenarios}).
\Dnull includes a wide range of vacuum magnetic configurations,
while \Dfinite covers a distribution of plasma profiles for a fixed magnetic configuration.
To find better models,
and to take the human out of the loop,
a Bayesian \gls{HP} search is performed (see~\cref{sec:pipeline}).

Since neural networks poorly extrapolate beyond the expressiveness of training data,
a large and experimentally relevant data set is essential for good out-of-sample performance (see~\cref{sec:dataset}).
Training runs are sampled as employed in the Bayesian scientific modeling framework Minerva~\cite{Svensson2007,Svensson2010},
aiming to reduce the covariate shift between the training and test data set.
The magnetic configurations are sampled from a large hyper-rectangle around the nine \gls{W7X} reference configurations~\cite{Andreeva2002},
while the plasma profiles are modeled as \glspl{GP}~\added[id=AM]{\mbox{\cite{Svensson2011}}},
and domain knowledge is embedded in the training data \replaced[id=AM]{through}{trough} \textit{virtual observations}\replaced[id=AM]{~\mbox{\cite{Svensson2004,Ford2010,Pavone2019b,Kwak2020a}}}{~\mbox{\cite[Section~8.2]{Pavone2019b}}}.

Since \gls{VMEC} assumes nested magnetic flux surfaces, magnetic islands in the equilibrium field are not included by design.
Furthermore, an ideal coil geometry (\ie, no coil misalignment or \glsentrylong{EM} deformations) is considered,
while ideal coil currents and plasma profiles (\ie, error-free measurements) are assumed.
The relaxation of these assumptions is not in the scope of this paper.

In the past,
Sengupta et.\ al successfully regressed single \glspl{FC} of the \gls{VMEC} output magnetic field,
using \gls{FP} with quadratic or cubic polynomials for vacuum~\cite{Sengupta2004} and finite beta~\cite{Sengupta2007} magnetic configurations.
The regression of the full \gls{VMEC} output was broken down into subproblems,
where a \gls{FP} model was derived for each \gls{FC},
leading to many free parameters to learn.
In this work,
on top of the previously mentioned components (\ie, physics-like plasma profiles and \gls{HP} search),
the learning task is to infer the full magnetic field geometry with a single \gls{MIMO} model,
where all the \gls{VMEC} output \glspl{FC} are regressed at once (see~\cref{sec:inputs-and-outputs}).
Using a single model drastically reduces the number of free parameters to learn,
and it forces the \gls{NN} to efficiently share them among the outputs.
Contrary to \gls{FP},
it is well-known that sufficiently wide or deep \glspl{NN} can approximate a broad class of functions~\cite{Cybenko1989a,Hornik1991a,Pinkus1999,Eldan2016,Lu2017}.
In addition,
\glspl{CNN},
as powerful tools of current deep learning methods,
are better suited to extract and reproduce translation-invariant spatial features from grid data,
and to share their free parameters between the features while reducing overfitting.
Furthermore,
from the user standpoint,
a single model can more easily be improved, adapted, and deployed.

For real-time plasma control,
having access to low-cost magnetic equilibria can improve traditional strategies,
and enable completely new data-driven approaches (\eg, \gls{RL} based control).
In fusion research,
the use of \gls{NN} models to compute the plasma topology~\cite{VanMilligen1995,Tribaldos1997,Joung2020} and to speed up slow workflows~\cite{Citrin2015a,Meneghini2017,VandePlassche2020,Pavone2020,Piccione2020} is not a novel idea,
nevertheless,
to our knowledge this paper represents the first which effectively addresses the 3D \gls{MHD} physics in \gls{W7X} scenarios.

\section{Methods}
\label{sec:method}

In the following relevant concepts and employed methodologies are described.

\subsection{The \gls{VMEC} code}\label{sec:vmec}

The equilibrium problem under the ideal-\gls{MHD} model is characterized by the force balance equation,
Ampere's and Gauss's law

\begin{gather}
  \vec{J} \times \vec{B} = \vec{\nabla} p \\
  \vec{\nabla} \times \vec{B} = \mu_0 \vec{J} \\
  \vec{\nabla} \cdot \vec{B} = 0.
\end{gather}

\gls{VMEC} uses a variational principle to solve the \textit{inverse} formulation,
which computes the mapping $f: \vec{\zeta} \rightarrow \vec{x}$ between flux coordinates $\vec{\zeta} = (s, \theta, \varphi)$,
normalized toroidal flux ($s = \nicefrac{\Phi}{\Phi_{\text{edge}}}$, where $\Phi(s)$ is the toroidal magnetic flux enclosed between the magnetic axis and the flux surface labeled $s$), poloidal and toroidal angle, respectively,
and real space cylindrical coordinates $\vec{x} = (R, \varphi, Z)$,
major radius, azimuth and height above mid-plane, respectively.
\gls{VMEC} adopts a spectral representation of $\vec{x}$ along the poloidal and toroidal angles.
Assuming stellarator symmetry,
the cylindrical coordinates can be expressed as

\begin{gather}
  R( s, \theta, \varphi ) = \sum\nolimits_{mn} R_{mn} ( s ) \cos ( m \theta - n \gls{Nfp} \varphi ) \label{eq:R}\\
  Z( s, \theta, \varphi ) = \sum\nolimits_{mn} Z_{mn} ( s ) \sin ( m \theta - n \gls{Nfp} \varphi ),\label{eq:Z}
\end{gather}

where $\gls{Nfp} \in \naturals$ is the number of field periods.
Furthermore,

\begin{gather}
  \lambda( s, \theta, \varphi ) = \sum\nolimits_{mn} \lambda_{mn} ( s ) \sin ( m \theta - n \gls{Nfp} \varphi ) \label{eq:lambda}
\end{gather}

is an angle renormalization parameter such that $\theta^* = \theta + \lambda(s, \theta, \varphi)$ represents the poloidal angle for which magnetic field lines are straight in $(s, \theta^*, \varphi)$~\cite{Hirshman1983}.
The equilibrium magnetic field $\vec{B}$ can be written in contravariant form

\begin{gather}
    \vec{B} = B^s \hat{e}_s + B^{\theta} \hat{e}_{\theta} + B^{\varphi} \hat{e}_{\varphi} = B^{\theta} \hat{e}_{\theta} + B^{\varphi} \hat{e}_{\varphi}, \label{eq:b}
\end{gather}

where $\vec{B} \cdot \vec{\nabla} p = B^s = 0$ under the assumption of nested magnetic flux surfaces.
The non-zero components are given by~\cite{Hirshman1983}

\begin{gather}
  B^{\theta} = \frac{1}{\sqrt{g}} \Phi^{\prime} (\gls{ibar} - \frac{\partial \lambda}{\partial \varphi}) \label{eq:b-theta-up}\\
  B^{\varphi} = \frac{1}{\sqrt{g}} \Phi^{\prime}(1 + \frac{\partial \lambda}{\partial \theta}), \label{eq:b-phi-up}
\end{gather}

where \gls{ibar} is the rotational transform,
the prime denotes $\partial / \partial s$,
and $\sqrt{g} = (\vec{\nabla} s \cdot \vec{\nabla} \theta \times \vec{\nabla} \varphi)^{-1}$ is the Jacobian of the coordinate transformation $f$.

The covariant representation of $\vec{B}$ can be obtained from~\cref{eq:b-theta-up,eq:b-phi-up} and the metric tensor $g_{ij} = \hat{e}_i \cdot \hat{e}_j = \frac{\partial \vec{x}}{\partial \zeta_i} \cdot \frac{\partial \vec{x}}{\partial \zeta_j}$ as follows

\begin{gather}
  B_{\theta} =  \vec{B} \cdot \hat{e}_{\theta} = B^{\theta} g_{\theta \theta} + B^{\varphi} g_{\varphi \theta}, \label{eq:b-theta-down}\\
  B_{\varphi} = \vec{B} \cdot \hat{e}_{\varphi} = B^{\theta} g_{\theta \varphi} + B^{\varphi} g_{\varphi \varphi}. \label{eq:b-phi-down}
\end{gather}

Finally,
the magnetic field vector strength is given by

\begin{gather}
  B^2 = \sum\nolimits_{i}{B^i B_i = (B^{\theta})^2 g_{\theta \theta} + 2 B^{\theta} B^{\varphi} g_{\theta \varphi} + (B^{\varphi})^2 g_{\varphi \varphi}}. \label{eq:b-squared}
\end{gather}

As in case of $\vec{x}$, the magnetic field strength is described by \gls{VMEC} using a spectral representation:

\begin{gather}
  B( s, \theta, \varphi ) = \sum\nolimits_{mn} B_{mn} ( s ) \cos ( m \theta - n \gls{Nfp} \varphi ), \label{eq:b-mn}
\end{gather}

Like in~\cite{Hirshman1983},
$\vec{x}$ is redefined as $\vec{x} = (R, \lambda, Z)$,
where the angle renormalization parameter $\lambda$ replaces the toroidal angle $\varphi$.

\subsection{Data set generation}\label{sec:dataset}

To generate a large and \gls{W7X} relevant data set of magnetic configurations and plasma profiles,
Minerva\added[id=AM]{~\cite{Svensson2007,Svensson2010}} is used.
Within Minerva,
models are described as directed, acyclic graphs.
Each node can be deterministic (\eg, a diagnostic model or a physics code)
or probabilistic (\eg, plasma parameters or diagnostic observed quantities).
The edges define the dependencies between nodes.
Model free parameters can be described via probabilistic nodes,
where the node a priori distribution encodes the domain knowledge on the parameter.
In the \textit{forward mode},
observed quantities can be computed,
while in the \textit{inverse mode},
the model free parameters can be inferred with different inversion techniques (\eg, \gls{MAP} and \gls{MCMC} methods).

Using Minerva to generate physics relevant samples for \glspl{NN} training has already been explored~\cite{Pavone2018}.
Here,
a \gls{VMEC} node is included in a Minerva model.
Free parameters are represented by the magnetic configuration and the plasma profiles.
The model is relatively simple and can be built as a stand-alone object,
yet Minerva allows embedding domain knowledge \added[id=AM]{(\ie, the prior distribution of the model free parameters)} in the \gls{NN} surrogate \deleted[id=AM]{model }by reducing the covariate shift between the training data set \D and the target application data set $\D_{\text{target}}$.
\replaced[id=AM]{
  This approach is similar to that described in~\cite{Ho2021},
  where experimental data have been used to populate the training data set.
  However,
  in this work,
  experimental data are not used directly,
  but simulated data drawn from experimentally validated distributions are used instead.
  This allows a dense coverage of the input parameter space,
  while restricting its extension to physically relevant regions only.
}{
  In particular,
  the plasma profiles are carefully restricted to physically relevant profiles only.
}

\gls{W7X} possesses a $\gls{Nfp}=5$-fold stellarator symmetry,
\ie,
the main coil system comprises five identical modules,
each of which is point symmetric towards the module center (see \Cref{sec:vmec}).
The resulting magnetic field has a five-fold symmetry along the toroidal direction.
Each half module includes five different non-planar and two planar coils.
The vacuum field depends only on the currents $I_{1 \ldots 5}$ and $I_{A, B}$,
respectively,
the currents in the non-planar and planar coils.
Except for a scaling of the magnetic field strength,
the vacuum magnetic configuration does not depend on the absolute values of
the coil currents but only on their ratios with respect to $I_1$,
$i_{2 \ldots 5}$ and $i_{A, B}$.
The current ratios are uniformly sampled from a hyper-rectangle whose boundaries are provided in \Cref{tab:hp-boundaries}.
These boundaries cover the nine reference configurations of \gls{W7X}~\cite{Andreeva2002},
while extending to a larger set of conceivable configurations.
To obtain a magnetic field strength of approximately \SI{2.5}{\tesla} on axis, at $\varphi = 0$ and for the standard configuration,
the normalization coil current $I_1$ is set to $I_1$ = \SI{13770}{\ampere}.

The plasma profiles cover a broad range of \gls{W7X} discharge scenarios, and include
plasma pressure on axis up to \SI{200}{\kilo\pascal},
corresponding to volume-averaged \averagePlasmaBeta of approximately \SI{5}{\percent},
and a net toroidal current ranging from \SIrange{-10}{10}{\kilo\ampere}.
The profiles are defined as a function of the normalized toroidal flux $s$:
$p(s)$ is the pressure of the plasma at the flux surface labeled $s$,
and $I(s)$ is the enclosed toroidal current flowing inside the surface $s$. 
With this definition,
$I(s = 1)$ is the total toroidal current in the plasma, which we will refer to as \gls{Itor}.

Theoretically,
all the possible continuous functions for $s \in [0, 1]$ should be sampled.
The exploitation of domain knowledge obtained from experience with \gls{W7X} discharges allows us to restrict the function space of the profiles,
and to sample the region of interest denser as compared to unconstrained parametrization.
The profile shapes are modeled via \glspl{GP}~\cite{Rasmussen2004},
stochastic processes whose joint distribution of every finite, linear combination of random variables is a multivariate Gaussian.
\glspl{GP} are usually employed in the modeling context,
as they can be seen as distributions of functions.
For example,
a one dimensional function $f: s \in \real \rightarrow \real$ with a \gls{GP} prior is

\begin{gather}
  f(s) \sim \mathcal{GP}(\mu(s), \Sigma(s, s^{\prime})),
\end{gather}

where $\mu(s)$ is the mean function, and $\Sigma(s, s^{\prime})$ is the covariate function.
Then,
for a set $\mathcal{S_*} := \{s \in \real\}$, the corresponding $\mathcal{F}_{*}$ are distributed as

\begin{gather}
  \mathcal{F}_* \sim \mathcal{N}(\mu(S_*), \Sigma(S_*, S_*)),
\end{gather}

where the covariate function matrix is computed element-wise.

In Minerva,
plasma profiles are usually specified as \glspl{GP} with zero mean,
which does not restrict the mean of the posterior process to be zero~\cite{Rasmussen2004},
and squared exponential covariance function~\cite{Kwak2020}.
In particular, since profiles can have substantially different gradients in the core and edge regions~\cite{Asdex1989},
a non-stationary covariance function~\cite{Higdon1999} is used~\cite{Chilenski2015}.
Here, the \gls{GP} mean and covariance functions are

\begin{gather}
  \mu(s) = 0, \\
  \Sigma(s_i, s_j) = \sigma_f^2 \sqrt{ \frac{2 \sigma_x (s_i) \sigma_x (s_j)}{ \sigma_x^2 (s_i) + \sigma_x^2 (s_j) } } \notag \\
  \times \text{exp} \left( - \frac{ (s_i - s_j)^2 }{ \sigma_x^2 (s_i) + \sigma_x^2 (s_j) } \right) + \sigma_y^2 \delta_{ij},
\end{gather}

where $\sigma_y$ is usually fixed to $\sigma_y = 10^{-3} \sigma_f$~\cite{Pavone2019b}, and $\sigma_x$, which represents the length scale function, is a hyperbolic tangent function

\begin{gather}
  \sigma_x(s) = \frac{\lCore + \lEdge}{2} - \frac{\lCore - \lEdge}{2} \text{tanh} \left( \frac{s - s_0}{s_w} \right) ,
\end{gather}

where $\lCore$ and $\lEdge$ are the core and edge length scale,
respectively.
$s_0$ is the transition location and $s_w$ represents the length scale for the transition.
\replaced[id=AM]{
  The domain knowledge on the plasma profiles is encoded via the \glspl{HP} of the \gls{GP} used to represent them,
  which define the distributions from where the profiles are drawn.
  T}{To include diverse profiles in the data set, t}
he values of the \gls{GP} \replaced[id=AM]{\glspl{HP}}{parameters} are uniformly sampled from a hyper-rectangle,
whose boundaries are given in \Cref{tab:hp-boundaries}.
\added[id=AM]{
  These values are adapted from previous works where the plasma profiles in \gls{W7X} are modeled via \glspl{GP}~\cite{Pavone2019b,Pavone2020,Kwak2020,Kwak2021a}. 
}

\begin{figure*}[!htb]
  \centering
  \begin{subfigure}{\subfigureTwo \subfigureWidth}
    \igraph[]{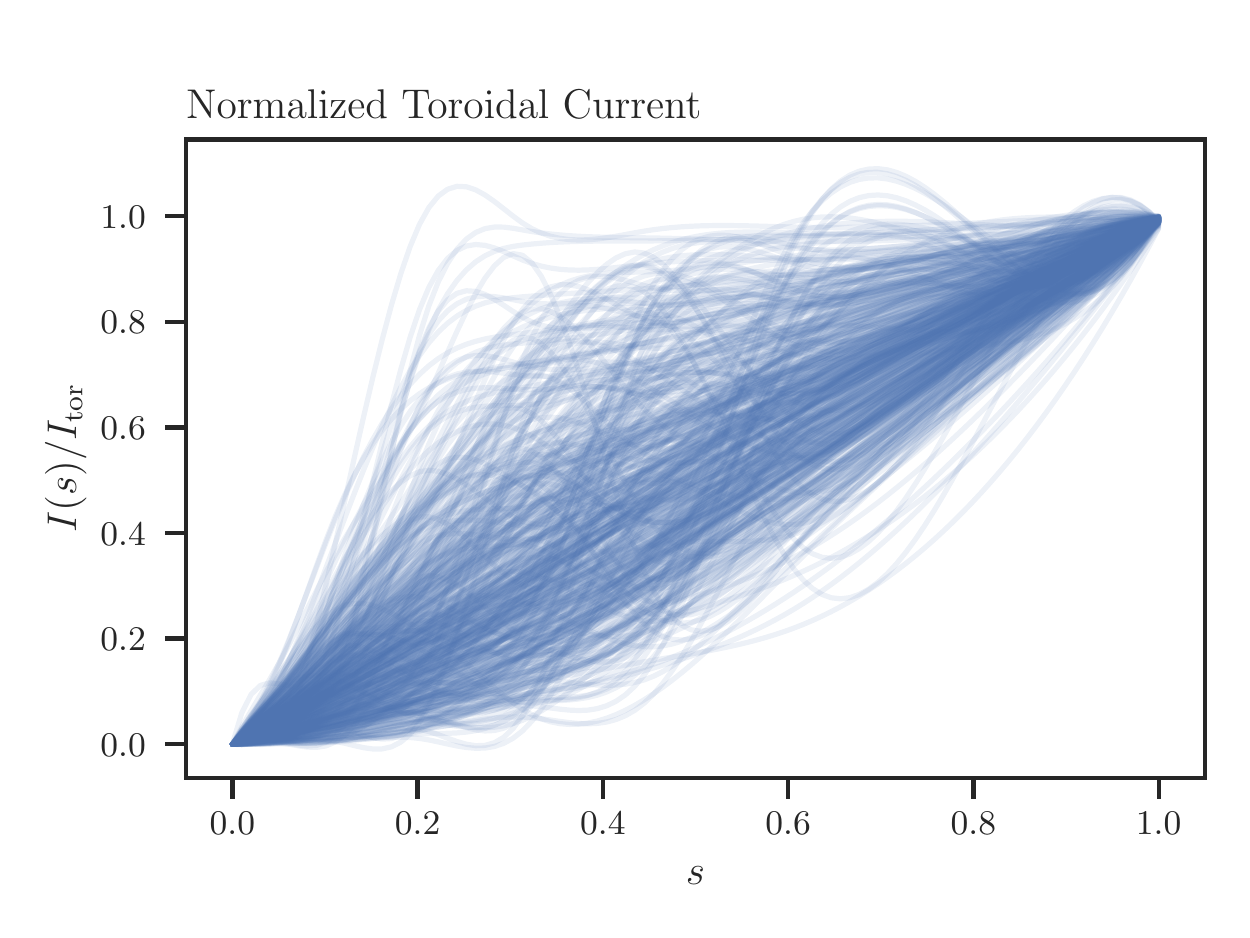}
  \end{subfigure}
  \begin{subfigure}{\subfigureTwo \subfigureWidth}
    \igraph[]{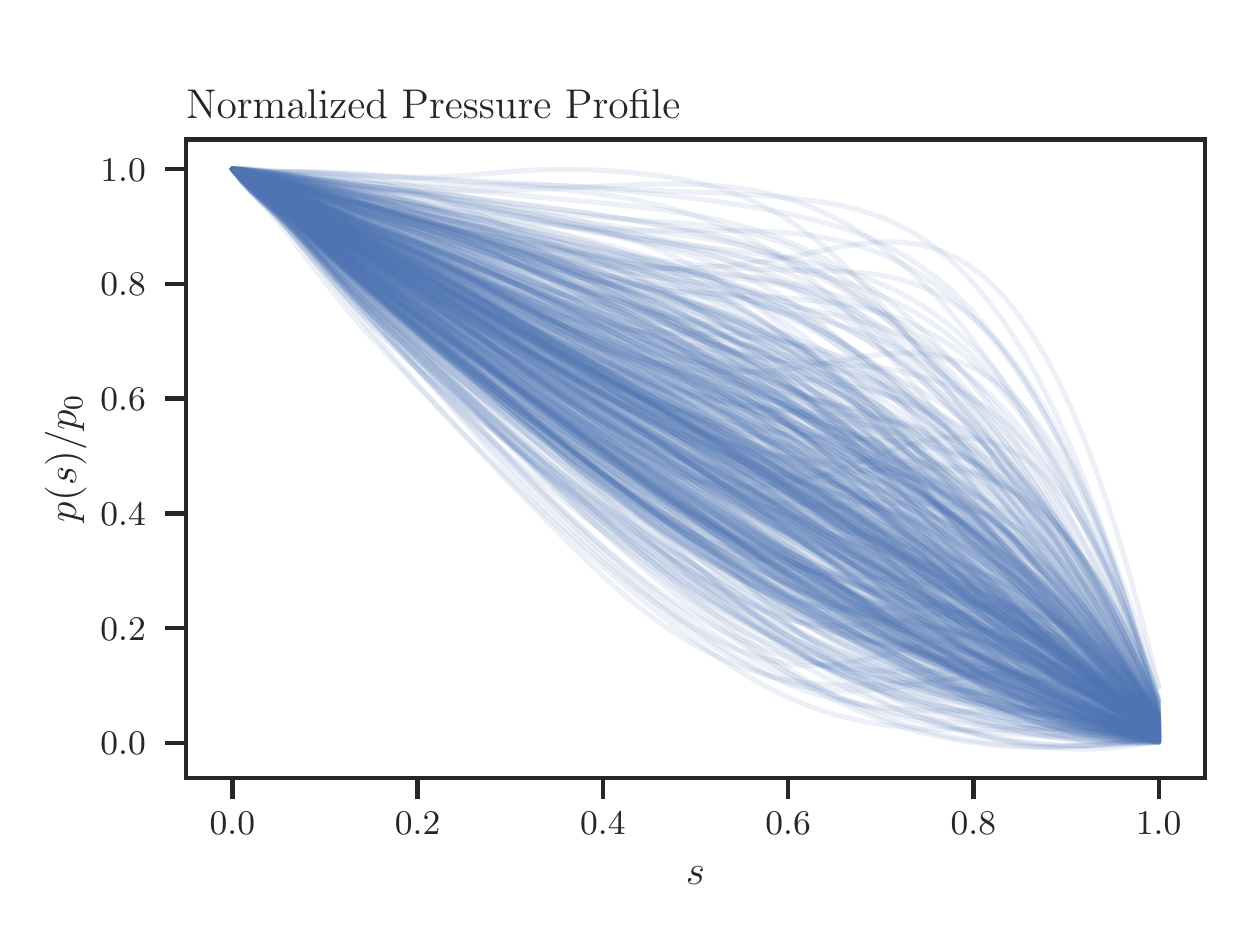}
  \end{subfigure}
  \caption{
    Subset of normalized plasma profiles included in the data set as a function of the flux radial coordinate $s$.
    \added[id=AM]{Only plasma profiles which resulted in a valid \gls{VMEC} equilibrium are depicted.}
  }%
  \label{fig:plasma-profiles}
\end{figure*}

The profiles are further constrained by the use of \textit{virtual observations}~\cite{Pavone2019b},
such that the \gls{GP} prior is refined with ``virtual diagnostic measurements'',
described by a normal distribution.
As usually observed in \gls{W7X} experiments,
the electron and ion density and temperature profiles are peaked\footnote{A globally decreasing function of the radial profile, not to be confused with a high ``peaking factor'' as used in the fusion community.} in the core~\cite{Drews2019,Klinger2019,Wolf2019}.
Therefore, the normalized pressure profile is constrained to \num{0} at the \gls{LCFS}
and \num{1} on axis.
Contrarily,
the normalized toroidal current profile  is set to \num{0} on axis,
and \num{1} at the \gls{LCFS}.
\Cref{fig:plasma-profiles} shows a subset of the normalized plasma profiles, which are independently sampled from the two refined \glspl{GP}.
Finally,
the profiles are scaled to the desired values:
the pressure profile is multiplied by $\gls{p}_0$,
the pressure value on axis,
and \gls{Itor},
which is the total toroidal current enclosed by the plasma,
is provided as input parameter to \gls{VMEC}.

\begin{table}
  \caption{Hyper-rectangle boundaries for the vacuum magnetic configurations, pressure and toroidal current profile included in the data set. Each parameter is uniformly sampled.}%
  \label{tab:hp-boundaries}
  \centering
  \begin{tabular}{lcccc}
    \toprule
    \multicolumn{4}{c}{Magnetic configuration} \\
    Free parameter & Min & Max & Unit \\
    \midrule
    $\Phi_{\text{edge}}$ & -2.5 & -1.6 & \si{\weber} \\
    $i_{[1 \ldots 5]}$ & 0.6 & 1.3 & - \\
    $i_{[A,B]}$ & -1.0 & 1.0 & - \\
    \bottomrule
    \multicolumn{4}{c}{Pressure profile} \\
    Free parameter & Min & Max & Unit\\
    \midrule
    $\gls{p}_0$ & 0 & 200 & \si{\kilo\pascal} \\
    $\sigma_f$ & 2.0 & 4.0 & -\\
    $l_{\text{core}}$ & 2.0 & 3.0 & -\\
    $l_{\text{edge}}$ & 1.0 & 2.0 & -\\
    $s_0$ & 0.7 & 0.9 & -\\
    $s_w$ & 0.3 & 0.4 & -\\
    \bottomrule
    \multicolumn{4}{c}{Toroidal current profile} \\  
    Free parameter & Min & Max & Unit\\
    \midrule
    \gls{Itor} & -10 & 10 & \si{\kilo\ampere} \\
    $\sigma_f$ & 2.0 & 3.0 & -\\
    $l_{\text{core}}$ & 2.0 & 3.0 & -\\
    $l_{\text{edge}}$ & 3.0 & \replaced[id=AM]{5.0}{2.0} & -\\
    $s_0$ & 0.1 & 0.6 & -\\
    $s_w$ & 0.01 & 0.1 & -\\
    \bottomrule
  \end{tabular}
\end{table}

All \gls{VMEC} calculations are performed in free boundary mode,
where the confined region is characterized with the total enclosed magnetic toroidal flux,
$\Phi_{edge} = \Phi (s = 1)$.
Given the large input space,
\gls{VMEC} runs which are not relevant for \gls{W7X}, \eg,
runs which did not converge or
exhibit values for the plasma volume and minor radius outside the boundaries given in~\Cref{tab:bounds},
are discarded.

\begin{table}
  \caption{Plasma volume and minor radius boundaries of valid \gls{VMEC} runs included in the data set.}%
  \label{tab:bounds}
  \centering
  \begin{tabular}{lccc}
    \toprule
    Variable & Min & Max & Unit\\
    \midrule
    $\gls{V}_p$ & 22.0 & 38.0 & \si{\meter^3} \\
    $a_{eff}$ & 45 & 60 & \si{\centi\meter} \\
    \bottomrule
  \end{tabular}
\end{table}

\subsubsection{Training scenarios}\label{sec:beta-scenarios}

To decouple the regression complexity of the 3D ideal-\gls{MHD} equilibrium from the vacuum field computation,
the problem is broken down in two different scenarios:
a null and finite-\averagePlasmaBeta cases,
which lead to two independent data sets,
$\Dnull$ and $\Dfinite$.
\Dnull is populated with vacuum magnetic configurations,
\ie,
pressure and plasma current profiles are constant \num{0}.
This scenario targets two applications:
discharges with low \averagePlasmaBeta values which could be effectively studied with a vacuum field,
and further investigations of the properties of the vacuum configurations of \gls{W7X}.
In particular,
the use of a slightly modified model is envisioned to further explore the richness of the vacuum magnetic configurations of \gls{W7X}, searching for optimized equilibria in terms of, \eg, neoclassical transport via the effective helical ripple amplitude $\epsilon_{eff}$~\cite{Nemov1999}
or ideal \gls{MHD} stability via the magnetic well~\cite{Nuhrenberg2016}.
In $\Dfinite$,
the standard magnetic configuration (EJM+252)~\cite{Andreeva2002} is fixed,
and the data set is populated with plasma profiles as described in~\Cref{sec:dataset}.
This scenario covers discharges with volume-averaged \averagePlasmaBeta up to \SI{5}{\percent} and net toroidal current up to \SI{10}{\kilo\ampere}.

\added[id=AM]{
  The number of \gls{VMEC} simulations for the two scenarios are \num{11360} and \num{11709}, respectively.
  Of these, only \num{10339} and \num{9675} converged.
}
\replaced[id=AM]{Finally,
  after filtering out the equilibria based on the ranges given in~\Cref{tab:bounds},
  t}{T}he data sets contain \replaced[id=AM]{$|\Dnull| = \num{9589}$}{$|\Dnull| = \num{10339}$} and \replaced[id=AM]{$|\Dfinite| = \num{9332}$}{$|\Dfinite| = \num{9675}$} valid runs,
respectively.

In this work,
the two dimensions to characterize a \gls{W7X} magnetic configuration,
the vacuum field geometry and the plasma profiles,
are independently explored in \Dnull and \Dfinite.
Given the large vacuum magnetic configuration space probed in \Dnull and the relatively low values of \averagePlasmaBeta included in \Dfinite, 
the spread of the \glspl{FC} describing the equilibrium field is expected to be higher in \Dnull than in \Dfinite.
Furthermore,
\gls{W7X} is an optimized stellarator where the plasma influence on the magnetic configuration has been strongly reduced by the minimization of the bootstrap current and the Shafranov shift~\cite{Beidler1990}.
Hence,
the equilibrium field coefficients are expected to be smooth functions of the main parameters characterizing the plasma,
$\gls{p}_0$ and \gls{Itor},
in contrast to the flexibility of the vacuum magnetic configurations of \gls{W7X}.
In the scope of the next steps of this proof of concept,
working models in these two extreme cases can give valuable insights on the use of \glspl{NN} for the regression of the equilibrium magnetic field in a arbitrary finite-\averagePlasmaBeta configuration.

\subsubsection{Models inputs and outputs}\label{sec:inputs-and-outputs}

In $\Dnull$,
the inputs are represented by $\Phi_{edge}$ and the six independent coil current ratios,
while in $\Dfinite$,
$\Phi_{edge}$,
$\gls{p}_0$,
\gls{Itor},
and the normalized pressure and toroidal current profiles are used.
In both scenarios the regressed outputs are the iota profile,
$\gls{ibar}( s )$,
the Fourier series of the flux surface coordinates,
represented by $R_{mn} ( s )$, $\lambda_{mn} ( s )$ and $Z_{mn} ( s )$
and the Fourier series of magnetic field strength, $B_{mn} ( s )$.
The output \glspl{FC} are regressed instead of the real space values for the following reasons:
first and foremost,
the Fourier series profiles are a compressed representation of the magnetic field,
thus letting the network learn a reduced number of independent outputs.
Furthermore,
we seek to replace \gls{VMEC} with similar input and output signature as the original code such that our application can serve as a drop-in replacement for existing use cases.
For example, in the context of the application of this work in \replaced[id=AM]{the inference of plasma parameters}{a \gls{BSM} framework},
the flux surface coordinates are needed to map real space diagnostic measurements to flux coordinates~\cite{Langenberg2019,Kwak2020},
and the magnetic field strength plays a crucial role in the analysis of many diagnostics (\eg, \gls{ECE}~\cite{Hoefel2019a}).

\begin{figure}[!htb]
    \centering
    \igraph[width=\linewidth]{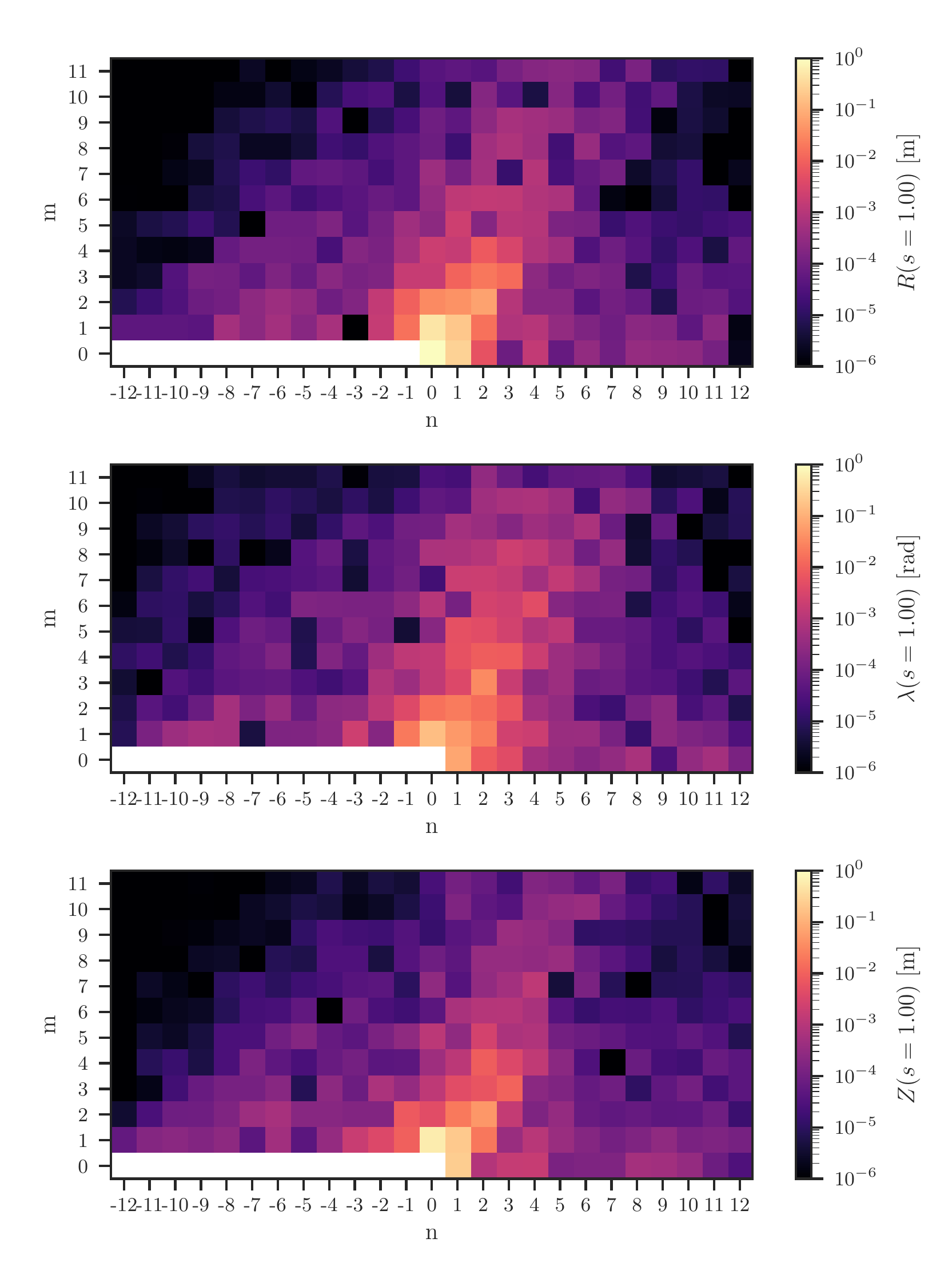}
    \caption{
      \glspl{FC} of the cylindrical coordinates evaluated at the \gls{LCFS}.
      The Fourier series have poloidal modes $m < \mpol $ and toroidal modes $|n| \leq \ntor $.
      A logarithmic colormap is used to show the span in orders of magnitude expressed by the data.
    }%
    \label{fig:fourier-map}
\end{figure}

For the generation of the data set, the resolution of the \gls{VMEC} output is set to $\Ns = 99$ flux surfaces
and $\mpol = \ntor = 12$
, where $|m| < \mpol$ and $|n| \leq \ntor$ are the poloidal and toroidal Fourier modes respectively.
Since all the outputs are real quantities,
$o_{mn} = (o_{-m,-n})^*$ for $o \in \{ R, \lambda, Z, B \}$.
This limits the independent \glspl{FC} to a subplane (usually $m \geq 0$).
Despite this symmetry consideration, still \num{28512} coefficients remain per output\footnote{The number of \glspl{FC} per coordinates scales as \orderof{\Ns \cdot \mpol \cdot \ntor}.}.
\Cref{fig:fourier-map} shows the \glspl{FC} of the three coordinates for one sample in the data set,
evaluated at the \gls{LCFS}.
However,
it has been argued that $\mpol = \ntor = 6$ modes are sufficient to represent the magnetic field in case of \gls{W7X} configurations~\cite{Sengupta2007}.
In this work,
the sufficient Fourier resolution is further investigated.
For the radial profile, a subset of $\hatNs$ flux surfaces is selected,
while up to $\hatmpol$ and $\hatntor$ poloidal and toroidal modes are used for the \glspl{FC}.
To more densely cover the plasma region near the axis,
the flux surfaces are selected such that their radial locations $s$ follow a quadratic progression in $[0, 1]$.
To compute the loss of information due to the downscaling,
the reduced representation is upscaled to match the full resolution by asserting $x_{mn} = 0$ for $x \in \{ R, \lambda, Z \}$ if $m \ge \hatmpol$ or $|n| > \hatntor$.
Then, the outputs $R$, $\lambda$, $Z$ and $B$
are evaluated with \cref{eq:R,eq:lambda,eq:Z,eq:b-mn} on a grid along the $\theta$ and $\varphi$ angles,
using $N_{\theta} = 18$ poloidal and $N_{\varphi} = 9$ toroidal points per period.
Finally,
the full radial resolution is recovered by cubic interpolation along $s$.
Similarly,
a reduced resolution of the iota profile is investigated,
using $\hatNs$ flux surfaces (the same as those employed for $\vec{x}$ and B).
To compare the reduced to the full resolution,
the iota profile is then upscaled via cubic interpolation.

Given a set $\mathcal{Y} = \{y \in \real^K\}$ of generic quantities $y$ with true or reference value $y^*$
, the \gls{rmse} between $y$ and $y^*$ is computed as

\begin{gather}
    \text{rmse}_{\mathcal{Y}} = \frac{1}{K} \sum\limits_{k=1}^{K} \sqrt{\frac{1}{|\mathcal{Y}|} \sum\limits_{i=1}^{|\mathcal{Y}|} (y_{ki} - y^{*}_{ki})^2} \ . \label{eq:rmse}
\end{gather}

Here it is used to compare the two resolutions,
where for each output, $y$ is the reduced output representation of $y^*$,
and $K$ is the number of evaluation points:
$K_{\gls{ibar}} = \hatNs$,
and $K_R = K_{\lambda} = K_Z = K_B = \hatNs N_{\theta} N_{\varphi}$.  
\Cref{fig:resolution-scan-results} shows the \gls{rmse} for different values of the resolution parameters.

\begin{figure}[!htb]
  \centering
  \begin{subfigure}[t]{0.70\columnwidth}
    \centering
    \hspace*{-1.80cm}
    \igraph[]{content_figures_iota_resolution_scan_rmse.pgf}
    \vspace*{-2mm}
    \subcaption{
      \gls{rmse} of the reduced representation of the iota profile,
      for different radial resolutions.
      The reference value \rmseStarIota = \num{e-4} is marked by the dashed line.
    }%
    \label{fig:iota-scan}
  \end{subfigure}
  \begin{subfigure}[t]{0.85\columnwidth}
    \centering
    \vspace*{2mm}
    \igraph[]{content_figures_surfaces_resolution_scan_rmse.pgf}
    \vspace*{-2mm}
    \subcaption{
      \gls{rmse} of the reduced representation of the flux surfaces coordinates,
      for different radial and Fourier resolutions.
      The dashed line marks the reference values \rmseStarRZ = \SI{e-3}{\meter} and \rmseStarLambda = \SI{e-3}{\radian}.
    }%
    \label{fig:surfaces-scan}
  \end{subfigure}
  \begin{subfigure}[t]{0.77\columnwidth}
    \centering
    \hspace*{-1.4cm}
    \vspace*{2mm}
    \igraph[]{content_figures_bcos_resolution_scan_rmse.pgf}
    \vspace*{-2mm}
    \subcaption{
      \gls{rmse} of the reduced representation of magnetic field strength,
      for different radial and Fourier resolutions.
      The dashed line marks the reference values \rmseStarB = \SI{e-3}{\tesla}.
    }%
    \label{fig:bcos-scan}
  \end{subfigure}
  \caption{
    Analysis of the \gls{rmse} between the full and reduced representation of the iota profile, flux surface coordinates and magnetic field strength.
    In \cref{fig:surfaces-scan,fig:bcos-scan} the truncated Fourier resolutions are ordered based on the total number of \glspl{FC} used, which scales as \orderof{\hatmpol \cdot \hatntor}.
    In the case of the flux surface coordinates and magnetic field strength, and for the Fourier truncated resolution of interest (\ie, $\hatmpol \approx 6$ and $\hatntor \approx 6$), an increased radial resolution of 20 flux surfaces does not significantly differ from using only 10 flux surfaces.
  }%
  \label{fig:resolution-scan-results}
\end{figure}

In case of the iota profile,
$\hatNs = 20$ flux surfaces are sufficient for a deviation of approximately $\text{\gls{rmse}}_{\gls{ibar}}^* = \num{e-4}$,
$\hatNs = 10$, $\hatmpol = 6$ and $\hatntor = 4$ are needed for the flux surfaces coordinates to achieve $\rmseStarRZ = \SI{e-3}{\meter}$ and $\rmseStarLambda = \SI{e-3}{\radian}$
, and $\hatNs = 10$, $\hatmpol = 6$ and $\hatntor = 12$ are used for $B$ to obtain $\rmseStarB = \SI{e-3}{\tesla}$.
These choices result in \num{20} locations for the iota profile,
while \num{1500} \glspl{FC} describe the flux surface coordinates and the magnetic field strength\footnote{\added[id=AM]{Each output is described by $N_{FCs} = N_o \hatNs [\hatmpol (2 \hatntor + 1) - \hatntor]$ \glspl{FC}, where $N_o$ is the output dimension ($N_o = 3$ for $\vec{x}$ and $N_o = 1$ for $B$).}}.
This resolution represents a practical trade-off between the complexity of the regression task and the reconstruction fidelity.
It is important to note that \rmseStarIota, \rmseStarRZ, \rmseStarLambda and \rmseStarB represent a lower bound of the reconstruction error that can be achieved by using the models presented in this work.

Given the two data sets, \Dnull \ and \Dfinite,
and the three output quantities,
\gls{ibar}, $\vec{x}$ and $B$,
six independent regression tasks are defined:
\nullIota and \finiteIota, \nullSurfaces and \finiteSurfaces, and \nullB and \finiteB.
In the \nullIota and \finiteIota tasks,
a \gls{NN} is trained to compute the reduced resolution iota profile,
using respectively \Dnull \ and \Dfinite \ as data set.
Similarly,
in the \nullSurfaces, \finiteSurfaces, \nullB and \finiteB tasks,
a \gls{NN} is trained to compute the \glspl{FC} of the reduced resolution $\vec{x}$ or magnetic field strength $B$,
using \Dnull \ and \Dfinite,
respectively.

Considering the scope of this paper which attempts to develop a \gls{VMEC} proof-of-concept surrogate model,
it is useful to investigate the performance on independent subproblems.
In future works,
a single \gls{NN} could be trained to compute all outputs and to handle both vacuum and finite-\averagePlasmaBeta \ runs.

\cleardoublepage

\subsection{\gls{NN} architectures}\label{sec:architectures}

In general,
given two quantities $\vec{\psi} \in \real^K$ and $\vec{\gamma} \in \real^D$,
and a set of $N$ observations $(\vec{\psi}_i, \vec{\gamma}_i)$ sampled from a fixed but unknown distribution $p(\vec{\psi}, \vec{\gamma})$,
a \gls{NN},
parametrized with a set of free parameters $\vec{w}$,
can be employed to learn a mapping $\tilde{f} : \real^K \rightarrow \real^D$ which minimizes
the empirical loss $\frac{1}{N} \sum\limits_{i} l(\vec{\gamma}_i, \tilde{f}(\vec{\psi}_i; \vec{w}))$,
where $l : \real^D \times \real^D \rightarrow \real$ is a given loss function.
In this work,
the expressive power of \glspl{NN} is exploited to learn a low-cost approximation of a known function,
using observations sampled from a known distribution.
A \gls{NN} usually employs successive layers of \textit{artificial neurons} to create the mapping $\tilde{f}$,
where each neuron computes a non-linear transformation of the neurons from the previous layer.
The \gls{NN} free parameters $\vec{w}$ are derived during the training process to minimize the empirical loss on the given training set.
For a detailed introduction on \glspl{NN} please refer to~\cite{Bishop1995}.

Two \gls{NN} architectures are adopted herein.
One is a \gls{FFFC},
which is composed of a sequence of dense blocks,
each comprising a dense layer with $L^2$ regularization and a non-linear activation function.
The number of hidden units is halved for each successive block.
The activation function for the last block is the identity.
\Cref{fig:dense-model} illustrates the architecture for the \nullIota task,
where a network with five of such blocks is shown.

The \gls{FFFC} architecture is used on the iota reconstruction,
where the regressed output is composed of only 20 elements.
However,
its number of free parameters grows linearly with the dimensionality of the output.
Thus,
more efficient architectures are needed for the surfaces and magnetic field strength reconstruction,
where each sample has 1500 output elements.
Hence,
3D \glspl{CNN}~\cite{Ji2013,LeCun2015} and encoder-decoder like architectures~\cite{Cho2015,Ronneberger2015,Badrinarayanan2017} are explored.
In these architectures an encoder processes variable-length input features and generates a fixed-length,
flattened representation.
Conditioned on the encoded representation,
the decoder then builds the required outputs.

\begin{figure}
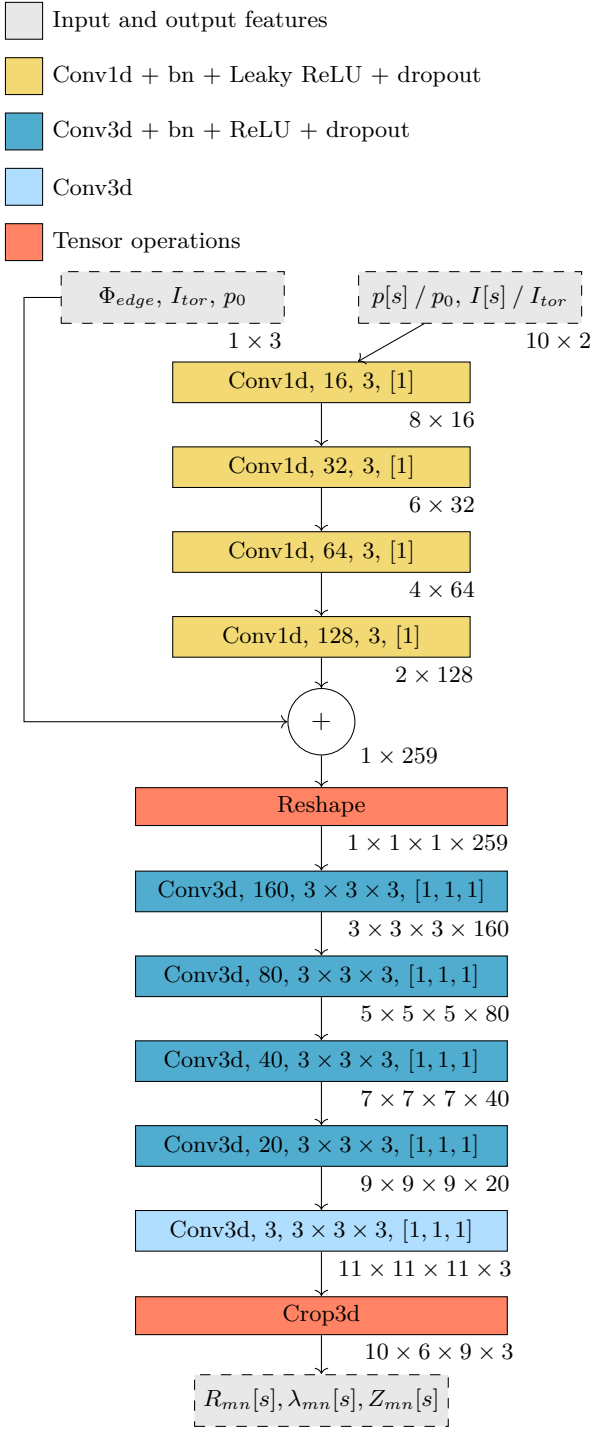

    \centering
    \igraph[width=\linewidth]{content_figures_conv.pgf}
    \caption{
      The 3D \gls{CNN} architecture for the \finiteSurfaces task.
      The grey blocks represent the input and output features,
      the yellow ones the 1D convolutions,
      the blue ones the 3D convolutions,
      and the red ones tensor operations.
      For each block\added[id=AM]{,}
      the output dimension is indicated on the bottom right\added[id=AM]{%
        , where the last number is always the number of features,
        and the antecedent ones the feature dimension
        (\eg, $\num{10} \times \num{6} \times \num{9} \times \num{3}$ refers to 3 features of size $\num{10} \times \num{6} \times \num{9}$).
      }.
      For the convolutional blocks,
      the number of filters,
      kernel size,
      and stride (in bracket) are indicated in sequence.
      The use of Batch Normalization is indicated via bn.
      The values of the \gls{HP} of the best performing model in the \finiteSurfaces task are shown here (see~\protect\cref{sec:pipeline,sec:hp-values}).
    }%
    \label{fig:conv3d-model}
\end{figure}

For the these tasks the $\vec{x}$ coordinates are stacked.
\Cref{fig:conv3d-model} displays an example of such architecture for the \finiteSurfaces task,
where a \gls{CNN} architecture with transposed convolution is used.
In the \textit{encoder} tree,
high-level features are extracted from the plasma profiles via consecutive 1D convolutional blocks
and concatenated back with the scalar inputs into a flattened representation.
Then, a \textit{decoder} tree gradually builds up the output via consecutive transposed convolutional blocks.
Finally, the output shape is matched via a 3D cropping operation.
Each \textit{encoder} block comprises a 1D convolutional layer,
batch normalization~\cite{Ioffe2015a},
and a non-linear activation function with dropout~\cite{Srivastava2014a}.
Similarly, a \textit{decoder} block is composed of a 3D transpose convolutional layer, 
batch normalization,
and a non-linear activation function with dropout.
For the last block, batch normalization is not included
and the identity activation function is used.
For each block the number of filters in the \textit{encoder} tree is doubled,
while halved in the \textit{decoder} tree.
Convolutional layers with stride are employed over up-sampling operations,
as suggested by~\cite{Springenberg2015}.

The stacking of consecutive convolutional layers acting on inputs of different length scales,
in conjunction with a scaling of the feature channels,
is a common approach in modern \gls{DCNN} architectures.
This structure decreases the number of free parameters by forcing the model to learn a hierarchical representation of high- and low-level features,
while imposing a regularizing effect during training.
\added[id=AM]{%
  A subset of the \gls{NN} architecture \glspl{HP} is not fixed a priori,
  but optimized via \gls{HP} search.
  The lists of the explored \glspl{HP} (\eg, the layer non-linear activation function) are provided in~\protect\cref{sec:hp-values}.
}

In both architectures all weights are uniformly initialized as suggested by~\cite{Glorot2010a},
while the bias terms,
where present,
are initialized to zero.
The weights are then optimized via the \textit{Adam} optimizer~\cite{Kingma2015},
while reducing the learning rate by a fixed multiplier factor once a validation loss plateau is reached.
Early stopping~\cite{Morgan1989} is employed during training.
\deleted[id=AM]{A subset of the \gls{NN} architecture hyper-parameters is not fixed a priori,
but optimized via hyper-parameters search.}
The \gls{NN} models are built, trained and evaluated via the open source software package \textit{Tensorflow}~\cite{Abadi2016} on a single NVIDIA RTX8000P virtual \gls{GPU}.

\begin{figure}
  \centering
  \igraph[width=0.6\linewidth]{content_figures_dense.pgf}
  \caption{
    \gls{FFFC} architecture for the \nullIota task.
    The grey blocks represent the input and output features,
    where the dimension is indicated on the right.
    A single block is composed of:
    a `Dense, $m$` layer with $m$ units and a non-linear activation function (\eg, the \gls{SeLU}).
    The last block uses the identity function.
    The values of the \gls{HP} of the best performing model in the \nullIota task are shown here (see~\protect\cref{sec:pipeline,sec:hp-values}).
    \added[id=AM]{
      The use of the \gls{SeLU} activation function,
      as discovered by \gls{HP} search,
      leads to whitened layer input distributions,
      which improve the convergence of the training process~\protect\cite{Klambauer2017}.
    }
  }%
  \label{fig:dense-model}
\end{figure}

\subsection{Training and evaluation pipeline}%
\label{sec:pipeline}

For each task defined in \Cref{sec:dataset},
the training and evaluation pipeline includes the following steps:

\begin{description}
\item[Data scaling]{
  It is known that \gls{NN} models converge faster during training if the input distributions are whitened~\cite{Wiesler2011},
  \ie, linearly transformed to have zero mean and unit variance.
  All scalar inputs are mapped to $[-1, 1]$,
  while non-scalar inputs and outputs
  (plasma profiles, \gls{ibar} profile and \glspl{FC}) are scaled to the inter-quartile range.
  These steps are performed via the open source software package \textit{Scikit-learn}~\cite{Pedregosa2011}.
}
\item[Bayesian \glspl{HP} search]{
The large number of \glspl{HP} and the significant training time of the considered \gls{NN} architectures make a manual model optimization procedure hardly effective.
Therefore,
to standardize the search of more performing models,
an automated approach to \glspl{HP} search is used in this work.
In particular,
the \gls{TPE}~\cite{Bergstra2011} algorithm,
provided via the open source software package \textit{hyperopt}~\cite{Bergstra2013},
is employed.
\gls{TPE} is a \gls{SMBO} algorithm,
where the true fitness function,
\eg,
the model training and evaluation,
is approximated with a low-cost model that is cheaper to evaluate.
The proxy model is then numerically optimized to retrieve new configurations to be evaluated.
Contrarily to other \gls{SMBO} strategies where the fitness function is directly learned,
\gls{TPE} models the distribution function of configuration values given classes of optimal and non-optimal fitness function values.
It then optimizes the \gls{EI} criterion~\cite{Jones2001} with a heuristic procedure.
Its main advantage over other \gls{HP} search approaches is the sampling efficiency on tree-structured configuration spaces~\cite{Bergstra2011},
\ie spaces in which not all dimensions are well-defined for all the configurations
(\eg, number of hidden units in the second layer of a single-layer \gls{FFFC} model).
For a detailed description of the algorithm, please refer to~\cite{Bergstra2011}.
On each learning task,
30 search iterations are performed.
The data set is split in \SI{20}{\percent} for testing, \SI{10}{\percent} for validation and \SI{70}{\percent} for training.
For each search iteration,
the training data is used to train the model,
while the validation data is used to \replaced[id=AM]{assess}{asses} the model regression error and inform the search strategy.
The best performing model is then adopted in the cross-validation scheme.
To ease the computational cost of the search,
a simple \gls{mse} loss is used for training and \gls{HP} validation.
}
\item[Repeated k-fold cross-validation]{
  To estimate the regression error on out-of-sample data,
  a five-fold cross-validation evaluation is repeated 10 times.
  \added[id=AM]{
    In a k-fold cross validation scheme~\cite{Stone1974,Geisser1975},
    the data set is partitioned into k-folds of equal cardinality.
    Then,
    for each fold,
    the training process is repeated k-times,
    using the selected fold as test set,
    and the remaining folds for the training and validation sets.
    The estimate of the regression error is the average of the test error on each fold.
    However,
    the cross-validation estimate of the regression error can be highly variable due to the single partition of the data set into the k-folds~\cite{Efron1997}.
    To overcome this limitation,
    in the repeated k-fold cross-validation scheme,
    the k-fold cross-validation scheme is repeated n-times,
    partitioning the data set into a different k-fold each time.
    The average of the test error on each fold is then used as the final estimate.
  }
}
\end{description}

\subsection{Data and code availability}
\label{sec:code}

The data sets and code needed to reproduce this work are available at \url{https://gitlab.com/amerlo94/vmecfastsurrogate}.

\section{Results}%
\label{sec:results}

\sisetup{separate-uncertainty=true,multi-part-units=single,round-mode=figures,round-precision=1}

The results achieved on each task are now presented.
It is important to remember that \Dfinite includes plasma profile for a fixed magnetic configuration (the standard configuration),
while \Dnull explores the rich space of \gls{W7X} vecuum magnetic configurations.
The changes in \Dfinite, induced by finite-beta effects,
are then small compared to those in \Dnull,
induced by coil currents (\ie, finite-beta effects span a space that only slightly expands the vacuum solution).
Therefore, the spread of the output data in the finite-beta cases is smaller than in the vacuum scenarios:
the coil system of \gls{W7X} has been designed to allow a large flexibility in the vacuum magnetic configuration space~\cite{Renner2002,Geiger2015},
while the W7-X optimization explicitly targeted robustness against changes in plasma profiles,
in particular pressure profiles~\cite{Beidler1990,Neuner2020}.
These features are expected to make the output data in the finite-\averagePlasmaBeta tasks more difficult to resolve because of the smaller spread.
Therefore,
to quantitatively compare the results across all tasks,
the \gls{nrmse} is used instead.

\begin{table*}[!htb]
  \caption{
    Main results across all learning tasks.
    The \gls{nrmse}, training and inference time mean and \SI[round-mode=places]{95}{\percent} confidence interval are evaluated with bootstrapping~\protect\cite{DiCiccio1996}.
    The inference time is conservatively estimated with a batch size of 1 on a single Intel Xeon Gold 6136 CPU.
    However, orders of magnitude in inference time can be gained by parallel computation,
    pre- and post-training\added[id=AM]{ optimizations (\eg, model pruning and quantization)}.
    The $\text{\gls{nrmse}}_{\text{best}}$,
    which refers to the error on the cross-validation fold used in \gls{HP} search,
    is within the \SI[round-mode=places]{95}{\percent} of the \gls{nrmse} distribution for all tasks,
    meaning that the model discovered in the \gls{HP} search is robust across the whole data set.
    }%
  \label{tab:all-results}
  \centering
\begin{tabular}{lccccc}
\toprule
           Task & \gls{nrmse} [\num{e-2}] & \tTrain [\SI{e2}{\second}] & \tPrediction [\SI{e-3}{\second}] &  $\text{\gls{NN}}_{\text{free parameters}}$ &       $\text{\gls{nrmse}}_{\text{best}} [\num{e-2}]$ \\
\midrule
      \nullIota &     \num{1.51 \pm 0.19} &        \num{1.45 \pm 0.26} &              \num{4.25 \pm 0.67} &                                        3436 &  \num[round-mode=places,round-precision=1]{1.403320} \\
    \finiteIota &     \num{4.77 \pm 0.50} &        \num{3.71 \pm 0.87} &              \num{5.51 \pm 0.80} &                                       14276 &  \num[round-mode=places,round-precision=1]{4.522424} \\
  \nullSurfaces &    \num{14.17 \pm 0.37} &          \num{9.0 \pm 1.9} &              \num{5.93 \pm 0.74} &                                      244989 & \num[round-mode=places,round-precision=1]{14.399000} \\
\finiteSurfaces &    \num{19.16 \pm 0.67} &         \num{13.1 \pm 2.7} &               \num{14.7 \pm 3.0} &                                     1607535 & \num[round-mode=places,round-precision=1]{19.501000} \\
         \nullB &     \num{3.39 \pm 0.27} &          \num{8.2 \pm 1.4} &              \num{7.23 \pm 0.88} &                                      316193 &  \num[round-mode=places,round-precision=1]{3.515700} \\
       \finiteB &     \num{9.87 \pm 0.35} &        \num{3.29 \pm 0.59} &                \num{8.7 \pm 1.3} &                                      541921 & \num[round-mode=places,round-precision=1]{10.215000} \\
\bottomrule
\end{tabular}

\end{table*}

Given $\mathcal{Y} = \{ y \in \real^K \}$ (see~\cref{sec:inputs-and-outputs}),
the \gls{nrmse} between the predicted $y$ and the true or reference $y^*$ is computed as:

\begin{gather}
  \text{nrmse}_{\mathcal{Y}} = \frac{1}{K} \sum\limits_{k=1}^{K} \sqrt{\frac{\sum\limits_{i=1}^{|\mathcal{Y}|} (y_{ki} - y_{ki}^*)^2}{\sum\limits_{i=1}^{|\mathcal{Y}|} (y_{ki} - \bar{y_k})^2}}
  \label{eq:nrmse}
\end{gather}

where $\bar{y_k} = \frac{1}{|\mathcal{Y}|} \sum\limits_{i=1}^{|\mathcal{Y}|} y_{ki}$.
The \gls{ibar} profile is evaluated along the radial profile with $\hatNs$ flux surfaces,
so $K_{\gls{ibar}} = 20$.
As employed in~\cref{sec:inputs-and-outputs},
an evaluation grid with $N_{\theta} = 18$ and $N_{\varphi} = 9$ is used for the flux surface coordinates and the magnetic field strength.
The use of the \gls{nrmse} allows us to aggregate the regression error on the three flux surface coordinates,
and to compare the results across all outputs and scenarios.

\Cref{tab:all-results} summarizes the results for all tasks.
As expected,
on each output,
the \gls{nrmse} in the vacuum scenario is lower as compared to the finite-\averagePlasmaBeta case.
Moreover,
a \gls{nrmse} below \SI{10}{\percent} is consistently achieved for the \gls{ibar} profile and the magnetic field strength.
In the flux surface coordinates tasks,
\gls{nrmse} values between \SI[round-mode=places]{14}{\percent} and \SI{20}{\percent} are achieved instead.

Given the relative small size of the data sets and of the \glspl{NN},
the model training time is on the order of magnitude of minutes but less than an hour.
More importantly,
the inference time,
even in the most conservative evaluation (\ie, with a single thread on \num{1} CPU core with a batch size of \num{1}) is on the order of few milliseconds.
However, parallel computation (\eg, batched inference and \gls{GPU} deployment),
pre- and post-training optimizations (\eg, model pruning and quantization),
are expected to deliver consistent orders of magnitude speed-up~\cite{Krishnamoorthi2018,Liang2021}.
These optimizations are out of the scope of this paper.

\Cref{tab:null-iota-hp,tab:finite-iota-hp,tab:null-surfaces-hp,tab:finite-surfaces-hp,tab:null-bcos-hp,tab:finite-bcos-hp} list the \gls{HP} values for the best performing models discovered via \gls{HP} search (see~\cref{sec:pipeline}).
As reported in~\cref{tab:all-results},
the \gls{nrmse} obtained during search is compatible with the \gls{nrmse} estimated via cross-validation (\ie, its value is within the \SI[round-mode=places]{95}{\percent} interval of the distribution).
This means that the \gls{HP} search procedure did not overfit\footnote{High variance of the model error on unseen data.} to the validation data,
but \gls{HP} values which perform well on the whole data set were found.

In the following,
the fidelity of the different \glspl{NN} is inspected in closer detail and the major influences on the regression error are identified.

\subsection{Iota regression}\label{sec:iota-regression}

\Cref{fig:iota-profile-rmse} shows the \gls{rmse} profile along the radial flux coordinates for the \nullIota and \finiteIota tasks.
Although the average \gls{nrmse} in the \finiteIota case is higher than in the \nullIota case,
the \gls{rmse} is on the order of \num{e-3} for both.
In the \nullIota scenario,
the \gls{rmse} increases from the axis to the edge.
This may be caused by the characteristic shear profile of \gls{W7X} magnetic configurations and the hence increasing spread of \gls{ibar} profile in the data from the axis to the edge.
Instead,
in the \finiteIota task,
the toroidal current (and partially the pressure) profile is the main parameter affecting \gls{ibar}.
By data set construction,
these have a larger spread at mid-radius (see~\cref{fig:plasma-profiles}).
The larger spread is reflected in the maximum at $s \approx 0.4$.
In both cases,
this work shows that even shallow, \gls{FFFC} \glspl{NN} can effectively regress the \gls{ibar} profile with high accuracy.

\begin{figure}[!htb]
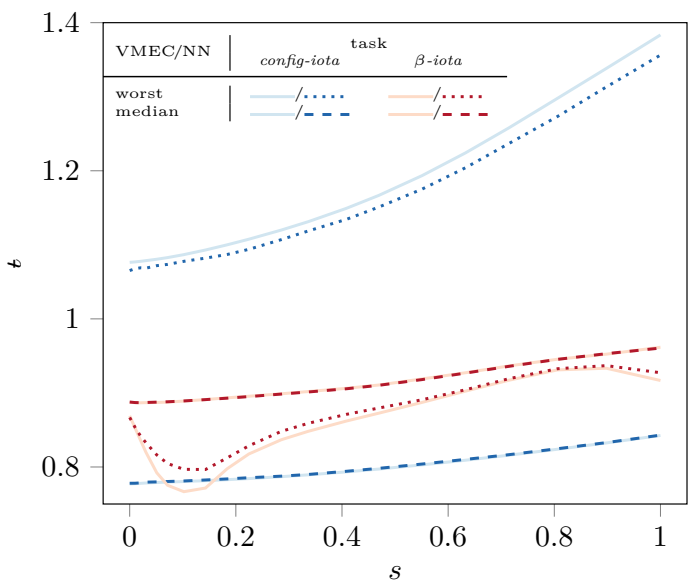

  \centering
  \igraph[width=\linewidth]{content_figures_iota_profile_sample.pgf}
  \caption{
    Worst and median predicted samples for the \nullIota (blue) and \finiteIota (red) tasks.
    The solid lines represent the true \gls{ibar} profiles as evaluated by \gls{VMEC},
    while the dotted (worst) and dashed (median) lines show the predicted profiles by the model.
    The results from the worst performing cross-validation fold are shown.
  }%
  \label{fig:iota-profile-sample}
\end{figure}

\begin{figure}[!htb]
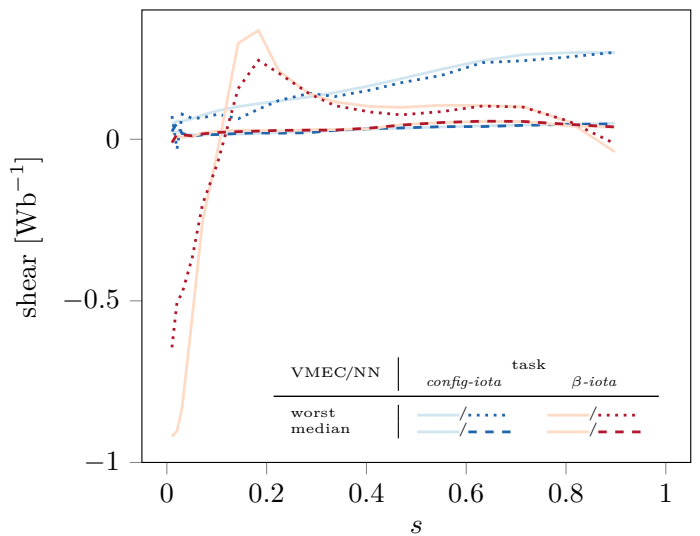

  \centering
  \igraph[width=\linewidth]{content_figures_shear_profile_sample.pgf}
  \caption{
    \added[id=AM]{
      Shear profiles of the worst and median predicted samples in the \nullIota (blue) and \finiteIota (red) tasks (same samples as in~\cref{fig:iota-profile-sample}).
      The solid lines represent the true shear profiles as evaluated from the \gls{VMEC} \gls{ibar} profiles,
      while the dotted (worst) and dashed (median) lines show the shear profiles derived from the model \gls{ibar} predicted profiles.
      The results from the worst performing cross-validation fold are shown.
      The shear profile is computed as $d\gls{ibar} / d\Phi$.
      Even in the case of the worst predicted samples,
      which feature a particular sheared profile,
      the \gls{ibar} shear is qualitatively regressed.
    }
  }%
  \label{fig:shear-profile-sample}
\end{figure}

The \replaced[id=AM]{qualitative}{qualitatively} fitness of the model can be visualized in~\cref{fig:iota-profile-sample},
which shows the worst and median predicted \gls{ibar} profiles for the worst performing cross-validation fold.
\replaced[id=AM]{
  In addition,
  as highlighted in~\cref{fig:shear-profile-sample},
  e
}{E}ven in case of the worst predicted sample in the worst performing cross-validation fold (\ie, the worst possible scenario included in the data set),
the model is still able to capture the main features of the \gls{ibar} profile (\eg, the \gls{ibar} shear).
  
\begin{figure}[!htb]
  \centering
  \igraph[width=\linewidth]{content_figures_iota_profile_rmse.pgf}
  \caption{
    Results for the \nullIota (blue) and \finiteIota (red) tasks.
    Lines show the \gls{rmse} mean and \SI[round-mode=places]{95}{\percent} confidence interval as a function of the radial coordinate $s$.
    While the \gls{rmse} in the \finiteIota task is generally lower than \num{e-3},
    the \gls{rmse} in the \nullIota scenario increases along the radial profile,
    following the characteristic vacuum \gls{W7X} \gls{ibar} profile.
  }%
  \label{fig:iota-profile-rmse}
\end{figure}

\subsection{Flux surfaces regression}\label{sec:flux-surfaces-regression}

\sisetup{round-mode=places,round-precision=2}

\begin{figure*}[!htb]
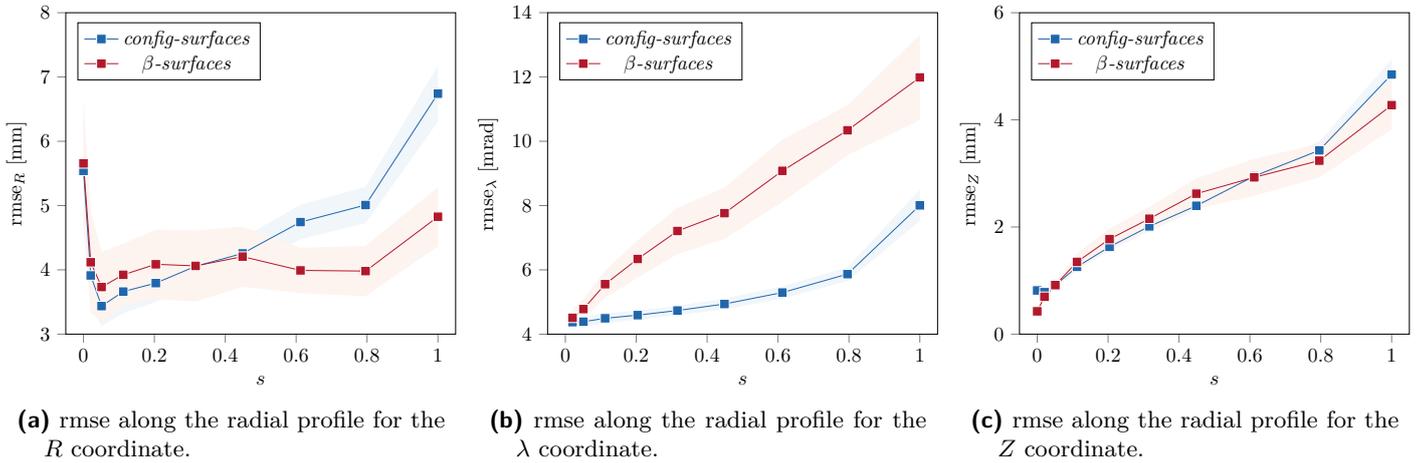

  \centering
  \begin{subfigure}{173.30093pt}
    \igraph[]{content_figures_rcos_profile_rmse.pgf}
    \subcaption{\gls{rmse} along the radial profile for the $R$ coordinate.}%
    \label{fig:rcos-profile-real-rmse}
  \end{subfigure}
    \begin{subfigure}{177.08842pt}
    \igraph[]{content_figures_lambdasin_profile_rmse.pgf}
    \subcaption{\gls{rmse} along the radial profile for the $\lambda$ coordinate.}%
    \label{fig:lambdasin-profile-real-rmse}
  \end{subfigure}
  \begin{subfigure}{173.30312pt}
    \igraph[]{content_figures_zsin_profile_rmse.pgf}
    \subcaption{\gls{rmse} along the radial profile for the $Z$ coordinate.}%
    \label{fig:zsin-profile-real-rmse}
  \end{subfigure}
  \caption{
    \gls{rmse} along the radial profile for flux surface coordinates in the \nullSurfaces and \finiteSurfaces tasks.
    The plotted values are the poloidal and toroidal average over each flux surface.
    The solid lines show the mean values for the cross-validation folds,
    while the shaded area the \SI{95}{\percent} confidence interval.
    The \gls{rmse} generally increases from the magnetic axis towards the edge on all tasks,
    apart for $R$ near the axis.
  }%
  \label{fig:surfaces-profile-real-rmse}
\end{figure*}

\Cref{fig:surfaces-profile-real-rmse} shows the \gls{rmse} broken down by flux surface coordinate along the radial profile.
The reported \gls{rmse} values are the poloidal and toroidal average on each flux surface,
on a grid as employed in \cref{sec:inputs-and-outputs}.
A solid line depicts the mean on the cross-validation folds,
while the shaded area represents the \SI{95}{\percent} confidence interval.
An initial decreasing \gls{rmse} from the magnetic axis till $s \approx 0.1$,
a plateau,
and a steep increase towards the edge can be observed in $R$ (see~\cref{fig:rcos-profile-real-rmse}).
Contrarily,
the \gls{rmse} for both $\lambda$ and $Z$ monotonically increases from the axis till the edge (see~\cref{fig:lambdasin-profile-real-rmse,fig:zsin-profile-real-rmse}).

\begin{figure*}[!htb]
  \centering
  \begin{subfigure}[t]{\subfigureTwo \subfigureWidth}
    \igraph[]{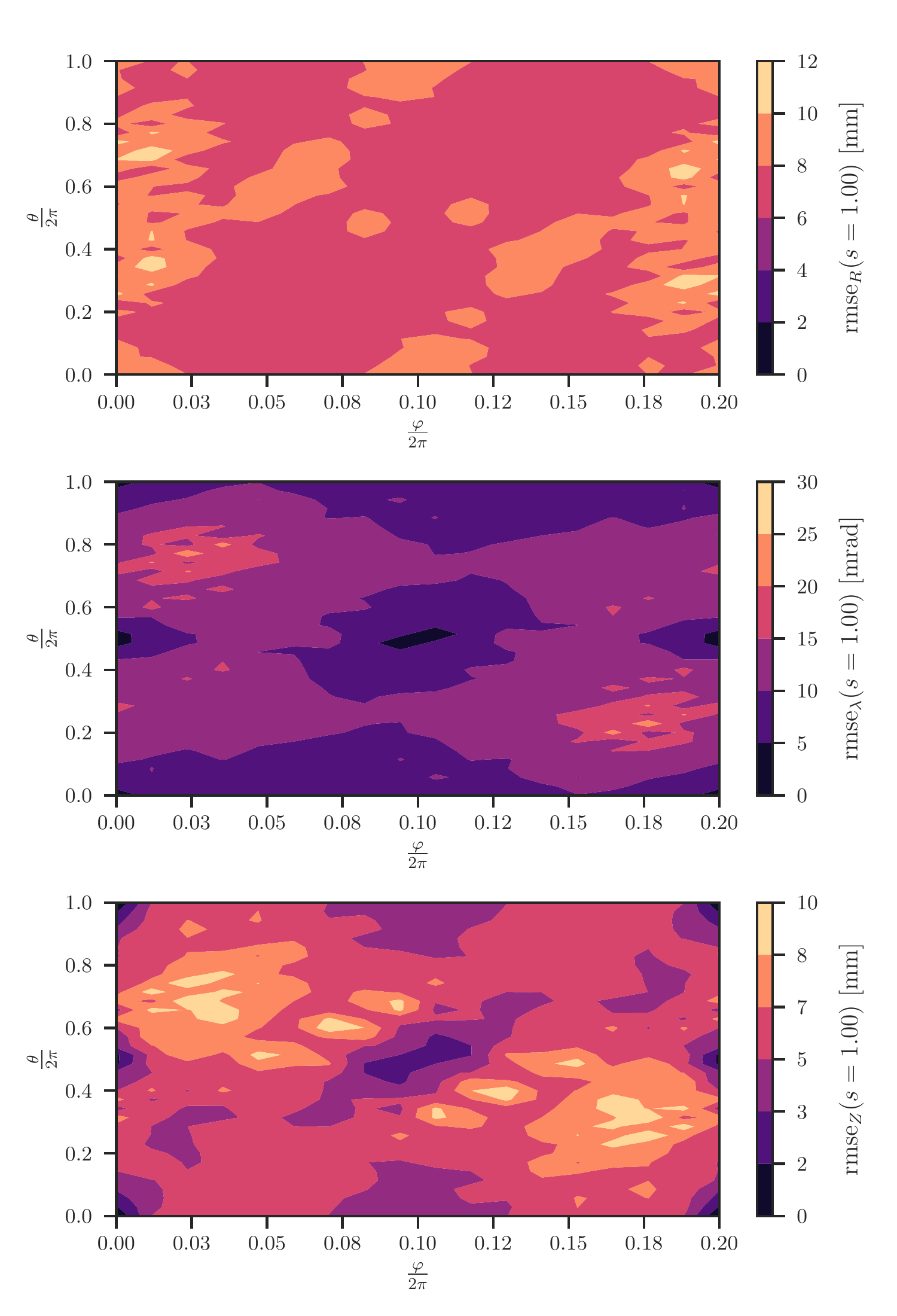}
    \subcaption{\gls{rmse} evaluated at the \gls{LCFS} on the \nullSurfaces task (\nullSquare).}%
    \label{fig:null-surfaces-map-real-rmse-lcfs}
  \end{subfigure}
  \begin{subfigure}[t]{\subfigureTwo \subfigureWidth}
    \igraph[]{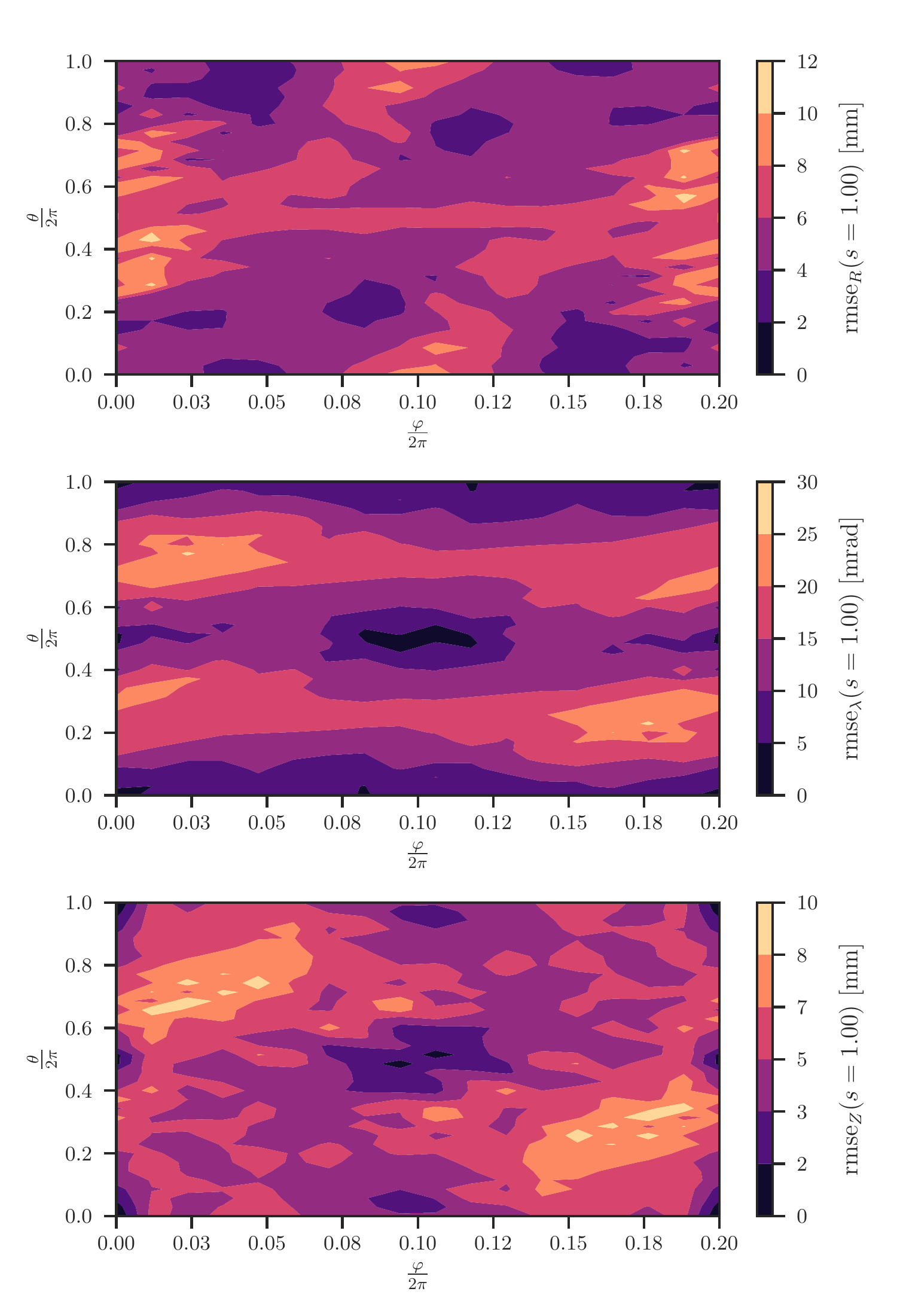}
    \subcaption{\gls{rmse} evaluated at the \gls{LCFS} on the \finiteSurfaces task (\finiteSquare).}%
    \label{fig:finite-surfaces-map-real-rmse-lcfs}
  \end{subfigure}
  \caption{
    \gls{rmse} for the surfaces tasks evaluated at the \gls{LCFS} on a grid with $N_{\theta} = 36$ poloidal and $N_{\varphi} = 18$ toroidal points per period.
    The results for the worst performing cross-validation fold are shown.
    In case of $R$,
    the bean-shape ($\varphi \approx \SI{0}{\radian}$) cross section exhibits the largest regression error.
    While for $\lambda$ and $Z$,
    the $\nicefrac{\varphi}{2 \pi} \approx 0.03)$ cross section has the largest \gls{rmse}.
  }%
  \label{fig:surfaces-map-real-rmse-lcfs}
\end{figure*}

In all coordinates,
apart from $R$ in the \finiteSurfaces task,
the \gls{rmse} is higher at $s=1$, \ie, the \gls{LCFS}.
We find it worth investigating this in more detail, and hence examine the poloidal and toroidal dependency of the \gls{rmse} specifically at the \gls{LCFS} with \cref{fig:surfaces-map-real-rmse-lcfs}.
In order to emphasize the error,
the worst performing cross-validation fold is shown,
and a grid with $N_{\theta} = 36$ poloidal and $N_{\varphi} = 18$ toroidal points per period has been used.
The error for $R$ is almost flat on the surface,
with maxima at $\varphi \approx \SI{0}{\radian}$,
representing the tips of the bean-shaped cross section.
On the other hand,
the error for $Z$ and $\lambda$ shows a $m = 1$, $n = 1$ dependency.
In the \nullSurfaces scenario,
at $\nicefrac{\varphi}{2 \pi} \approx 0.03$ (and at $\nicefrac{\varphi}{2 \pi} \approx 0.17$ following the symmetry)
a higher \gls{rmse} is observed.
In the \finiteSurfaces task,
while the \gls{rmse} for $Z$ still shows a poloidal and toroidal dependency similar to that observed in the vacuum case,
the dominant \gls{rmse} factor for $\lambda$ is a poloidal $m = 1$ term.

\begin{figure*}[!htb]
  \centering
  \begin{subfigure}[t]{\subfigureTwo \subfigureWidth}
    \igraph[]{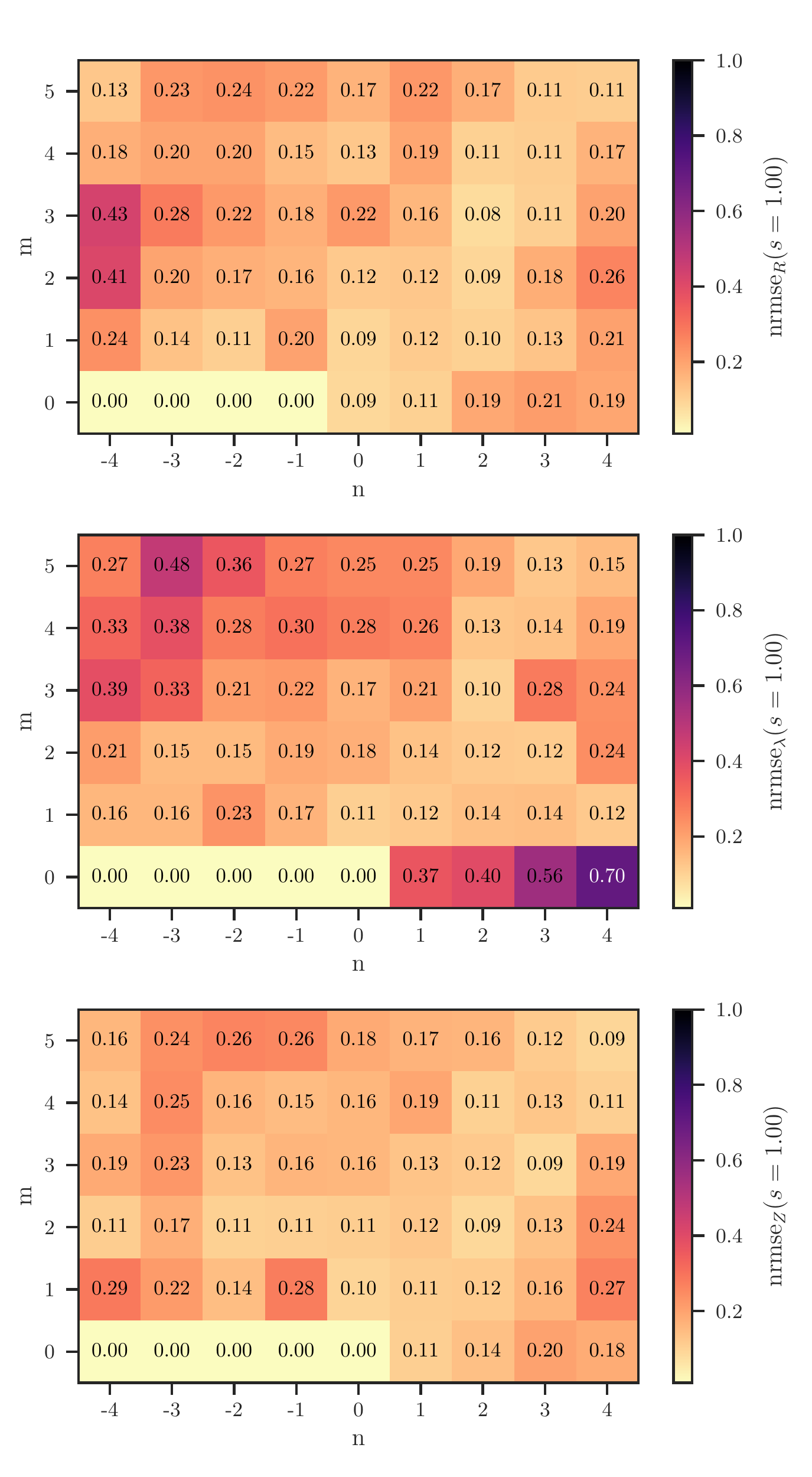}
    \subcaption{\gls{nrmse} for the \glspl{FC} of $\vec{x}$ at the \gls{LCFS} on the \nullSurfaces task (\nullSquare).}%
    \label{fig:null-surfaces-map-fourier-nrmse-lcfs}
  \end{subfigure}
  \begin{subfigure}[t]{\subfigureTwo \subfigureWidth}
    \igraph[]{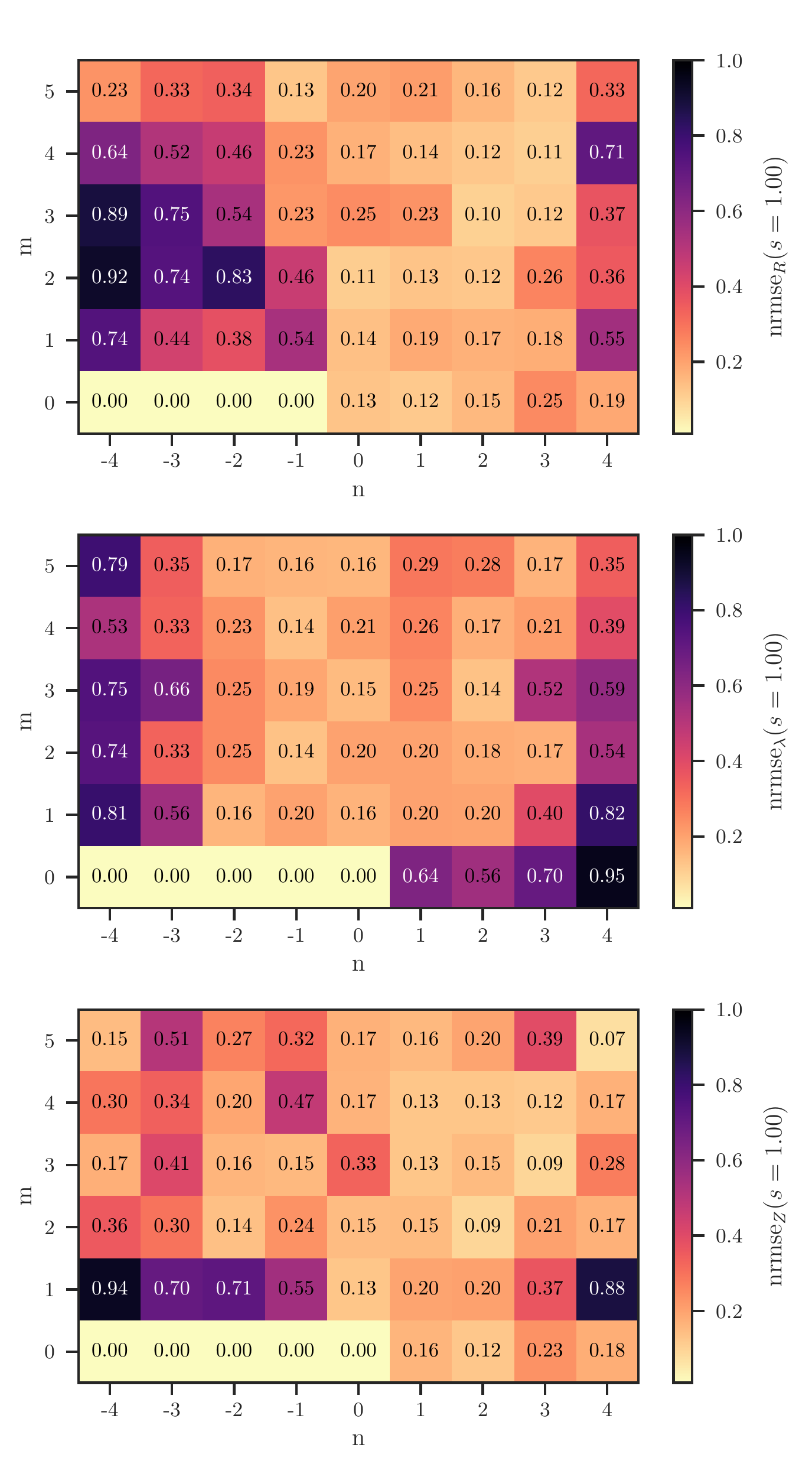}
    \subcaption{\gls{nrmse} for the \glspl{FC} of $\vec{x}$ at the \gls{LCFS} on the \finiteSurfaces task (\finiteSquare).}%
    \label{fig:finite-surfaces-map-fourier-nrmse-lcfs}
  \end{subfigure}
  \caption{
    \gls{nrmse} for the regressed \glspl{FC} in the \nullSurfaces and \finiteSurfaces tasks.
    For each FC, the \gls{nrmse} value is annotated at the $(m, n)$ location.
    The worst performing cross validation fold is shown.
  }%
  \label{fig:surfaces-map-fourier-nrmse-lcfs}
\end{figure*}

To further investigate the \gls{rmse} poloidal and toroidal dependency,
\cref{fig:surfaces-map-fourier-nrmse-lcfs} shows the regression error on the \glspl{FC} of $(R, \lambda, Z)$ evaluated at the \gls{LCFS}.
In this figure,
to effectively compare the error on both low-order and high-order modes (see \cref{fig:fourier-map}),
the \gls{nrmse} is used.
Again, the worst performing cross-validation fold is shown.
It is important to note that the \glspl{FC} are the actual quantities which the \gls{NN} learned.
In both cases,
the leading \glspl{FC} are regressed with a \gls{nrmse} below \SI{20}{\percent}.
However,
there are some regions in the $(m, n)$ space which the model struggles to reconstruct,
in particular in the \finiteSurfaces task (see~\cref{fig:finite-surfaces-map-fourier-nrmse-lcfs}).

\begin{figure*}[!htb]
  \centering
  \begin{subfigure}{\subfigureThree \subfigureWidth}
    \igraph[]{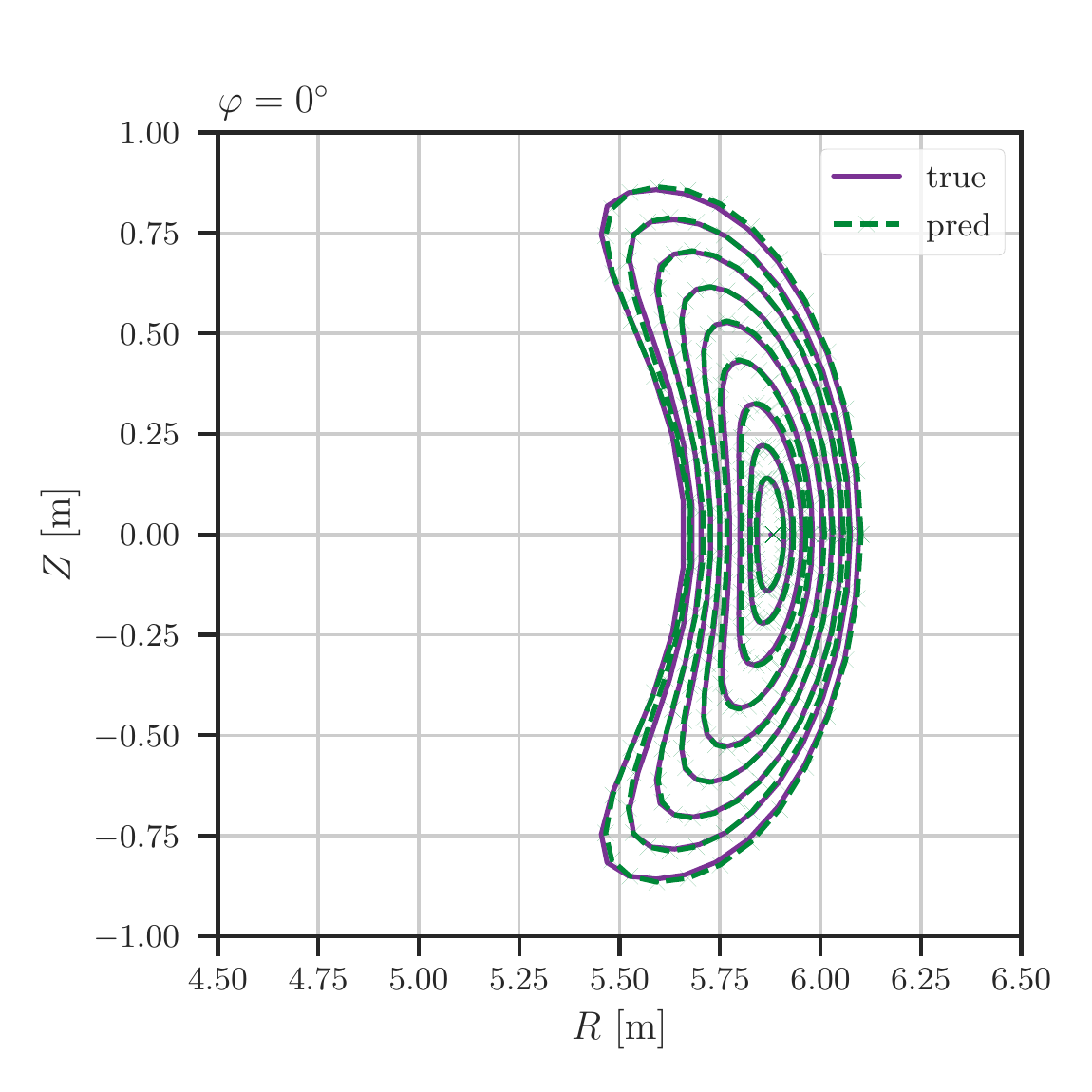}%
    \label{fig:null-surfaces-0-worst}
  \end{subfigure}
  \begin{subfigure}{\subfigureThree \subfigureWidth}
    \igraph[]{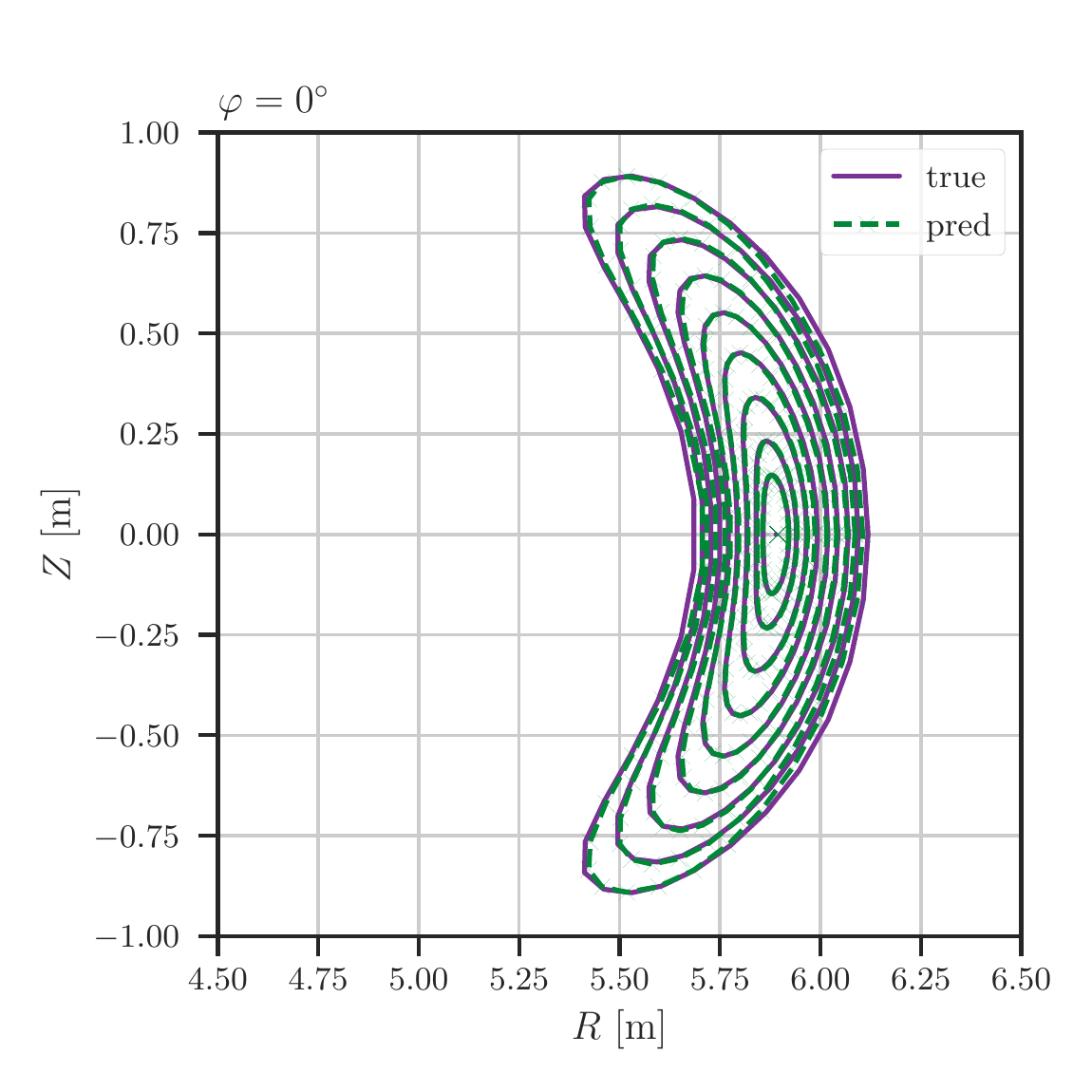}%
    \label{fig:null-surfaces-0-median}
  \end{subfigure}
  \begin{subfigure}{\subfigureThree \subfigureWidth}
    \igraph[]{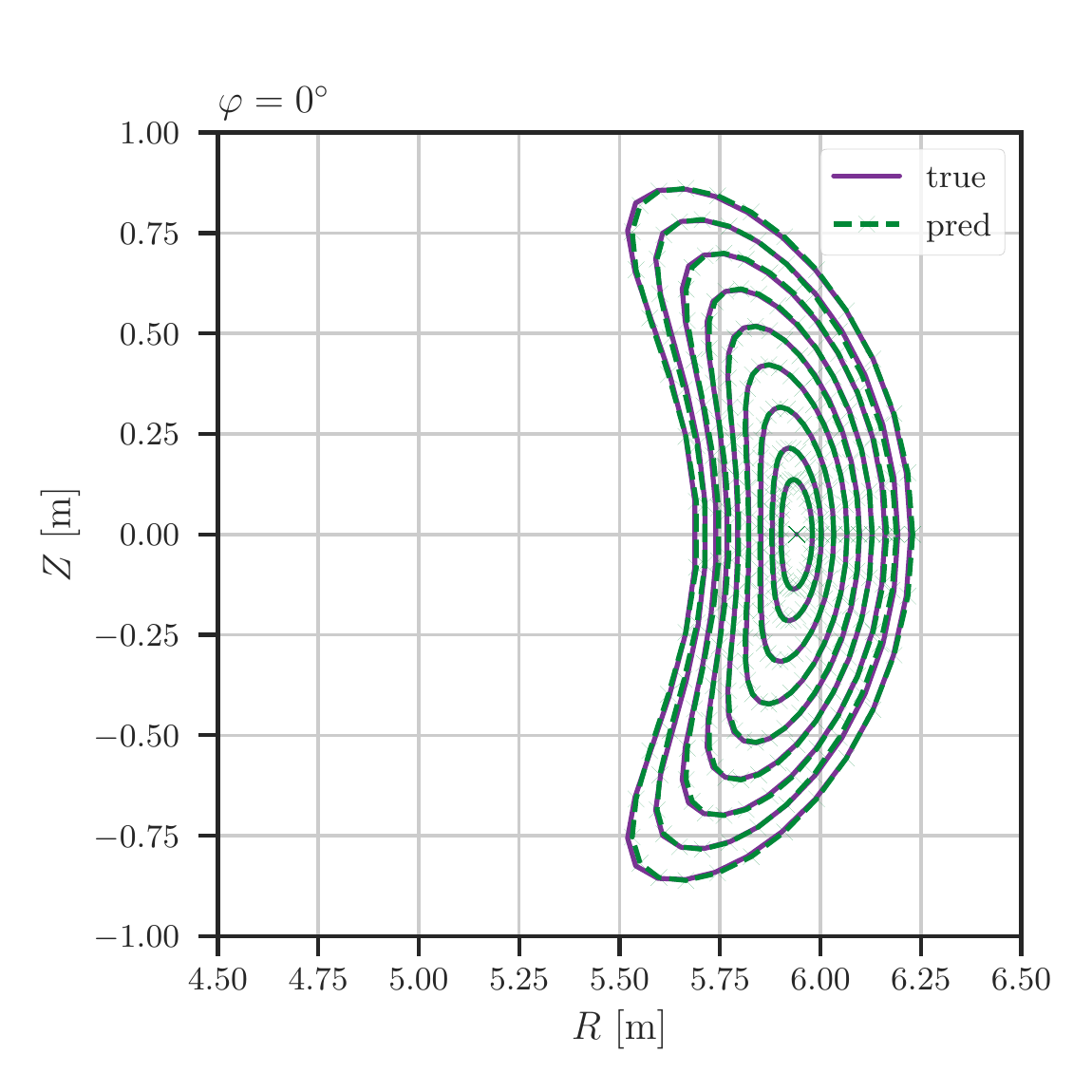}%
    \label{fig:null-surfaces-0-best}
  \end{subfigure}
  \begin{subfigure}{\subfigureThree \subfigureWidth}
    \igraph[]{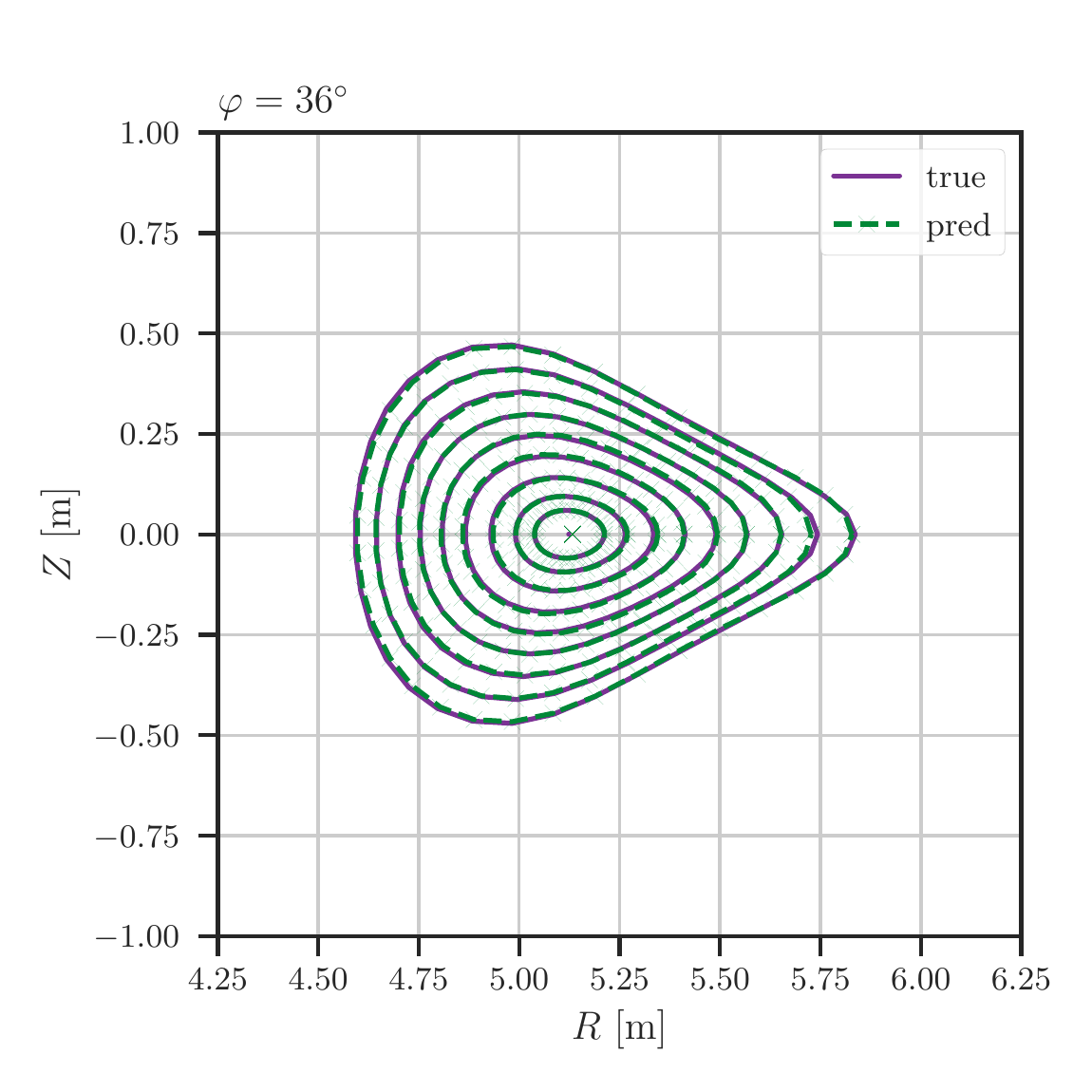}%
    \label{fig:null-surfaces-36-worst}
  \end{subfigure}
  \begin{subfigure}{\subfigureThree \subfigureWidth}
    \igraph[]{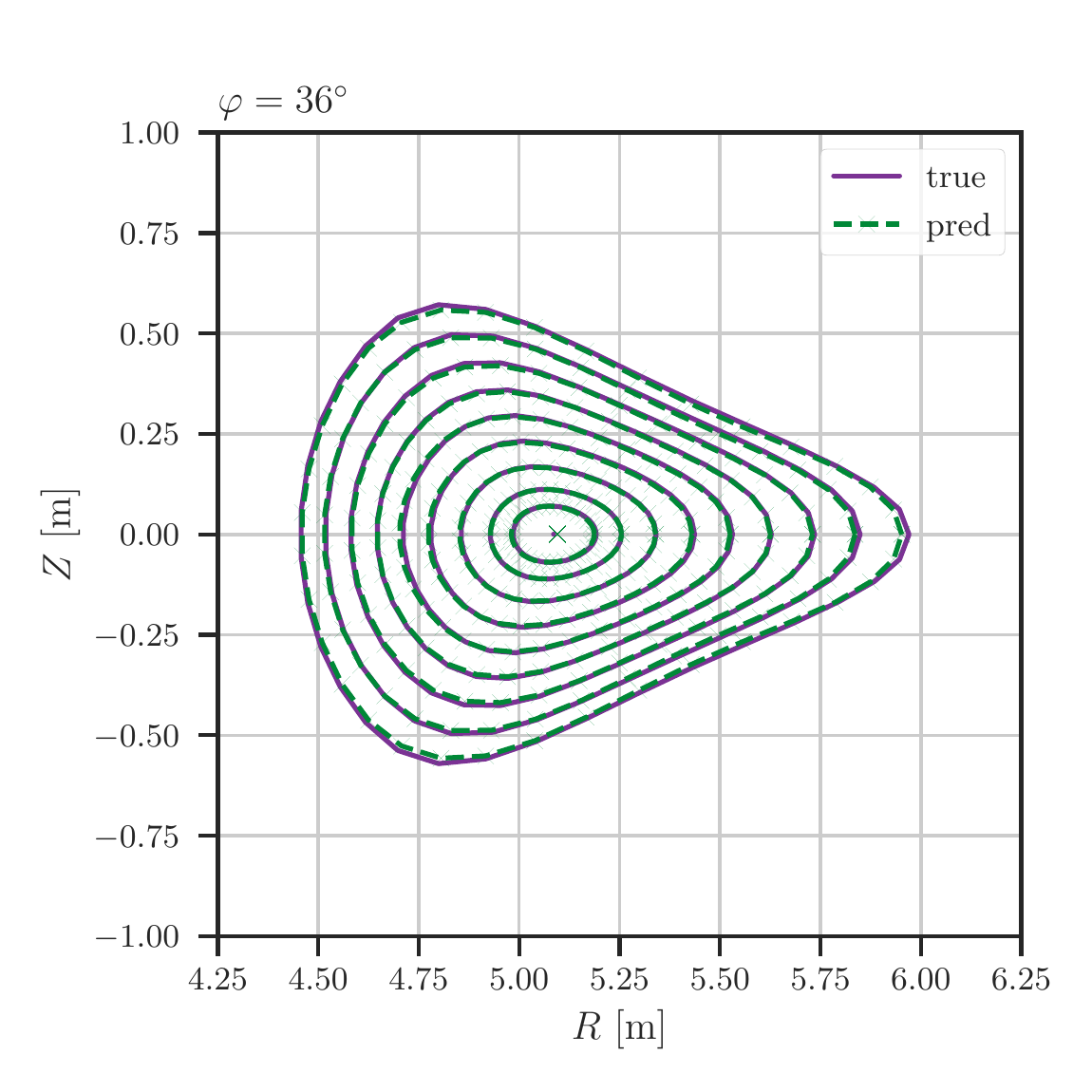}%
    \label{fig:null-surfaces-36-median}
  \end{subfigure}
  \begin{subfigure}{\subfigureThree \subfigureWidth}
    \igraph[]{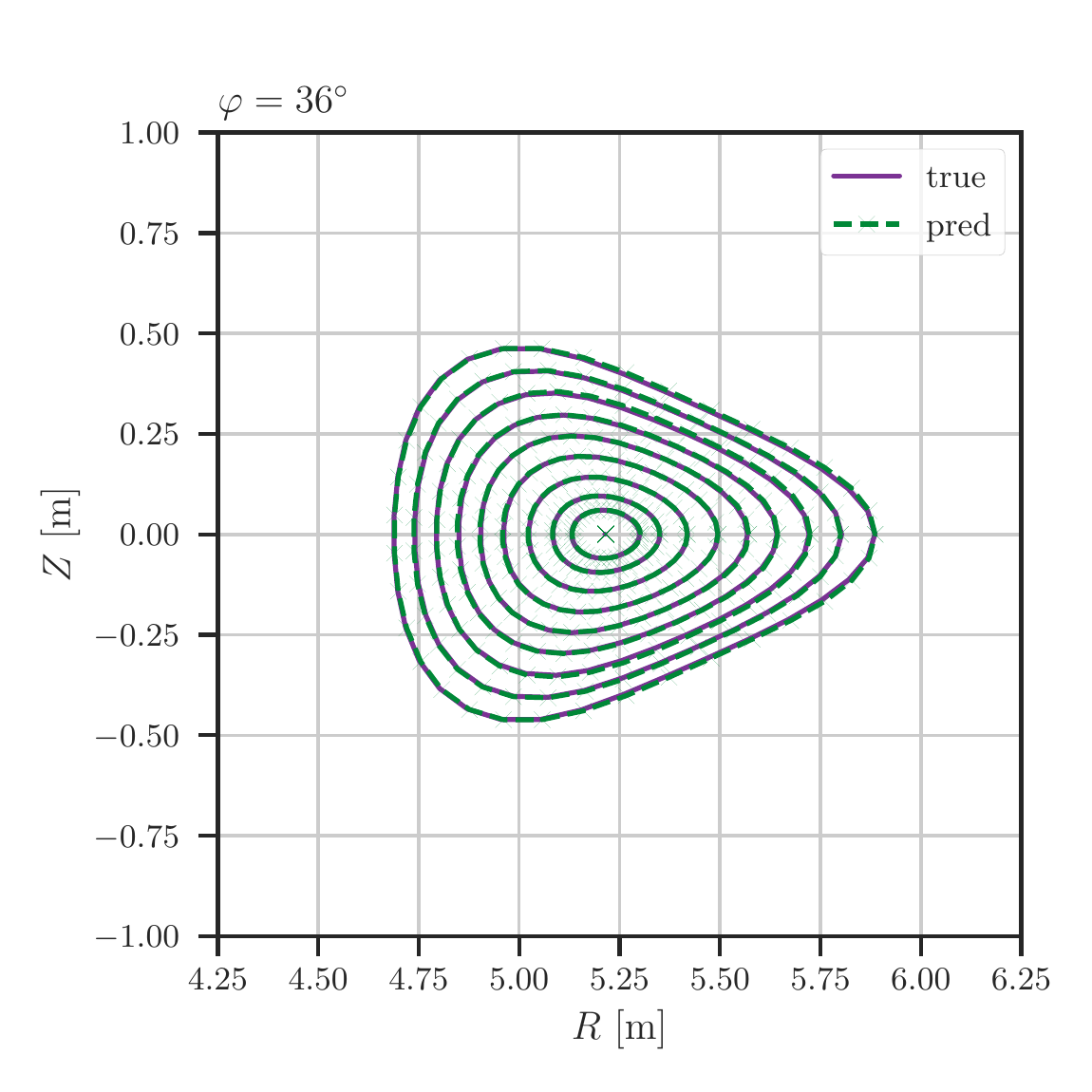}%
    \label{fig:null-surfaces-36-best}
  \end{subfigure}
  \caption{
    True (pink) and predicted (green) flux surfaces for the bean-shape (upper) and triangular (bottom) cross sections on the \nullSurfaces task (\nullSquare).
    The worst (left), median (center) and best (right) regressed samples are shown from the worst performing cross-validation fold.
  }
  \label{fig:null-surfaces}
\end{figure*}

\begin{figure*}[hb!]
  \centering
  \begin{subfigure}{\subfigureThree \subfigureWidth}
    \igraph[]{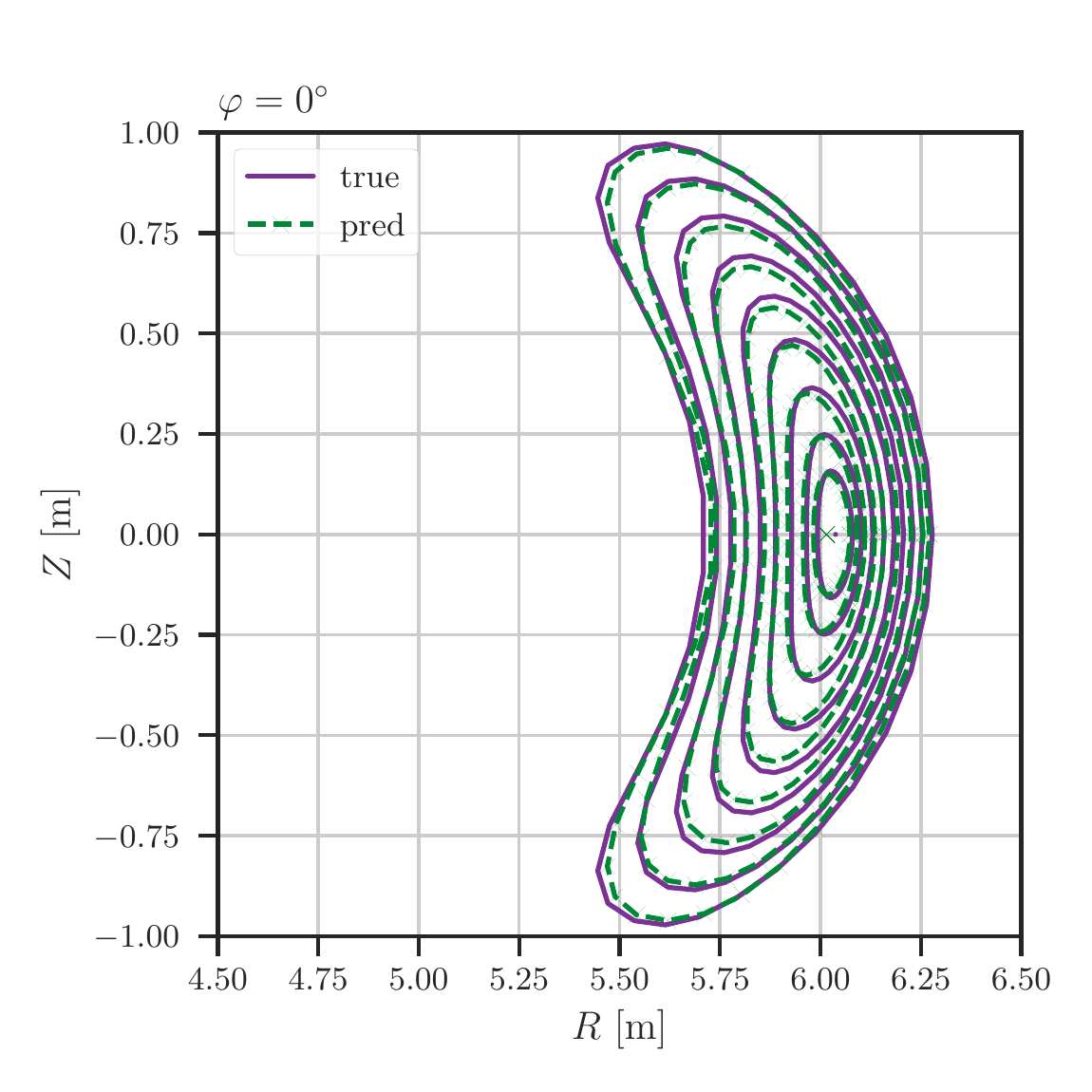}%
    \label{fig:finite-surfaces-0-worst}
  \end{subfigure}
  \begin{subfigure}{\subfigureThree \subfigureWidth}
    \igraph[]{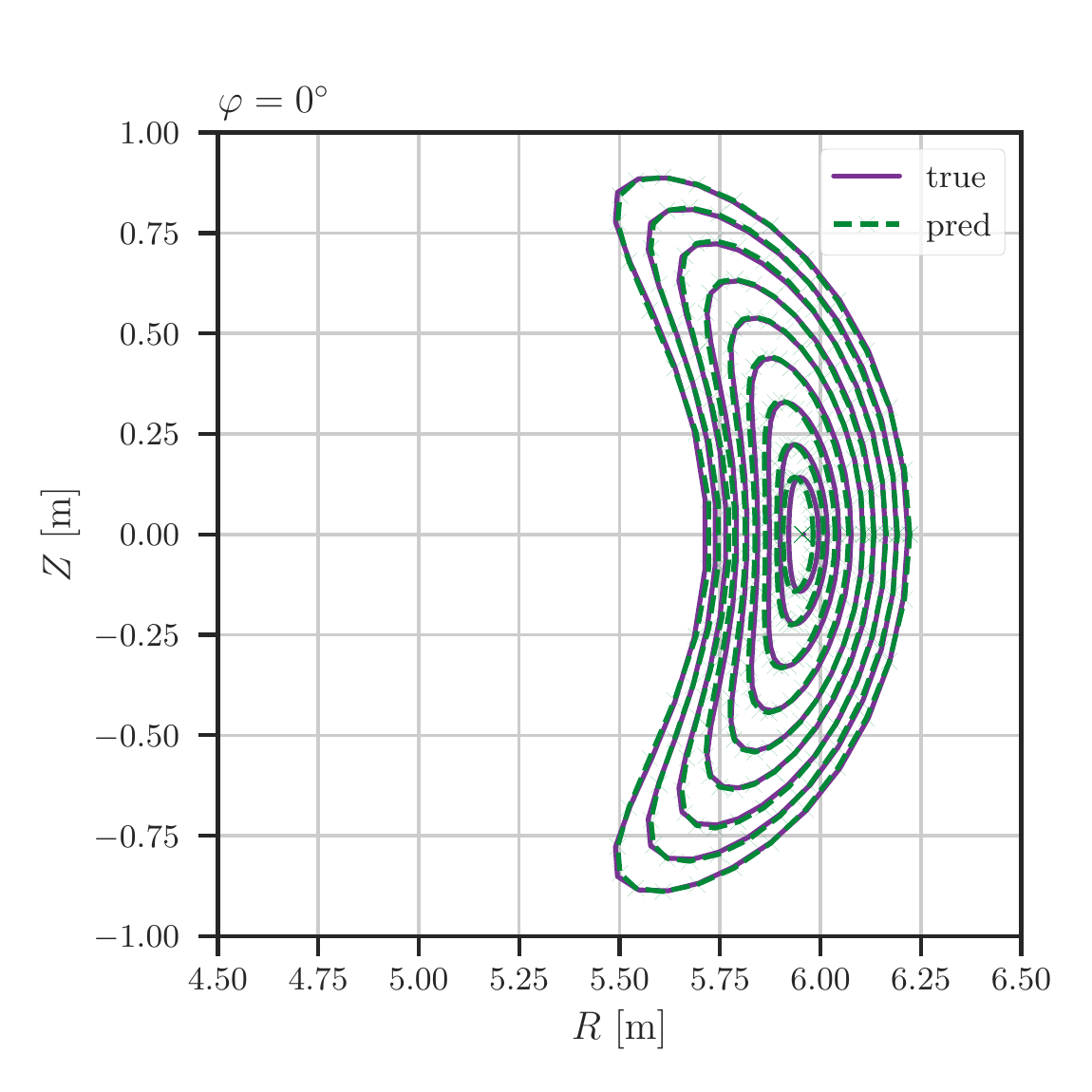}%
    \label{fig:finite-surfaces-0-median}
  \end{subfigure}
  \begin{subfigure}{\subfigureThree \subfigureWidth}
    \igraph[]{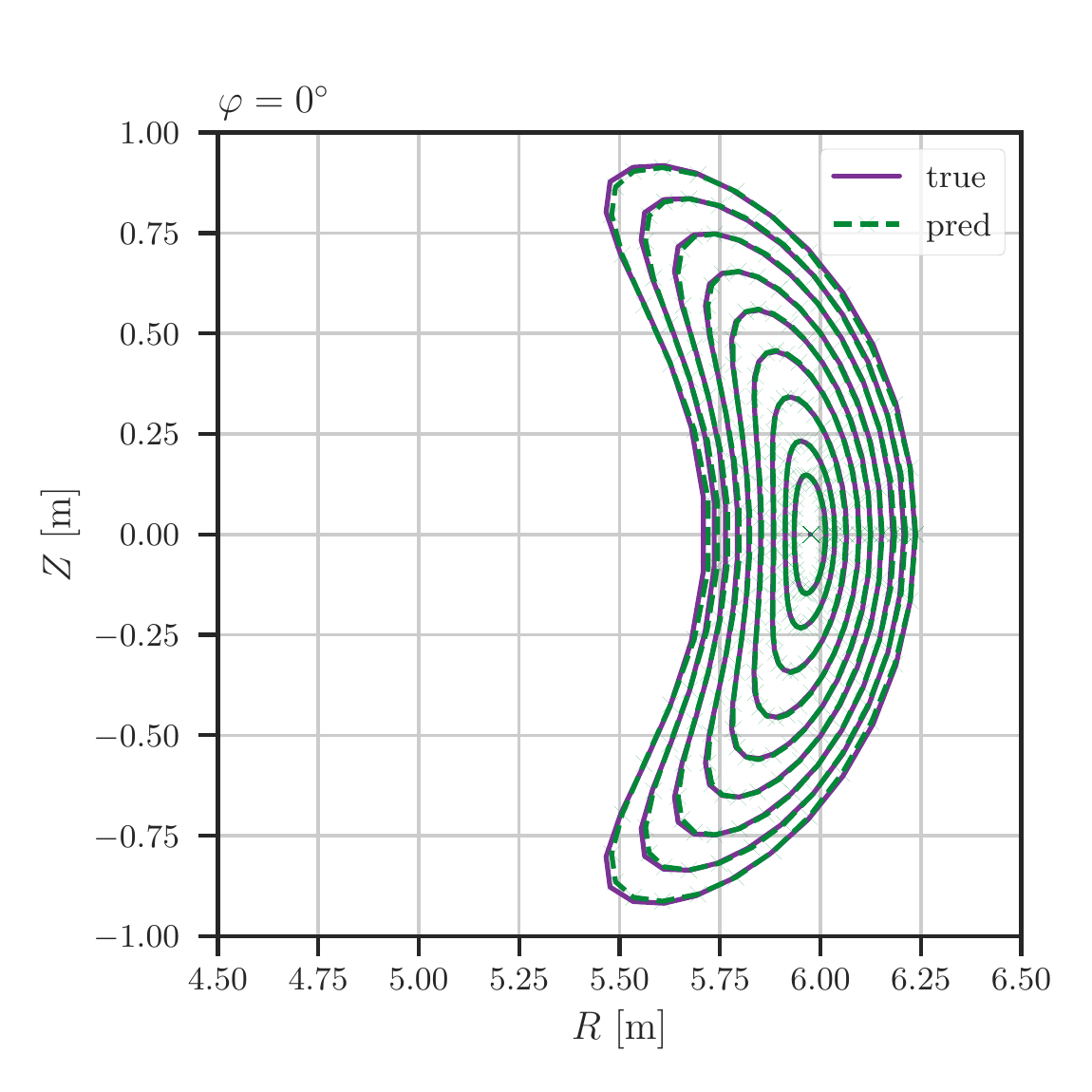}%
    \label{fig:finite-surfaces-0-best}
  \end{subfigure}
  \begin{subfigure}{\subfigureThree \subfigureWidth}
    \igraph[]{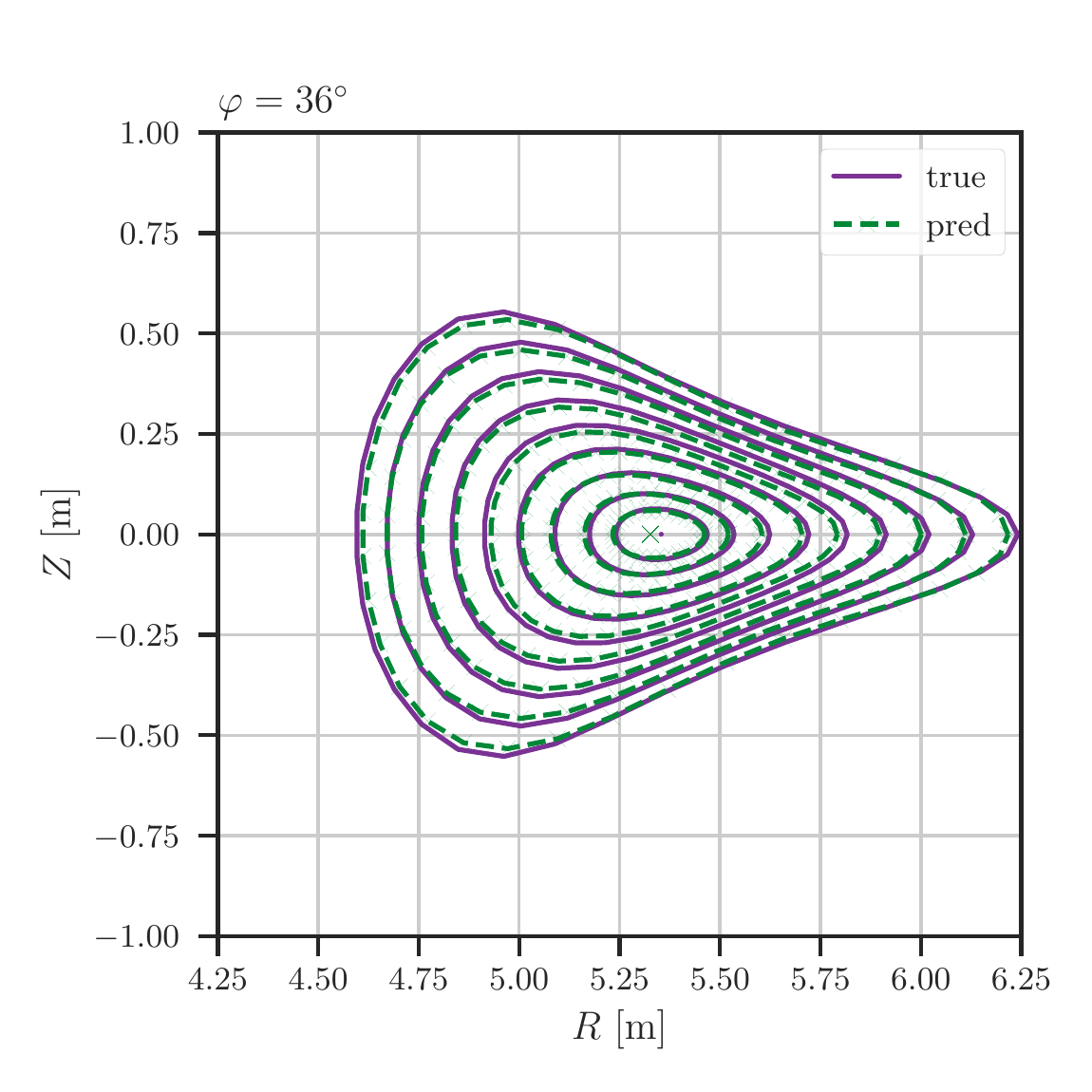}%
    \label{fig:finite-surfaces-36-worst}
  \end{subfigure}
  \begin{subfigure}{\subfigureThree \subfigureWidth}
    \igraph[]{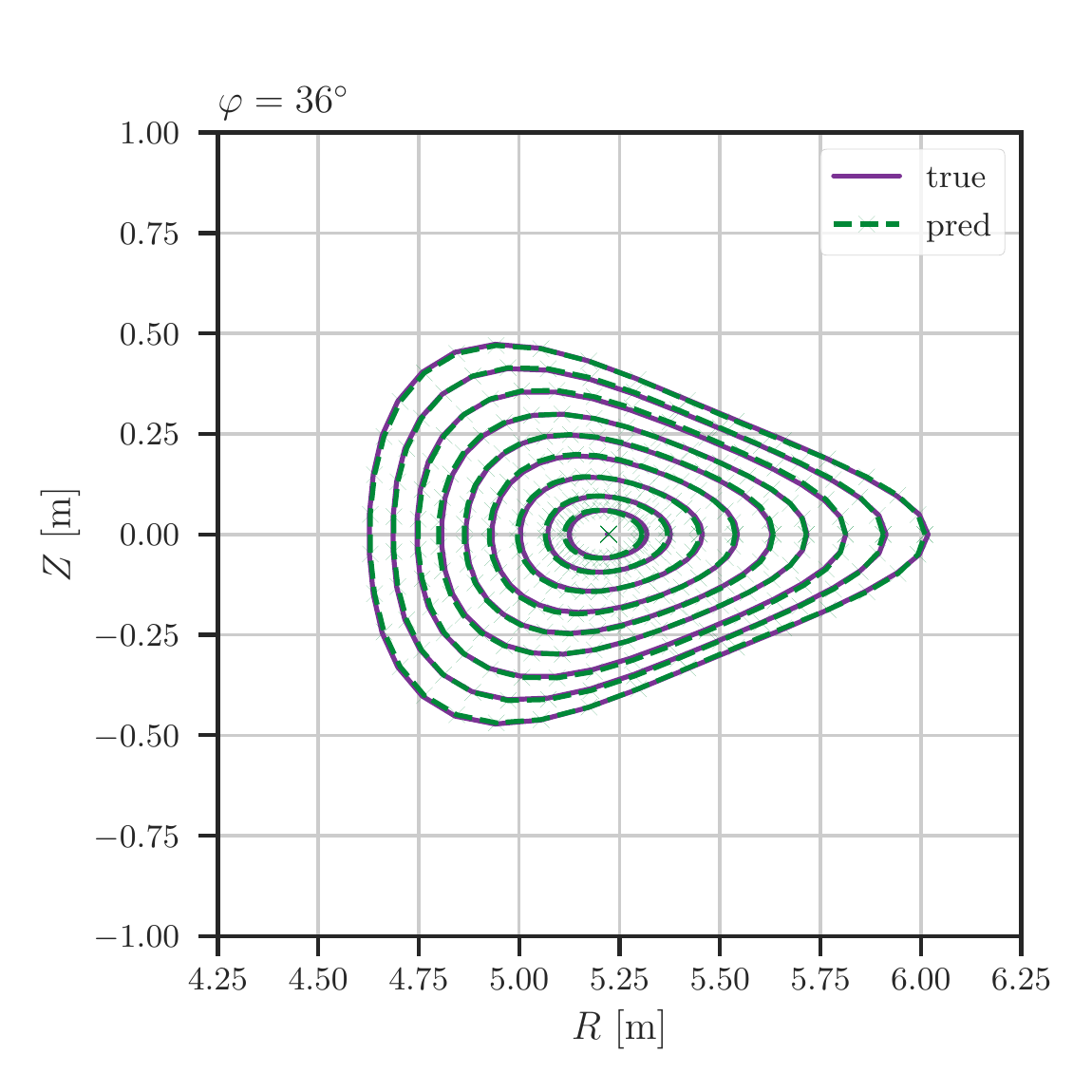}%
    \label{fig:finite-surfaces-36-median}
  \end{subfigure}
  \begin{subfigure}{\subfigureThree \subfigureWidth}
    \igraph[]{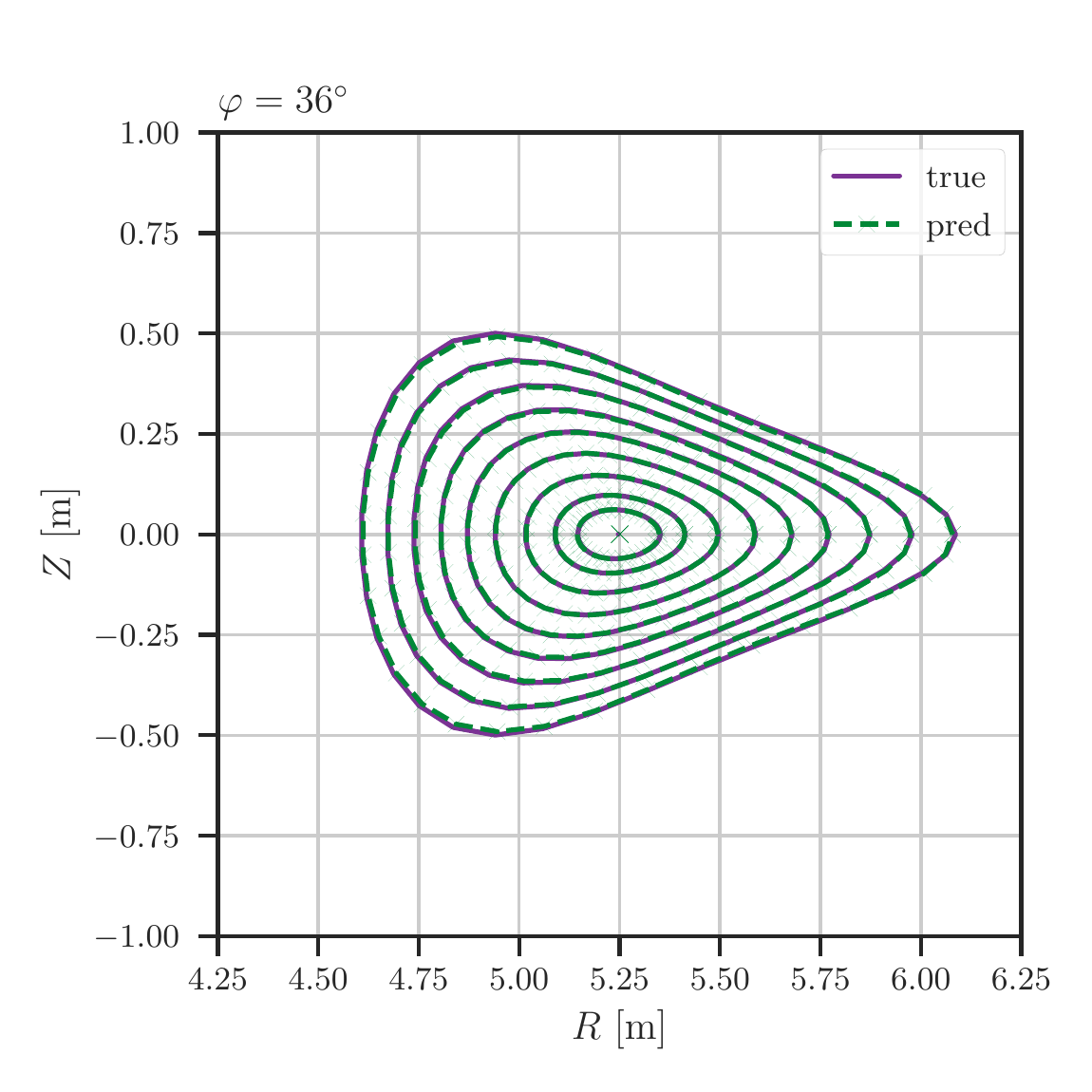}%
    \label{fig:finite-surfaces-36-best}
  \end{subfigure}
  \caption{
    True (pink) and predicted (green) flux surfaces for the bean-shape (upper) and triangular (bottom) cross sections on the \finiteSurfaces task (\finiteSquare).
    The worst (left), median (center) and best (right) regressed samples are shown from the worst performing cross-validation fold.
  }%
  \label{fig:finite-surfaces}
\end{figure*}

The regression performance is visualized in \cref{fig:null-surfaces,fig:finite-surfaces},
where the true and regressed flux surfaces at the bean-shape ($\varphi = \SI{0}{\radian}$) and triangular ($\nicefrac{\varphi}{2 \pi} = \num{0.1}$) cross sections are represented.
Worst (left),
median (center),
and best (right) regressed samples from the worst performing cross-validation fold are shown.
The \gls{LCFS} shows the largest inconsistency (as already observed in \cref{fig:surfaces-profile-real-rmse}),
and in particular the $R$ coordinate of the high-field side of the bean-shaped cross section (\ie, $\theta = \pi$) appears to be the most complicated feature to resolve (as previously observed in \cref{fig:surfaces-map-fourier-nrmse-lcfs}).

Of the three flux surface coordinates,
$\lambda$ is the most arduous to reconstruct.
Although not needed to compute the location of the flux surfaces,
it gives information on the direction of the magnetic field lines.
In particular,
the $\lambda_{0n}$ \glspl{FC} are hardly regressed in both flux surface tasks.
Earlier works have encountered similar challenges and the lack of spectral minimization for $\lambda$ in \gls{VMEC} is presumed to cause such difficulties~\cite{Sengupta2007}.

\subsection{Magnetic field strength regression}\label{sec:bcos-regression}

\begin{figure}[!hbt]
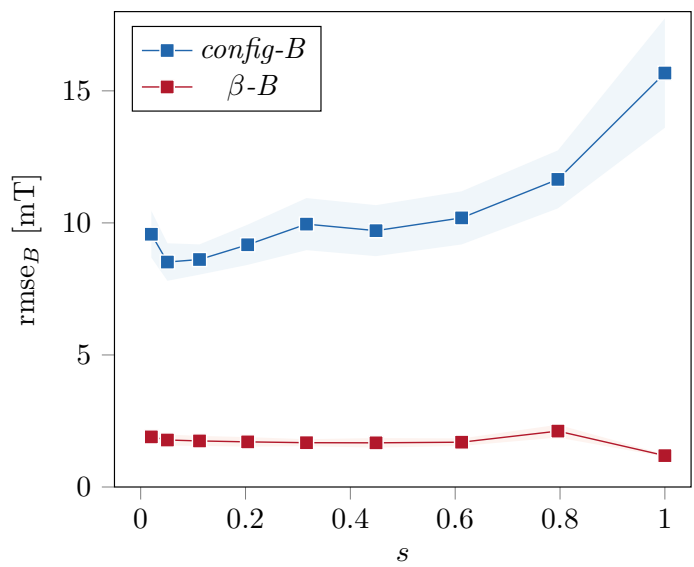

  \centering
  \igraph[width=\linewidth]{content_figures_bcos_profile_rmse.pgf}
  \caption{
    \gls{rmse} along the radial profile for the \nullB and \finiteB tasks.
    Lines show the \gls{rmse} mean and \SI[round-mode=places]{95}{\percent} confidence interval.
    The \gls{rmse} in the \finiteB task is considerable lower than in \nullB due to the smaller spread of $B$ in the data set.
  }%
  \label{fig:bcos-profile-rmse}
\end{figure}

The variance of the magnetic field strength contained in the vacuum and finite-\averagePlasmaBeta scenario data sets are on different orders of magnitude.
In \nullB,
the magnetic field strength exhibits an average spread of $\sigma_{\nullB} = \SI[round-mode=places,round-precision=0]{252.484953}{\milli\tesla}$,
while in \finiteB,
the spread is only $\sigma_{\finiteB} = \SI[round-mode=places,round-precision=0]{28.171711}{\milli\tesla}$.
This mainly derives from the rich vacuum magnetic configuration space of \gls{W7X} and the low impact of pressure and toroidal current on the equilibrium field~\cite{Geiger2015}.
Therefore,
the magnetic field strength in the \finiteB task is more difficult to resolve.
Indeed,
even though the achieved \gls{nrmse} for the \finiteB task is higher than in \nullB,
the \gls{rmse} in \finiteB is considerably lower than in \nullB,
as \cref{fig:bcos-profile-rmse} shows,
due to the smaller spread in the data set.
Additionally,
in \finiteB,
the \gls{rmse} does not seem to have any radial dependency,
while the regression error increases from the magnetic axis towards the edge for \nullB.

\begin{figure*}[!hbt]
  \centering
  \begin{subfigure}{\subfigureTwo \subfigureWidth}
    \igraph[]{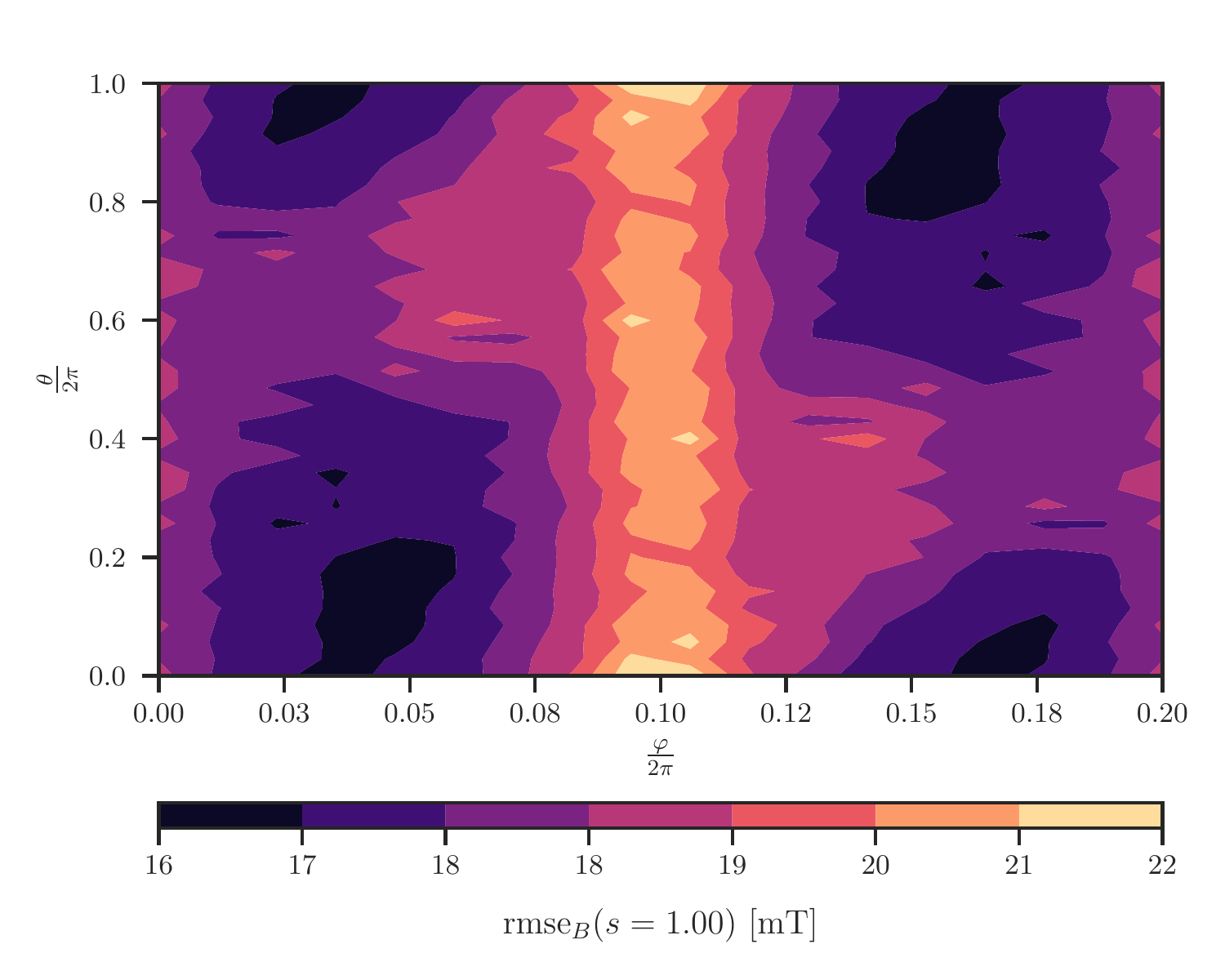}
    \subcaption{\gls{rmse} evaluated at the \gls{LCFS} on the \nullB task (\nullSquare).}%
    \label{fig:null-bcos-map-real-rmse-lcfs}
  \end{subfigure}
  \begin{subfigure}{\subfigureTwo \subfigureWidth}
    \igraph[]{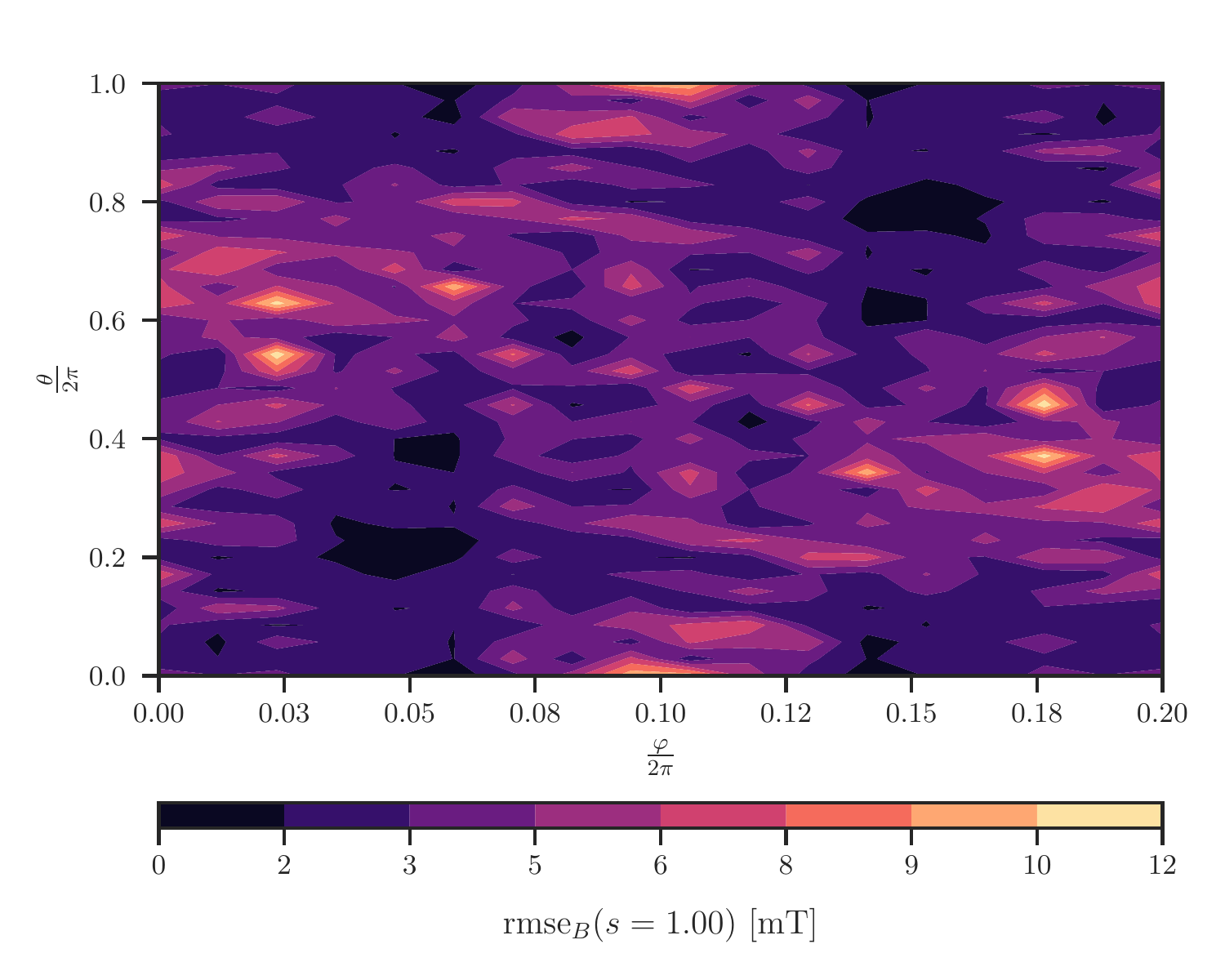}
    \subcaption{\gls{rmse} evaluated at the \gls{LCFS} on the \finiteB task (\finiteSquare).}%
    \label{fig:finite-bcos-map-real-rmse-lcfs}
  \end{subfigure}
  \caption{
    \gls{rmse} for the magnetic field strength tasks evaluated at the \gls{LCFS} on a grid with $N_{\theta} = 36$ poloidal and $N_{\varphi} = 18$ toroidal points per period.
    The worst performing cross-validation fold is shown.
    In \nullB,
    the errors on the $B_{00}$, $B_{01}$ and $B_{10}$ terms are the major contributors to the \gls{rmse}.
    In \finiteB instead,
    the \gls{rmse} is almost flat with a shallow, high-order structure.
  }%
  \label{fig:bcos-map-real-rmse-lcfs}
\end{figure*}

The topology of the regression error of the magnetic field strength at the \gls{LCFS}, for the worst performing cross-validation fold, is visualized in~\cref{fig:bcos-map-real-rmse-lcfs}.
In the \nullB task,
in addition to a non-zero baseline error (\ie, $m=0$ and $n=0$),
$n=1$ toroidal and $m=1$ poloidal terms are visible.
This stems from the fact that the main \glspl{FC} of \gls{W7X},
besides $B_{00}$,
are $B_{01}$ and $B_{10}$,
while in general the other $B_{mn}$ are much smaller~\cite{Geiger2015}.
Contrarily,
in the \finiteB task,
the \gls{rmse} surface is more indented and higher $B_{mn}$ error terms become the dominant influence on the regression error.

\begin{figure*}[!htb]
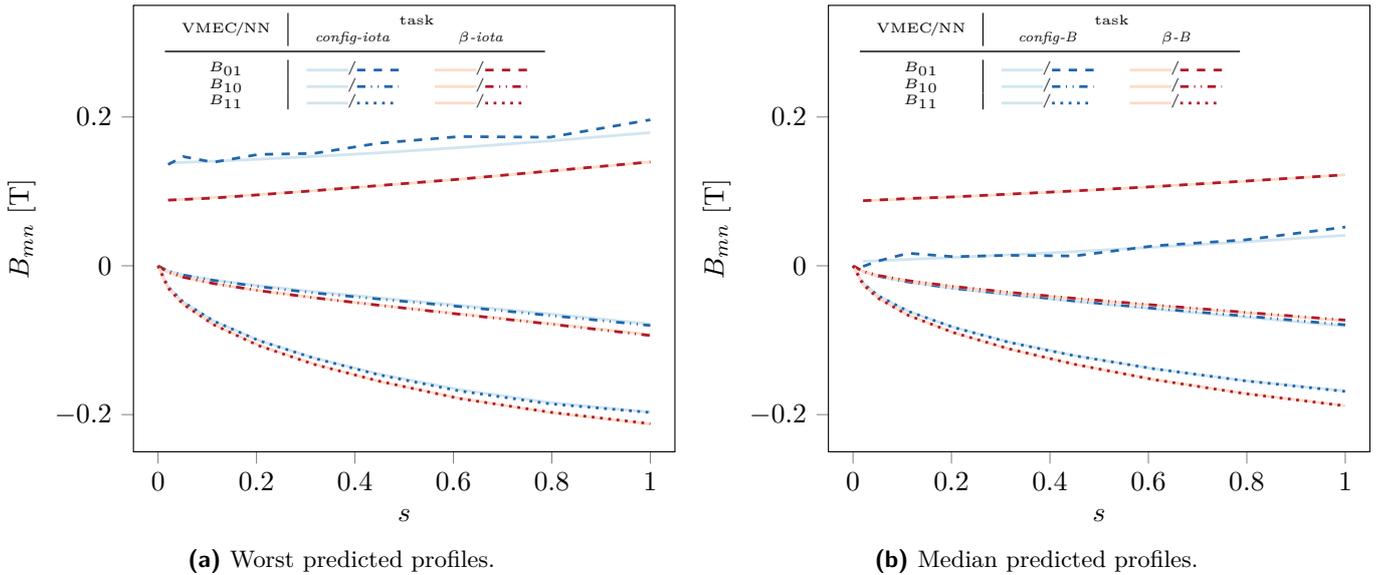

  \centering
  \begin{subfigure}{\subfigureTwo \subfigureWidth}
    \igraph[]{content_figures_bcos_profile_worst_sample.pgf}
    \subcaption{
      Worst predicted profiles.
    }%
    \label{fig:bcos-profile-worst-sample}
  \end{subfigure}
  \begin{subfigure}{\subfigureTwo \subfigureWidth}
    \igraph[]{content_figures_bcos_profile_median_sample.pgf}
    \subcaption{
      Median predicted profiles.
    }%
    \label{fig:bcos-profile-median-sample}
  \end{subfigure}
    \caption{
      The solid lines represent the true $B_{mn}$ profiles as evaluated by VMEC,
      while the marks show the profiles as predicted by the model.
      The results from the \nullB (blue) and \finiteB (red) tasks,
      and the worst performing cross-validation fold are shown.
  }%
  \label{fig:bcos-profile-sample}
\end{figure*}

\Cref{fig:bcos-profile-sample} qualitatively captures the regression of the leading \glspl{FC},
where the true and predicted \gls{FC} profiles are plotted in case of the worst and median samples.
The worst performing cross-validation fold is shown.
As observed in~\cref{fig:bcos-map-real-rmse-lcfs},
$B_{01}$ shows the largest discrepancy.

\section{Summary and outlook}\label{chap:conclusion}

This paper investigates the feasibility of building a fast surrogate \gls{NN} model of the \gls{MHD} equilibrium code \gls{VMEC} in \gls{W7X} magnetic configurations.
It extends earlier works~\cite{Sengupta2004,Sengupta2007} by using physics constrained plasma profiles, modern \gls{NN} architectures and workflows, and by employing single models to reconstruct multiple output quantities.
The decomposition of the problem into a vacuum and finite-\averagePlasmaBeta data set allows the independent study of the two limiting cases,
of which the viability is necessary for a future \gls{VMEC} surrogate model.

The reconstruction of the \gls{ibar} profile shows a \gls{nrmse} between \SI{1}{\percent} and \SI{5}{\percent}.
Regression of flux surface coordinates $(R, \lambda, Z)$ gives \gls{nrmse} values between \SI{14}{\percent} and \SI{20}{\percent},
where the $\lambda$ coordinate appears to be the most problematic to regress.
For the magnetic field strength $B$,
\gls{nrmse} values between \SI{3}{\percent} and \SI{10}{\percent} are obtained.
In almost all outputs,
the regression error increases from the magnetic axis towards the edge.
As expected,
the regression of the finite-\averagePlasmaBeta samples proves to be more challenging than the vacuum cases.
However,
the observed \gls{rmse} values were often similar for the two scenarios.
Limited to the investigated scenarios,
a relatively small data set (\eg, \SI{10}{\kilo\nothing} samples) seems to be adequate.

\added[id=AM]{%
  The promising results of this paper show that \glspl{NN} can be used to deploy a drop-in surrogate model for \gls{VMEC},
  although additional questions have to be investigated.
  First,
  the performance of such models to resolve both the vacuum magnetic configuration and the finite-beta effects on the equilibrium magnetic field has to be assessed by using a data set which comprises both vacuum and finite-\averagePlasmaBeta samples.
  Second,
  \replaced[id=AM]{%
    to define a quantitative required accuracy for the models,
    which strongly depends on the target application,
    the degree to which physics quantities of interest,
    such as \gls{MHD} stability or neoclassical transport rates,
    are faithfully reproduced has to be characterized.
    This verification represents a key metric to gauge the use of \gls{NN} models to provide fast,
    yet physics-preserving,
    \gls{MHD} equilibria.
  }{%
    broader \gls{HP} search and an ensemble of \glspl{NN} could further improve the performance over single base learner~\cite{Hansen1990},
    and optimization techniques,
    such as pruning and quantization,
    are expected to deliver improved inference times.
    Moreover, the results of this work suggest that the full Fourier resolution is in reach if larger \glspl{NN} and longer training time are accessible.}
}

Given such unexplored application,
several paths can still be investigated.
First,
multiple output quantities (\eg, \gls{ibar} and $\vec{x}$) can be regressed at once with a single model,
thus exploiting the correlation between those quantities.
\added[id=AM]{Second,
  to obtain self-consistent equilibrium magnetic fields and flux surfaces geometries,
  the magnetic field strength $B$ could be computed directly from the model's \gls{ibar} and $\vec{x}$ instead of being regressed (see~\cref{eq:b-theta-up,eq:b-phi-up,eq:b-theta-down,eq:b-phi-down,eq:b-squared}).
}
\added[id=AM]{%
  Third,
  domain knowledge and physics constraints could be embedded in both \gls{NN} architecture and training process~\cite{Raissi2019},
  and the coil system geometry could be extended to a generic device geometry,
thus opening up the possibility to use such a surrogate model in a generic stellarator optimization workflow.
}
\replaced[id=AM]{Fourth}{Second},
to reduce the dimensionality of the problem,
the radial dependency of the output quantities could be cast as an additional predictor,
also gaining the ability to compute analytical derivatives with respect to the radial coordinate.

\added[id=AM]{%
  Furthermore,
  broader \gls{HP} search and an ensemble of \glspl{NN} could further improve the performance over single base learner~\cite{Hansen1990},
  and optimization techniques,
  such as pruning and quantization,
  are expected to deliver improved inference times.
  Moreover,
  the results of this work suggest that the full Fourier resolution is in reach if larger \glspl{NN} and longer training time are accessible.
}

\deleted[id=AM]{The promising results of this paper show that \glspl{NN} can be used to deploy a drop-in surrogate model for \gls{VMEC},
although additional questions have to be investigated.
First,
the performance of such models to resolve both the vacuum magnetic configuration and the finite-beta effects on the equilibrium magnetic field has to be assessed by using a data set which comprises both vacuum and finite-\averagePlasmaBeta samples.
Second,
broader \gls{HP} search and an ensemble of \glspl{NN} could further improve the performance over single base learner~\cite{Hansen1990},
and optimization techniques,
such as pruning and quantization,
are expected to deliver improved inference times.
Moreover, the results of this work suggest that the full Fourier resolution is in reach if larger \glspl{NN} and longer training time are accessible.}

\deleted[id=AM]{%
Furthermore,
domain knowledge and physics constraints could be embedded in both \gls{NN} architecture and training process~\cite{Raissi2019},
and the coil system geometry could be extended to a generic device geometry,
thus opening up the possibility to use such a surrogate model in a generic stellarator optimization workflow.
}

Finally,
the use of \gls{MHD} fast surrogate models can impact multiple applications:
fast \replaced[id=AM]{Bayesian inference of plasma parameters}{\gls{BSM}} and equilibrium reconstruction workflows for intra-shot analysis\footnote{
  \replaced[id=AM]{
    If target physics quantities are not adequately reproduced by the surrogate model,
    a two-stage approach should be pursued:
    employ the model predictions to extensively explore the target input space,
    then,
    switch to a high-fidelity equilibrium computation to refine the solution.
    This may be applied to provide fast transformations for diagnostics,
    a broad exploration of the posterior probability distribution in a Bayesian framework,
    or good initial configurations for a more rapid convergence of equilibrium codes.
  }{
    The results obtained could be used to provide fast transformations for diagnostics, and may also serve to provide good initial configurations for a more rapid convergence of the equilibrium codes if they are needed.
  }
},
access to large and rich optimization spaces for present and future magnetic confinement devices,
milliseconds-range \gls{MHD} equilibrium computations for real-time plasma control,
and the generation of very large data sets of equilibrium computations necessary to investigate \gls{ML} control strategies (\eg, \glsentrylong{RL})\footnote{It is important to note that when a sufficiently large data set is accessible,
given the relative low training time,
the proposed \gls{NN} models could be trained to target specific data distributions expected for a use case application,
thus reducing the covariate shift between the training and test set.
}.

\section{Author Statement and Acknowledgement}
\label{sec:acknowledgement}

The contributions to this paper are described using the CRediT taxonomy~\cite{Brand2015}:
\begin{description}
\item[Andrea Merlo] Conceptualization, Ideas, Data Curation, Formal Analysis, Investigation, Methodology, Software, Visualization, Writing - Original Draft Preparation, Writing - Review \& Editing.
\item[Daniel Böckenhoff] Ideas, Methodology, Software, Supervision, Validation, Writing - Original Draft Preparation, Writing - Review \& Editing.
\item[Jonathan Schilling] \added[id=AM]{Data Curation, Methodology, }Software, Writing – Review \& Editing.
\item[Udo Höfel] \added[id=AM]{Ideas, Methodology, }Software.
\item[Sehyun Kwak] \added[id=AM]{Ideas, Methodology, }Software\added[id=AM]{, Writing – Review \& Editing}.
\item[Jakob Svensson] \added[id=AM]{Ideas, Methodology, }Software.
\item[Andrea Pavone] \added[id=AM]{Ideas, Formal Analysis, Methodology, }Software.
\item[Samuel Aaron Lazerson] Supervision\added[id=AM]{, Writing – Review \& Editing}.
\item[Thomas Sunn Pedersen] Conceptualization, Writing – Review \& Editing, Supervision.
\end{description}


We wish to acknowledge the helpful discussions on \gls{VMEC} and \gls{MHD} equilibrium with J. Geiger.
Furthermore,
we are indebted to the communities behind the multiple open-source software packages on which this work depends.

The data sets were generated on the \gls{MPCDF} cluster ``DRACO'',
Germany.
Financial support by the European Social Fund (ID: ESF/14-BM-A55-0007/19) and the Ministry of Education, Science and Culture of Mecklenburg-Vorpommern,
Germany via project ``NEISS'' is gratefully acknowledged.
This work has been carried out within the framework of the EUROfusion Consortium and has received funding from the Euratom research and training program 2014-2018 and 2019–2020 under Grant agreement No. 633053.
The views and opinions expressed herein do not necessarily reflect those of the European Commision.


\ifthenelseproperty{compilation}{appendix}{%
    \section{Appendix}
\label{sec:appendix}
\subsection{Hyper-parameters values}
\label{sec:hp-values}

\sisetup{round-mode=places,round-precision=1}

\Cref{tab:null-iota-hp,tab:finite-iota-hp,tab:null-surfaces-hp,tab:finite-surfaces-hp,tab:null-bcos-hp,tab:finite-bcos-hp} report the \gls{HP} values of the best performing model on each task discovered via \gls{HP} search.

\begin{table}
  \caption{
    Hyper-parameters values for the best \gls{FFFC} model on the \nullIota task.
    The float type hyper-parameters are reported with two significant digits.
  }%
  \label{tab:null-iota-hp}
  \centering
\begin{tabular}{lc}
\toprule
                            HP &            value \\
\midrule
                  dense layers &          \num{4} \\
      first layer hidden units &         \num{64} \\
           activation function &             SeLU \\
                    batch size &         \num{64} \\
                 learning rate & \num{6.5000e-04} \\
      learning rate decay rate & \num{4.0000e-01} \\
     learning rate decay steps &         \num{20} \\
   $L^2$ regularization factor & \num{1.6000e-05} \\
early stopping patience epochs &         \num{40} \\
\bottomrule
\end{tabular}

\end{table}

\begin{table}
  \caption{
    Hyper-parameters values for the best \gls{FFFC} model on the \finiteIota task.
    The float type hyper-parameters are reported with two significant digits.
  }%
  \label{tab:finite-iota-hp}
  \centering
\begin{tabular}{lc}
\toprule
                            HP &            value \\
\midrule
                  dense layers &          \num{4} \\
      first layer hidden units &        \num{128} \\
           activation function &             SeLU \\
                    batch size &         \num{32} \\
                 learning rate & \num{2.6000e-03} \\
      learning rate decay rate & \num{3.4000e-01} \\
     learning rate decay steps &         \num{20} \\
   $L^2$ regularization factor & \num{1.6000e-05} \\
early stopping patience epochs &         \num{40} \\
\bottomrule
\end{tabular}

\end{table}

\begin{table}
  \caption{
    Hyper-parameters values for the best 3D \gls{CNN} model on the \nullSurfaces task.
    The float type hyper-parameters are reported with two significant digits.
  }%
  \label{tab:null-surfaces-hp}
  \centering
\begin{tabular}{lc}
\toprule
                            HP &                 value \\
\midrule
                decoder layers &               \num{5} \\
   decoder first layer filters &             \num{112} \\
           decoder kernel size & $3 \times 3 \times 3$ \\
                decoder stride & $1 \times 1 \times 1$ \\
   decoder activation function &                  SeLU \\
               decoder dropout &      \num{2.1800e-02} \\
                    batch size &              \num{32} \\
                 learning rate &      \num{8.9000e-04} \\
      learning rate decay rate &      \num{2.5600e-01} \\
     learning rate decay steps &              \num{15} \\
   $L^2$ regularization factor &      \num{3.3500e-06} \\
early stopping patience epochs &              \num{35} \\
\bottomrule
\end{tabular}

\end{table}

\begin{table}
  \caption{
    Hyper-parameters values for the best 3D \gls{CNN} model on the \finiteSurfaces task.
    The float type hyper-parameters are reported with two significant digits.
  }%
  \label{tab:finite-surfaces-hp}
  \centering
\begin{tabular}{lc}
\toprule
                            HP &                 value \\
\midrule
                encoder layers &               \num{4} \\
   encoder first layer filters &              \num{16} \\
           encoder kernel size & $3 \times 3 \times 3$ \\
                encoder stride & $1 \times 1 \times 1$ \\
   encoder activation function &            Leaky ReLU \\
               encoder dropout &      \num{4.8000e-01} \\
                decoder layers &               \num{5} \\
   decoder first layer filters &             \num{160} \\
           decoder kernel size & $3 \times 3 \times 3$ \\
                decoder stride & $1 \times 1 \times 1$ \\
   decoder activation function &                  ReLU \\
               decoder dropout &      \num{9.9000e-02} \\
                    batch size &              \num{32} \\
                 learning rate &      \num{1.0000e-03} \\
      learning rate decay rate &      \num{3.5000e-01} \\
     learning rate decay steps &              \num{10} \\
   $L^2$ regularization factor &      \num{1.5000e-04} \\
early stopping patience epochs &              \num{45} \\
\bottomrule
\end{tabular}

\end{table}

\begin{table}
  \caption{
    Hyper-parameters values for the best 3D \gls{CNN} model on the \nullB task.
    The float type hyper-parameters are reported with two significant digits.
  }%
  \label{tab:null-bcos-hp}
  \centering
\begin{tabular}{lc}
\toprule
                            HP &                 value \\
\midrule
                decoder layers &               \num{3} \\
   decoder first layer filters &              \num{64} \\
           decoder kernel size & $5 \times 5 \times 5$ \\
                decoder stride & $1 \times 1 \times 2$ \\
   decoder activation function &                  ReLU \\
               decoder dropout &      \num{2.8000e-04} \\
                    batch size &              \num{32} \\
                 learning rate &      \num{5.2000e-04} \\
      learning rate decay rate &      \num{3.4000e-01} \\
     learning rate decay steps &              \num{25} \\
   $L^2$ regularization factor &      \num{8.2000e-05} \\
early stopping patience epochs &              \num{35} \\
\bottomrule
\end{tabular}

\end{table}

\begin{table}
  \caption{
    Hyper-parameters values for the best 3D \gls{CNN} model on the \finiteB task.
    The float type hyper-parameters are reported with two significant digits.
  }%
  \label{tab:finite-bcos-hp}
  \centering
\begin{tabular}{lc}
\toprule
                            HP &                 value \\
\midrule
                encoder layers &               \num{2} \\
   encoder first layer filters &              \num{16} \\
           encoder kernel size & $3 \times 3 \times 3$ \\
                encoder stride & $2 \times 2 \times 2$ \\
   encoder activation function &            Leaky ReLU \\
               encoder dropout &      \num{2.1600e-01} \\
                decoder layers &               \num{3} \\
   decoder first layer filters &              \num{64} \\
           decoder kernel size & $5 \times 5 \times 5$ \\
                decoder stride & $1 \times 1 \times 2$ \\
   decoder activation function &                  ReLU \\
               decoder dropout &      \num{2.1000e-02} \\
                    batch size &              \num{96} \\
                 learning rate &      \num{9.5000e-04} \\
      learning rate decay rate &      \num{3.2800e-01} \\
     learning rate decay steps &              \num{15} \\
   $L^2$ regularization factor &      \num{4.6000e-04} \\
early stopping patience epochs &              \num{45} \\
\bottomrule
\end{tabular}

\end{table}

\cleardoublepage

}{}

\ifthenelseproperty{compilation}{acknowledgement}{%
    \chapter{Acknowledgements}\label{sec:acknowledgement}
This work was not possible without the help of many people.
\TODO{Thanksgiving}
\par
\TODO{Maybe the following applies.}
This work has been carried out within the framework of the EUROfusion Consortium and has received funding from the Euratom research and training programme 2014-2018 and 2019-2020 under grant agreement No 633053. 
The views and opinions expressed herein do not necessarily reflect those of the European Commission.

}{}

\ifthenelseproperty{compilation}{affidavit}{%
    \thispagestyle{empty}
\ifthenelseproperty{compilation}{clsdefineschapter}{%
	\ifKOMA
		\addchap[Statutory declaration]{Statutory declaration}
	\else
    	\chapter[Statutory declaration]{Statutory declaration}
    \fi
}{%
	\ifKOMA
		\addsec[Statutory declaration]{Statutory declaration}
	\else
    	\section[Statutory declaration]{Statutory declaration}
    \fi
}
I hereby declare in accordance with the examination regulations that I myself have written this document, that no other sources as those indicated were used and all direct and indirect citations are properly designated, that the document handed in was neither fully nor partly subject to another examination procedure or published and that the content of the electronic exemplar is identical to the printing copy.

\Signature{\getproperty{document}{location}}{\textsc{%
    \ifluatex
        \IfSubStr{\getproperty{author}{firstname}}{TODO}{%
            \getproperty{author}{firstname}
        }{%
            \FirstWord{\getproperty{author}{firstname}}
        }
    \else
        \getproperty{author}{firstname}
    \fi
    \getproperty{author}{familyname}}}

}{}

\ifthenelseproperty{compilation}{lof}{%
    \disabledprotrusion{\listoffigures}
}{}

\ifthenelseproperty{compilation}{lot}{%
    \disabledprotrusion{\listoftables}
}{}

\ifthenelseproperty{compilation}{lol}{%
    \disabledprotrusion{\lstlistoflistings}
}{}

\ifthenelseproperty{compilation}{listofpublications}{%
    \begin{refsection}
	\setboolean{bibCV}{true}
	\newrefcontext[sorting=ndymdt]
	\nocite{*}
    \ifthenelseproperty{compilation}{clsdefineschapter}{%
		\ifKOMA
			\addchap[Publications as first author]{Publications as first author}\label{sec:publications_as_first_author}
		\else
	    	\chapter[Publications as first author]{Publications as first author}\label{sec:publications_as_first_author}
	    \fi
    }{%
		\ifKOMA
			\addsec[Publications as first author]{Publications as first author}\label{sec:publications_as_first_author}
		\else
	    	\section[Publications as first author]{Publications as first author}\label{sec:publications_as_first_author}
	    \fi
    }
	\printbibliography[
						keyword=firstAuthor,
						keyword=refereed,
						heading=subbibliography,
						title={Peer-reviewed articles},
						resetnumbers=true
					]
	\setboolean{bibCV}{false}
\end{refsection}
\begin{refsection}
	\setboolean{bibCV}{true}
	\newrefcontext[sorting=ndymdt]
	\nocite{*}
    \ifthenelseproperty{compilation}{clsdefineschapter}{%
		\ifKOMA
			\addchap[Publications as coauthor]{Publications as coauthor}\label{sec:publications_as_coauthor}
		\else
	    	\chapter[Publications as coauthor]{Publications as coauthor}\label{sec:publications_as_coauthor}
	    \fi
    }{%
		\ifKOMA
			\addsec[Publications as coauthor]{Publications as coauthor}\label{sec:publications_as_coauthor}
		\else
	    	\section[Publications as coauthor]{Publications as coauthor}\label{sec:publications_as_coauthor}
	    \fi
    }
	\setboolean{isCoauthorList}{true}
	\printbibliography[
						keyword=coAuthor,
						keyword=refereed,
						heading=subbibliography,
						title={Peer-reviewed articles},
						resetnumbers=true
					]
	\setboolean{bibCV}{false}
\end{refsection}

}{}

\ifthenelseproperty{compilation}{glossaries}{%
    \ifthenelseproperty{compilation}{acronyms}{%
	\printglossary[type=\acronymtype,style=mcoltree]%
}{}%
\ifthenelseproperty{compilation}{los}{%
	\setlength\extrarowheight{5pt}%
	\printglossary
			[
				title=List of Symbols,
				type=symbols,
				style=customListOfSymbols,
			]
	\setlength\extrarowheight{0pt}%
}{}%

}{}

\ifthenelseproperty{compilation}{bibliography}{%
    \printbibliography
}{}

    \typeout{----- END OF DOCUMENT -----}

\end{document}
\typeout{----- END OF MAIN -----}